Come alla Corte di Federico Secondo

# Napoli Tokyo Napoli

## Ovvero
## L'impensabile viaggio della biomeccanica del judo.

**Attilio Sacripanti**

**28/10/2010**

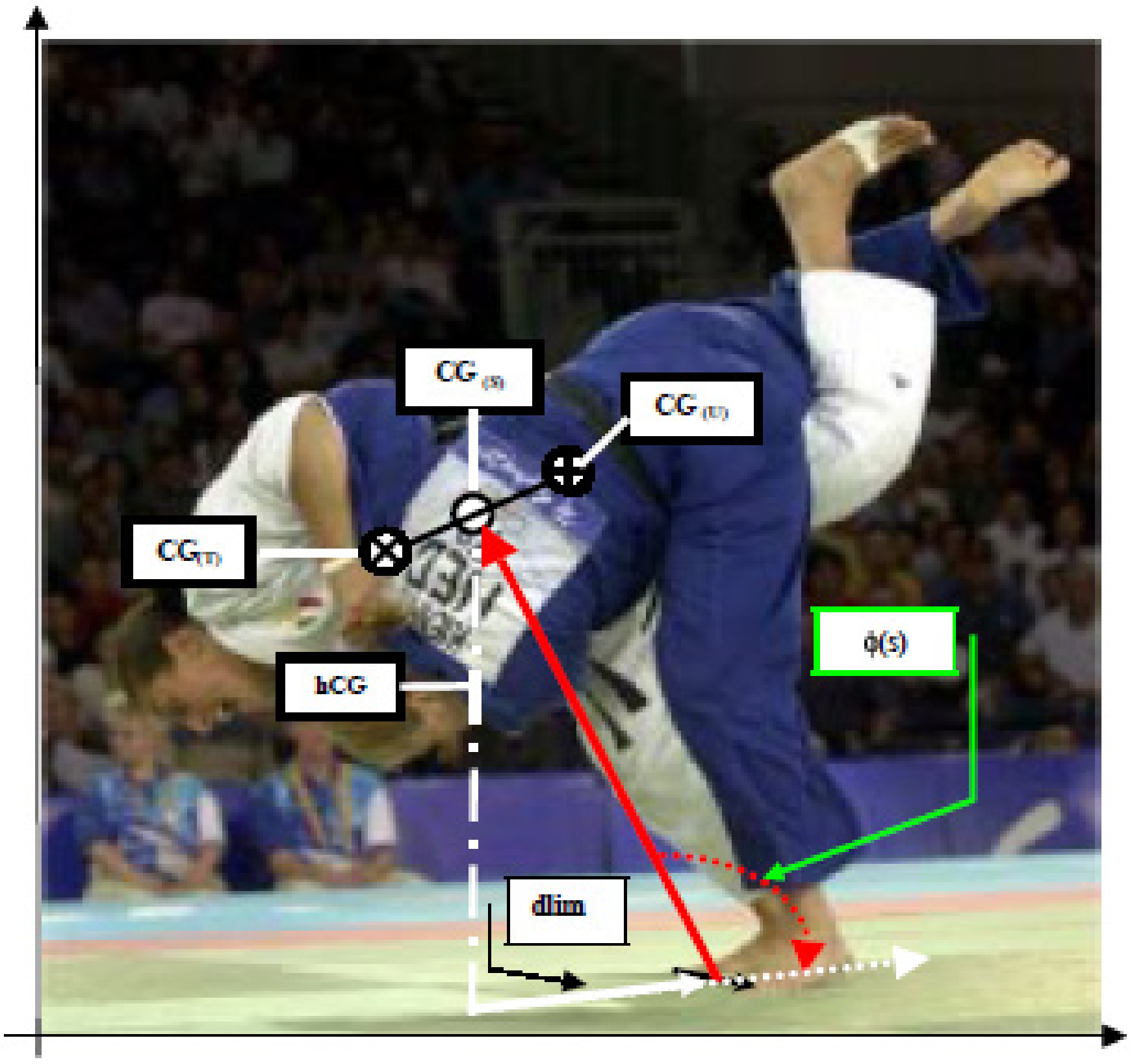

*In questa cronaca di viaggio si parla di un numero ricorrente il 28, di due grandi uomini e del piccolo nodo che li congiunse inconsapevolmente tre secoli dopo.*

## Indice

Attilio Sacripanti

# Napoli Tokyo Napoli

Ovver

L'impensabile viaggio della biomeccanica del judo.

*In questa cronaca di viaggio è mia intenzione parlare di un numero quasi ricorrente, il 28 , di due grandi uomini e del piccolo nodo che li congiunse inconsapevolmente tre secoli dopo.*

---

**Introduzione: l'Itinerario di viaggio**

Questo viaggio parte dalla nostra cara Napoli capitale di cultura e città europea per molti secoli.
Infatti nell'anno 1608 nasceva in Castel Nuovo oggi più noto come Maschio Angioino
Giovanni, Francesco, Antonio, Alfonso Borelli, (fisiologo fondatore della scuola Iatromeccanica, l'antenata della biomeccanica), fisico, matematico, astronomo, vulcanologo e forse anche medico allievo di Castelli e condiscepolo di Torricelli l'esempio paradigmatico delle versatilità che spesso alberga nei Napoletani più illuminati.
Il viaggio proseguirà giungendo alla prefettura di Hyogo in Giappone la cui capitale è Kobe, dove nel 1860 nasceva il secondo grande personaggio del nostro viaggio il Dr. Jigoro Kano laureato in scienze politiche ed economia, di fatto uno dei più importanti e profondi educatori del Giappone moderno.
Questo viaggio terminerà con un inusitato ritorno nella nostra città, Napoli, ove l'eredità di questi due grandi uomini fu congiunta nel 1985 in quella disciplina che oggi viene correntemente chiamata Biomeccanica del Ju-Do e che, rappresenta in modo sintetico la capacità interpretativa dei moderni concetti scientifici nei confronti degli intriganti sistemi complessi non lineari che presentano aspetti "chaotici" e "frattali".

---

## *NAPOLI- Alfonso Borelli: l'eclettico padre della Biomeccanica e Biofisica moderna.*

Il nostro uomo nacque a Napoli il **28** gennaio del 1608, primo di tre figli, di padre Spagnolo appartenente alla guarnigione di Castel Nuovo ( l'attuale Maschio Angioino) non si hanno molte notizie della sua fanciullezza, ma sicuramente il nostro giovane fu istruito da Tommaso Campanella che in quel periodo era prigioniero proprio in Castel Nuovo a Napoli.
La familiarità con il Campanella è confermata, anche, dalla notizia che Il fratello più giovane di Borelli, Filippo accompagnò Tommaso Campanella quando quest'ultimo dovette fuggire a Parigi nel 1634 ed assistendolo anche nella pubblicazione dei suoi lavori.
Ritroviamo, comunque Borelli a Roma negli anni successivi al 1628 allievo del padre benedettino Benedetto Castelli matematico ed idraulico, allievo di Galileo Galilei, con un illustre collega e condiscepolo il noto fisico Evangelista Torricelli.
Da Roma Alfonso Borelli si trasferì a Messina dove ebbe la cattedra di matematica nella locale Università, successivamente passò per Bologna dove impressionò per la sua grande preparazione il noto matematico, Bonaventura Cavalieri.
Borelli prima di ritornare a Messina si recò a Padova ed a Venezia.
Nel 1643 la sua reputazione era tanto cresciuta per l'intera penisola, sebbene non avesse ancora pubblicato nulla come autore singolo, che le sue opinioni in matematica, fisiologia ed astronomia erano fortemente condivise. Egli fu considerato il miglior matematico Italiano dopo Cavalieri e come suo successore Borelli attendeva la cattedra a Bologna, che però non gli fu mai conferita.
Ottenne invece la cattedra di matematica presso l'Università di Pisa, cattedra già di Galileo Galilei. Grande fu la sua attività in questa città, infatti appena giuntovi istituì, presso casa sua, un laboratorio anatomico con numerosi studenti ed iniziò una fertile collaborazione con un suo allievo, già laureato in Bologna, il grande medico Marcello Malpighi; collaborò anche con un altro famoso medico Lorenzo Bellini allo studio del rene pubblicando il *De renum usu judicum* nel 1664.
Usando il microscopio, come il Malpighi, Borelli individuò per primo i globuli rossi nel flusso sanguigna.
A lui l'amico Malpighi inviò sotto forma di lettere il *De Pulmonibus* che contenevano la relazione della scoperta della circolazione capillare . il *De Pulmonibus* fu pubblicato come testo in Bologna nel 1662.
Borelli partecipò assiduamente come componente di spicco alla famosissima *Accademia del Cimento*, il cui motto "Provando e Riprovando" metteva in risalto la necessità di una verifica rigorosamente sperimentale dei principi della filosofia naturale sostenuti fino ad allora, soprattutto sulla base dell'autorità di Aristotele.
Pur mancando un formale sistema di iscrizione all'Accademia, parteciparono con continuità ai suoi lavori Francesco Redi, Lorenzo Magalotti, che svolse le funzioni di Segretario, Vincenzo Viviani, il nostro Giovanni Alfonso Borelli e Carlo Renaldini. Durante le riunioni, che si tenevano solitamente in Palazzo Pitti, furono compiute numerose esperienze, soprattutto di termometria, barometria e pneumatica, utilizzando strumenti appositamente costruiti.
Il ritratto del Borelli mostrato in Fig.1 è preso da un affresco di Gaspero Martellini raffigurante una esperienza, riportata nei *Saggi di naturali esperienze fatte nell'Accademia del Cimento*.
L'esperienza fu effettuata per stabilire se il 'freddo' si rifletteva negli specchi come 'il caldo' delle braci accese e la luce. Gli accademici utilizzarono specchi concavi metallici per riflettere il calore. All'esperienza parteciparono Giovanni Alfonso Borelli, che nell'affresco copre lo specchio, Vincenzo Viviani che annuncia il risultato dell'esperimento, Francesco Redi che indica come il liquido termometrico risalga nel termometro il segretario dell'Accademia, Lorenzo Magalotti, prende nota dei risultati, mentre il Granduca Ferdinando II è seduto di fronte al principe Leopoldo de' Medici che, in piedi, assiste all'esperimento

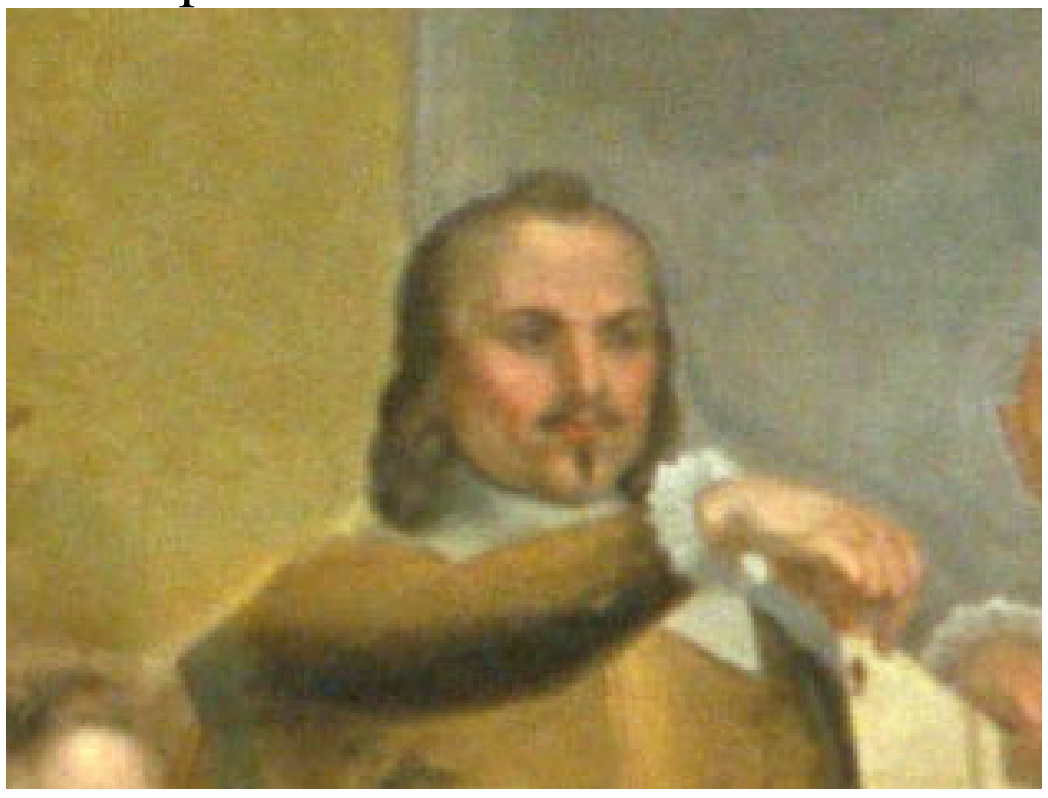

***Fig.1 Giovanni Alfonso Borelli***

Nei dieci anni trascorsi in Toscana il Borelli costituì uno dei punti di riferimento di tutta l'attività scientifico-sperimentale organizzata dal Principe Leopoldo de' Medici (1617-1675). Non vi fu praticamente esperimento dell'Accademia del Cimento che non recasse il suo apporto, pur trovandosi spesso in contrasto con Carlo Renaldini che Borelli definiva il "Simplicio" del Cimento.
Fu un periodo di grande attività e di successi continui dopo la pubblicazione nel 1649 del trattato "De *la cagione delle febbri maligne"* egli intraprese due grandi lavori, iniziò a lavorare ad un compendio dei quattro libri superstiti dei *Conici* di Apollonio (seconda metà III sec. a.C.), fino ad allora sconosciuti, in collaborazione con uno scolaro Maronita Abramo Ecchellensis, traducendo un manoscritto Arabo che si trovava nella biblioteca medicea; opera che avrebbe pubblicato molti anni più tardi (Roma, 1679) con il titolo :" *Apollonii Pergaei Conicorum libri v., vi. et vii";* mentre aveva già intrapreso la revisione degli *Elementi* di Euclide (sec. IV a.C.) – che stampò nel 1658 col titolo "*Euclides restitus, sive prisca geometriae elementa, brevius, & facilius contexta"* ed in cui presentava una nuova versione del postulato delle parallele.
Ma il contributo più importante dato dal Borelli durante il soggiorno toscano fu senza dubbio il lavoro compiuto sui satelliti di Giove, le *Theoricae Mediceorum Planetarum ex causis physicis deductae* (Firenze, 1666), un'opera destinata ad inserirsi efficacemente nelle discussioni cosmologiche europee. In essa in un quadro di riferimento dichiaratamente copernicano, il Borelli giunse ad ipotizzare un moto planetario ellittico animato da due forze: quella centrifuga e quella di attrazione solare, la cui composizione permetterebbe l'equilibrio dei pianeti nell'etere.
Newton ebbe conoscenza del lavoro di Borelli apprezzandone sia l'originalità dell'approccio che la complessità del moto ellittico risultante,e lo cita chiaramente nei *Principia*.
Ma ben oltre, il Borelli, sotto lo pseudonimo di Pier Maria Mutoli, pubblicò nel 1665 "*Del movimento della cometa di Decembre 1664*" in cui l'autore , fu uno dei primi astronomi a proporre , sulla base di dati sperimentali e calcoli che, le comete si muovevano lungo orbite paraboliche.
Infatti, i lavori precedenti, tra cui quelli di Keplero, avevano sempre considerato le comete come visitatori transitori del sistema solare che si muovevano lungo traiettorie rettilinee.
Nel 1667 il Borelli si congedò dai Medici e tornò a Messina; nello stesso anno faceva stampare a Bologna il "*De vi percussionis"*, che raccoglieva, ampliandole, le ricerche fisiche effettuate del Cimento, ma su questo testo torneremo poi.
Tornato a Messina nel Maggio 1669 Borelli fu invitato da Henry Oldenburg a fornire alla Royal Society di Londra uno stato dello sviluppo scientifico in Sicilia ed una cronaca scientifica dell'eruzione dell' Etna; Borelli scrisse così quello che è considerato come il primo trattato di vulcanologia moderna *Historia et meteorologia incendii Aetnaei anni 1669* e ne inviò alla Royal Society 16 copie insieme a diverse copie del *De motionibus naturalibus a gravitate pendentibus'* .
Nella sua esistenza partecipò instancabilmente ed attivamente alla vita scientifica del nostro paese fu membro dell'Accademia della Fucina a Messina , dell'Accademia del Cimento in Firenze, come ricordato, ed infine dell'Accademia degli investiganti in Napoli.
Pochi anni dopo fuggì da Messina, per cause politiche oscure durante l' insurrezione antispagnola. Ritornò così a Roma, dove fu, tra l'altro, uno dei membri fondatori dell'Accademia dell'Esperienza patrocinata dalla Regina Cristina di Svezia. Purtroppo le sue condizioni finanziarie sempre più precarie lo costrinsero nel 1677 ad accettare l'ospitalità dei Padri Scolopi.
Visse gli ultimi anni dando lezioni di matematica ad alcuni novizi dell'Ordine ed ultimando quello che è considerato il suo capolavoro, il *De motu animalium* pubblicato postumo dalla Regina Cristina. Fig 2.

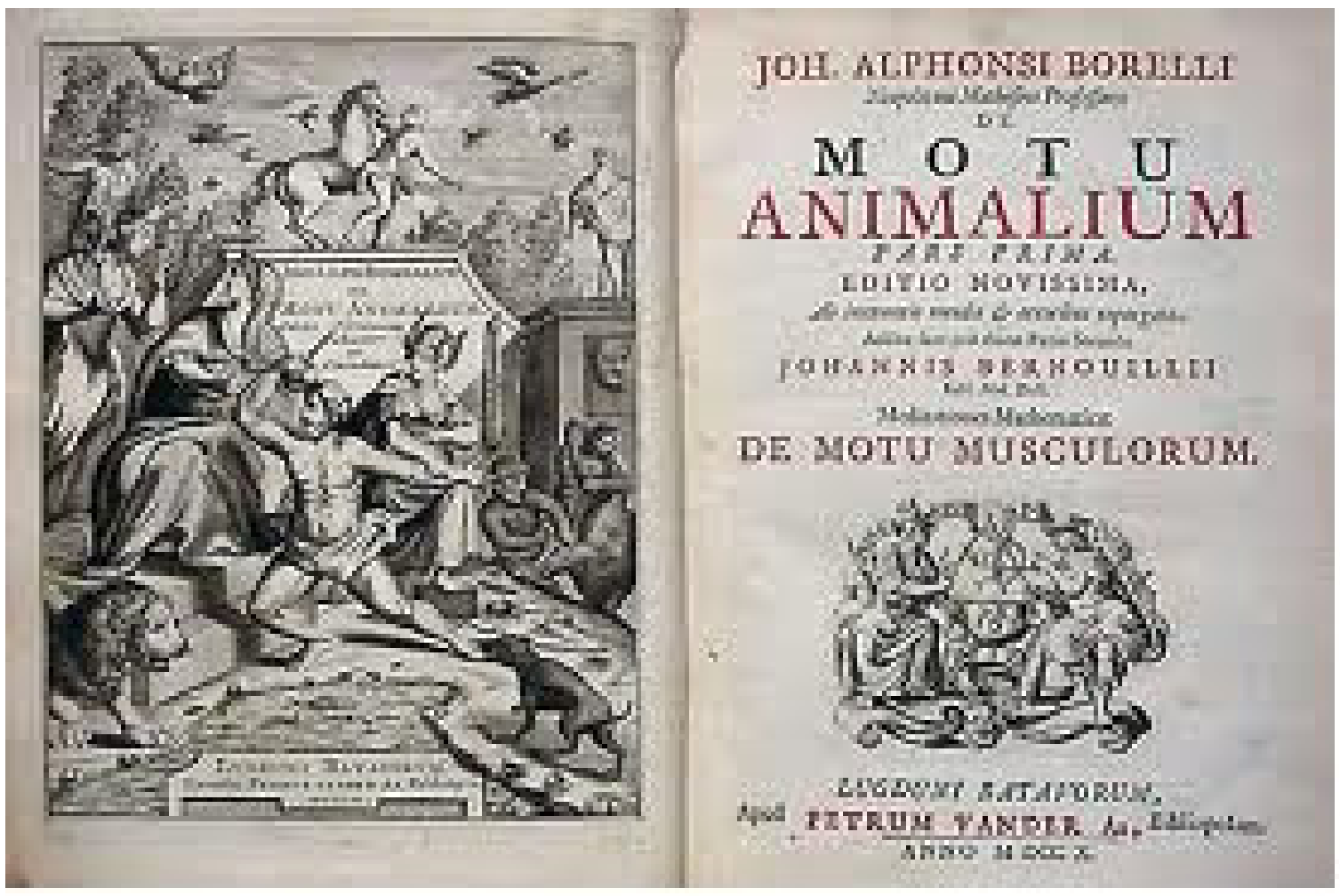

***Fig. 2 Frontespizio del De Motu Animalium conservato all'Università di Ghent ( Olanda)***

In esso  Borelli cercò di spiegare il movimento dei corpi animali e dell'uomo , il moto muscolare e le altre funzioni del corpo in termini di meccanica. I suoi studi fisiologici erano basati per la prima volta su solidi principi meccanici.
Il suo lavoro incluse l'analisi dei muscoli e l'illustrazione matematica dei movimenti come: il salto, il nuoto, ecc. Egli cercò di chiarire le ragioni della fatica muscolare, spiegò le secrezioni degli organi e fece interessanti ipotesi sul concetto di dolore.
Nell'opera in due volumi , Borelli trattò nel primo i moti esterni degli uomini e degli animali come: locomozione, corsa, salto, volo e nuoto dei pesci; il secondo volume invece fu dedicato allo studio dei moti interni ( oggi diremo fisiologia) come respirazione, circolazione sanguigna, funzione dei reni, ecc.
Uno degli aspetti biomeccanici più intriganti notati dal Borelli  fu che i muscoli agiscono con bracci di leva molto corti, in modo tale che l'articolazione interessata trasmetta una forza maggiore del peso che deve spostare. Borelli  superò con l'evidenza sperimentale  il vecchio concetto dell'azione muscolare che affermava, che i lunghi bracci delle leve ossee permettevano ai muscoli  deboli di muovere oggetti pesanti.  Infatti egli scrive " *Galeno  afferma che un tendine è come una leva Egli infatti ritiene conseguentemente che la piccola forza di un animale non può muovere un peso grande.  Questa opinion generale non è condivisa dalle mie esperienze, ma sorprendentemente non è stata messa in dubbio da alcuno.  Infatti appare stupido pensare ad una macchina che usi una forza grande per muovere carichi leggeri, o pensare di usare una macchina non per risparmiare forza ma per consumarla. Questo può apparire strano ed illogico e contro il buon senso, sono d'accordo, ma io posso convincentemente dimostrare che ciò che dico avviene all'interno del corpo e che, mi sia permesso, i sostenitori delle argomentazioni opposte errano*."
Questo aspetto della biomeccanica muscolare non fu compreso addirittura fino al 1935 anno in cui Friedrich Pauwels lo riscoprì (di nuovo!!).
Le immagini che seguono rappresentano una misura della sua eredità come padre della Biomeccanica Moderna e della Biofisica.
In esse Borelli applica le leggi della geometria Euclidea e della meccanica alle sue osservazioni sul moto dei muscoli, alla ricerca del baricentro del corpo umano. In questo testo egli descrive correttamente la contrazione cardiaca derivante da stimolazioni elettriche e postula che il moto dei muscoli derivi da reazioni chimiche a livello cellulare.
Quanta capacità di analisi e di deduzione è presente in questo suo ultimo capolavoro interdisciplinare, imperituro esempio, per i posteri, della potenza investigativa del metodo sperimentale Galileano, capace di fugare sovrastrutture e false credenze con l'evidenza della valutazione analitico -sperimentale.

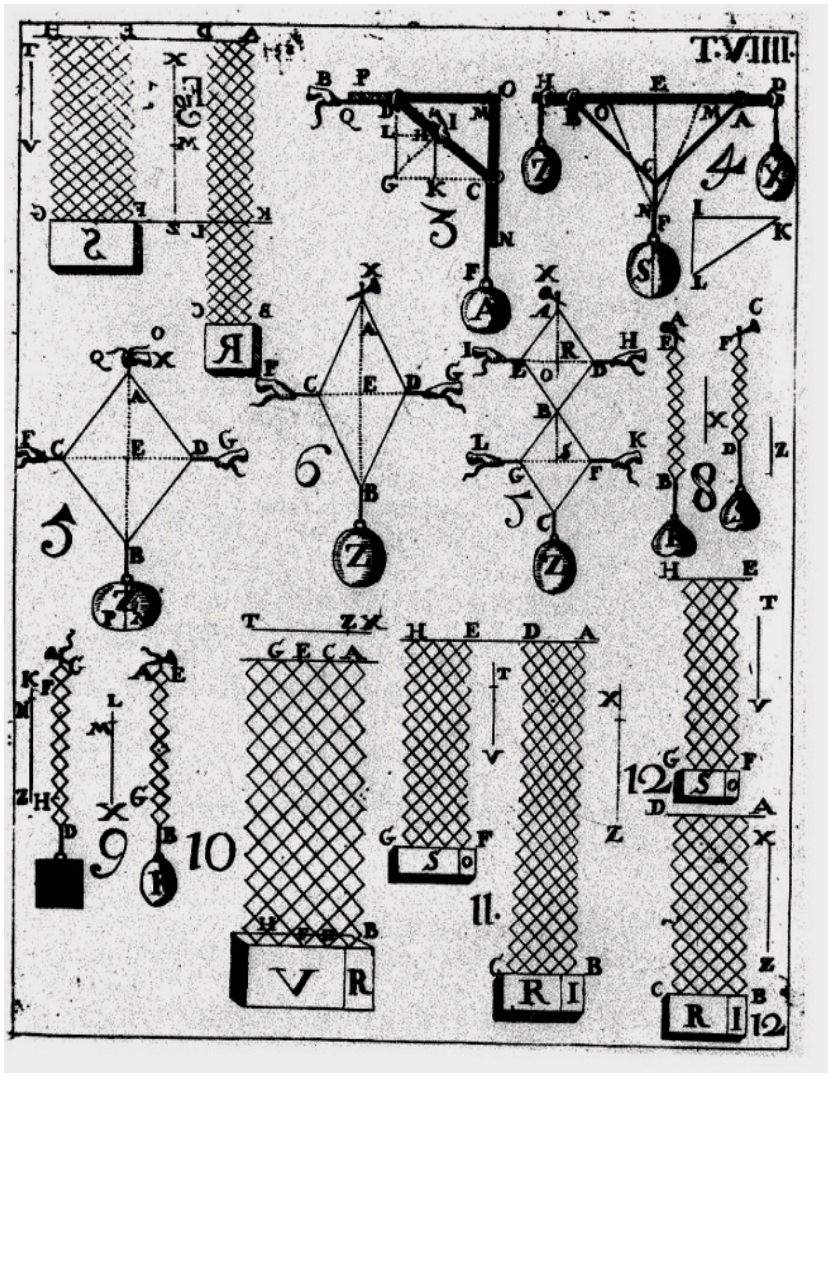

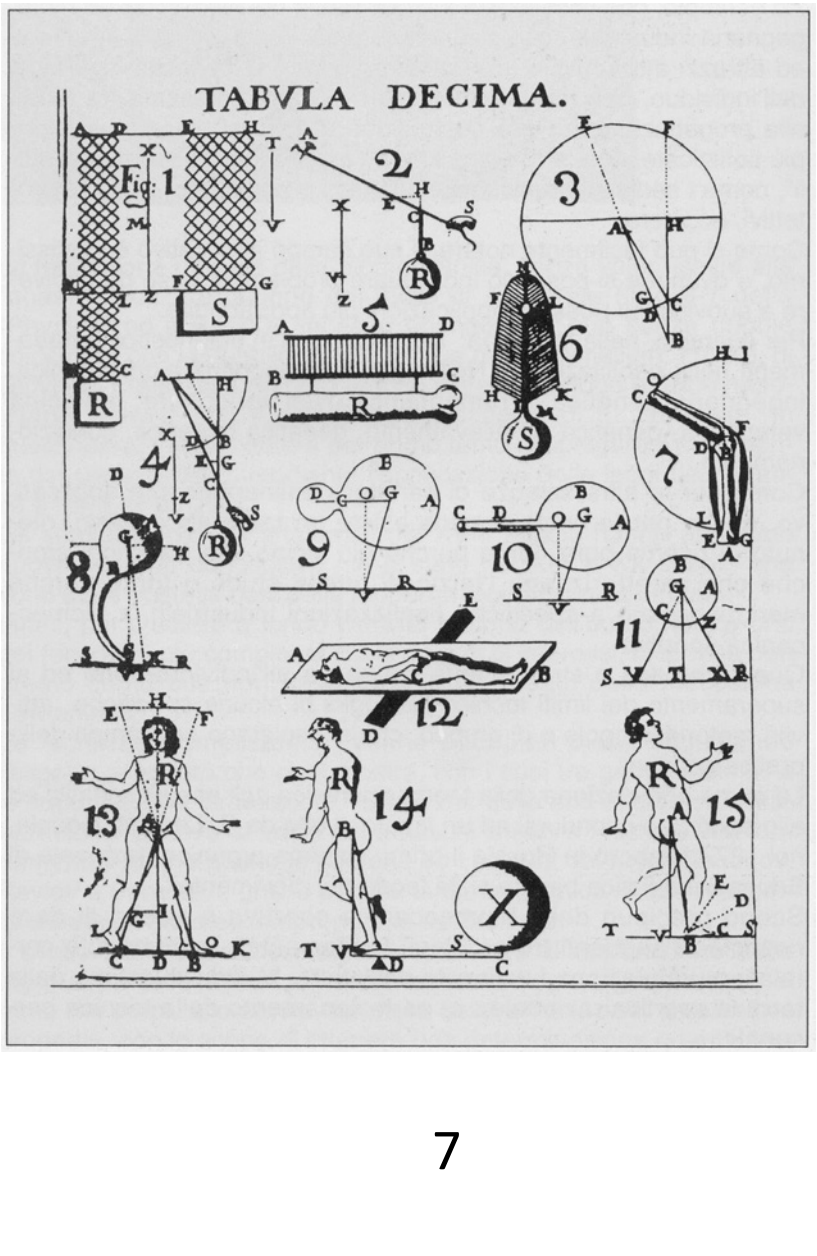

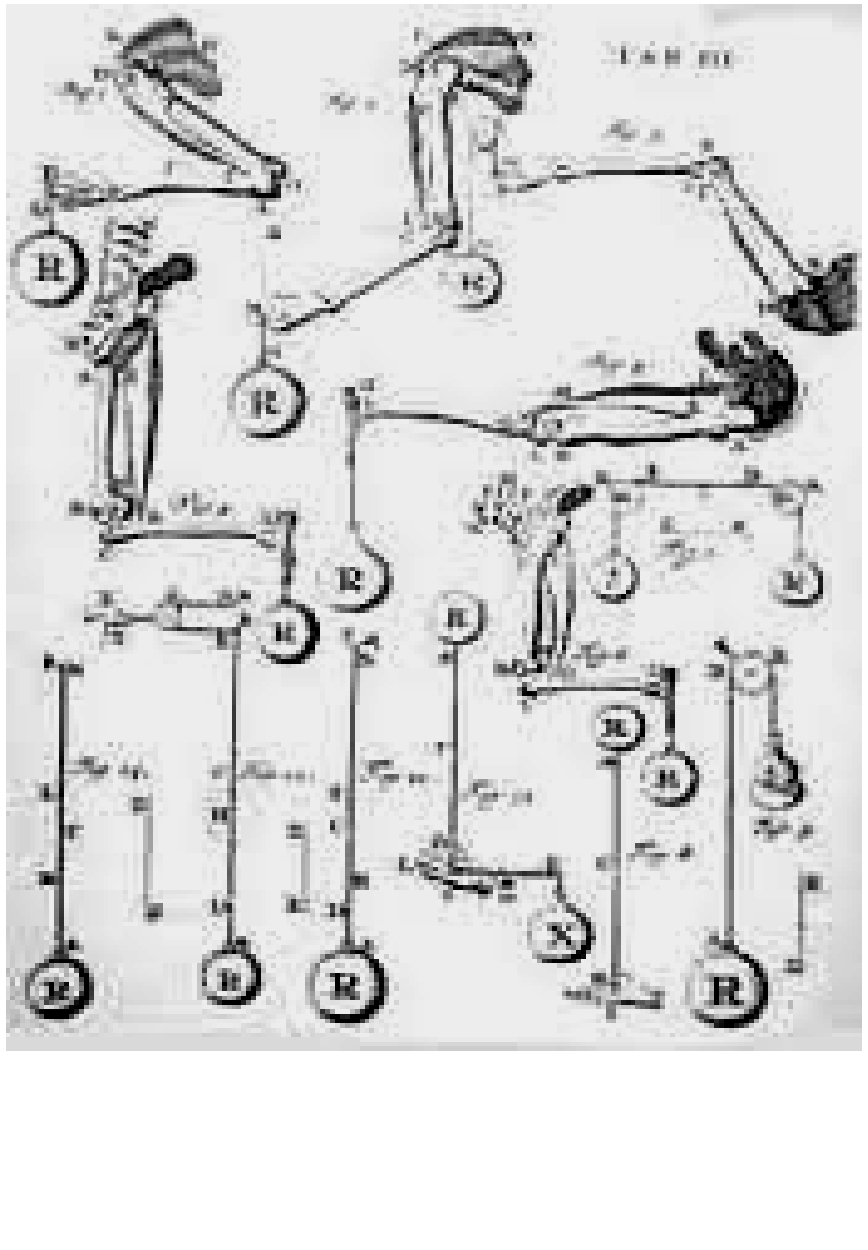

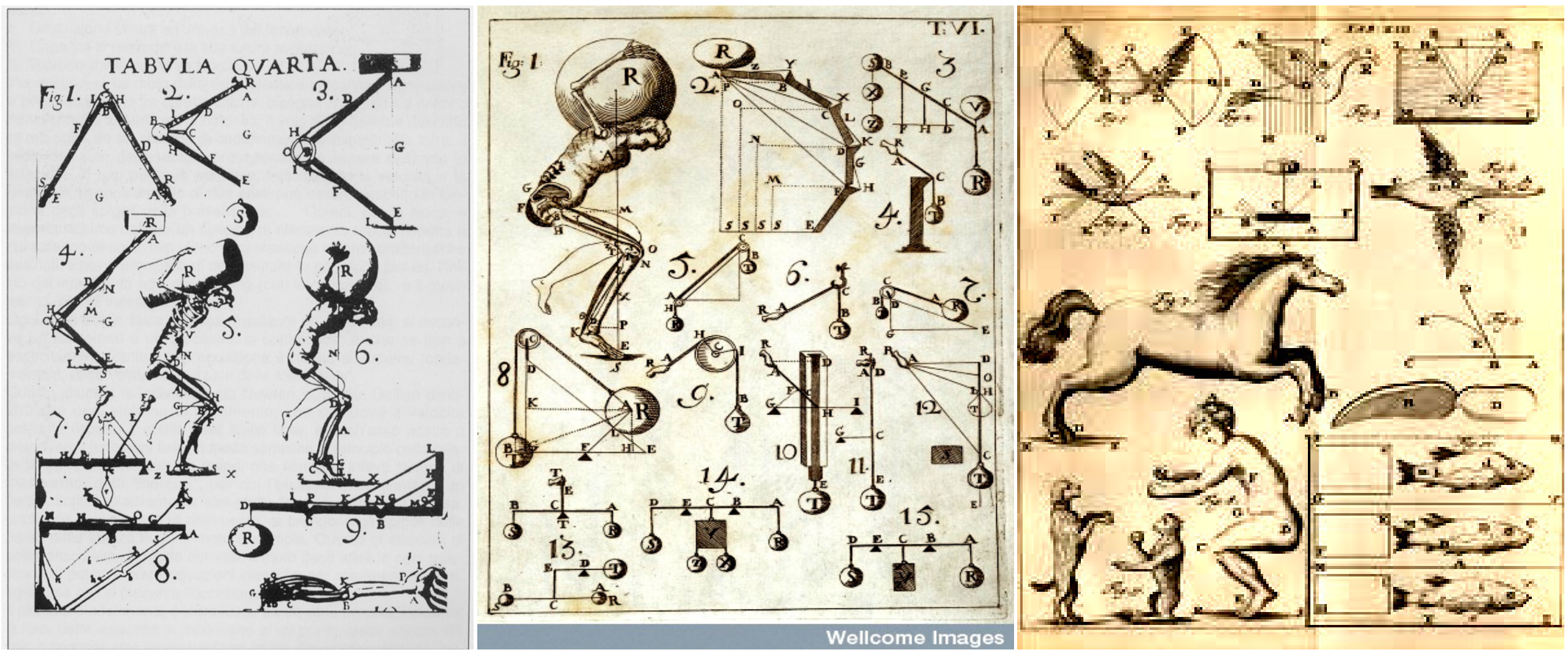

***Figg 3-9 Tavole dal De Motu Animalium di A. Borelli***
***indicanti la contrazione muscolare, il baricentro del corpo umano, lo studio delle catene cinetiche, studi sul volo ed analisi comparata.***

## *TOKYO - Jigoro Kano: un'audace ipotesi, la Motricità come mezzo d'educazione*

***Nulla sotto il sole è più importante dell'educazione. L'insegnamento di una persona virtuosa può influenzarne molte altre. Ciò che è stato profondamente compreso da una generazione può essere tramandato a centinaia di generazioni successive.*** "Jigoro Kano da un discorso del 1934"

Jigoro Kano nacque il 28 ottobre 1860 ( un altro 28!!!) nella cittadina di Mikage, che si affaccia sul mare presso Kobe (oggi è stata inglobata nella città). Il padre di Kano, Jirosaku (Kano), fu un sacerdote Shinto, ma anche un importante ufficiale governativo incaricato di una ditta di trasporti per spedizioni navali oltre oceano, mentre la madre Sadako Kano figlia più grande di una famiglia di industriali nel campo del sakè , senza discendenza maschile.
Pertanto per preservare il casato della moglie il padre Jirokasu accettò di divenire figlio adottivo della famiglia e tramandarne il cognome.
Jigoro Kano fu il terzo ed ultimo figlio della sua famiglia, egli, morta la madre, si trasferì a Tokyo con la famiglia nel 1871. La restaurazione Meji avvenuta nel 1868 aveva prodotto grandi mutamenti sociali in Giappone, tutti gli intellettuali erano bramosi delle novità culturali europee, mentre nuovi venti politici soffiavano per la prima volta in Giappone, con l'idea di democrazia e proprio in questo periodo unico nella storia del Giappone il giovane Kano si tuffò nella cultura e nella vita effervescente che vibrava a Tokyo.

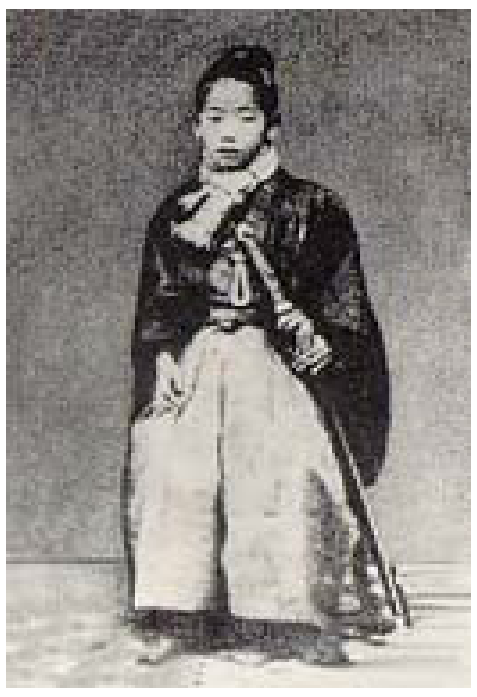

***Fig.10 il giovane Kano a 10 anni***

Nel 1877 entrò nell'università di Tokyo, la prima reimpostata secondo criteri occidentali, ciò gli consentì di sfuggire al controllo del padre e di dedicarsi allo studio del jujutsu. Kano, con molte difficoltà poiché anche in una città come Tokyo era difficile trovare un dojo, riuscì, grazie all'aiuto di Teinosuke Yagi,un anziano maestro non più praticante, ad iscriversi alla scuola del maestro Fukuda di Tenshin shin'yo, il quale restò ammirato dalla dedizione del suo giovane allievo. Purtroppo anche Fukuda era anziano e, deceduto questi, Kano dovette cercare un nuovo maestro; lo trovò in Mataemon Iso, anch'egli di Tenshin shin'yo, con il quale completò lo studio di questa Ryu (scuola), ricevendo il grado di Shian (maestro), nonché il libro segreto che gli fu lasciato in eredità.
A questo punto Jigoro Kano iniziò lo studio di un'altra scuola di jujitsu, questa volta Kito, che apprese sotto la guida di Ikubo Tsunetoshi. Questa scuola era famosa per le sue tecniche di proiezione dell'avversario e per praticare il randori (pratica libera), a differenza della quasi totalità delle altre scuole che fondavano l'insegnamento attraverso i Kata (forme preordinate).
La sua dedizione e l'impegno profuso lo portarono a conquistare il grado Shian anche in questo stile del tutto diverso dal precedente. Lo studio del jujitsu non gli impedì comunque di laurearsi in Scienze Politiche ed Economiche nel 1881. Tentò la vita politica che abbandonò presto per intraprendere gli studi di Estetica e Morale ed anche di Calligrafia. Jigoro Kano è conosciuto e ricordato in Giappone non come il fondatore di una scuola di jujitsu o di uno sport, ma come un grande educatore.
Questo processo si sviluppò consapevolmente in lui che, propose come ipotesi innovativa ed audace per i suoi tempi : ***la Motricità come mezzo di educazione* ,** ma la motricità di Kano non era semplice movimento degli arti, o qualcosa di simile all'attuale wellness, ma qualcosa di più profondo e complesso.
Infatti mentre progrediva nel suo studio del ju jitsu, intuì che le tecniche erano molto razionali ed armoniche (dal punto di vista della dinamica del movimento e anche in termini di relazione con il corpo) e concluse che tale razionalità ed armoniosità sarebbero state appropriate e molto vantaggiose per educare i giovani del Giappone, che a quel tempo si stava aprendo alle culture occidentali sviluppandosi sia culturalmente sia socialmente.
Collegando inoltre questi principi, ai principi etici Confuciani Kano organizzò i tre aspetti dell'educazione - fisico, intellettuale e morale- in un metodo educativo completamente nuovo.
Decise di rinominare le arti marziali tradizionali conosciute come "jujutsu" ( le tecniche della cedevolezza) in "judo" ( la via della cedevolezza) via intesa in senso morale.
In traslato, potremo così ritenere che per Kano le antiche scuole di Ju Jitsu dovevano esser considerate come: l'arte dell'auto-protezione, mentre il suo Ju Do come l'arte dell' auto-perfezione.
Si trattava di un'innovazione di grandissima importanza nella storia dell'educazione del Giappone e divenne parte della locale teoria dell'educazione.
Con l'intento di adottare il più ampiamente possibile tale pensiero filosofico nell'ambito dell'educazione accademica, Kano si rivolse a Sasaburo Takano, la figura più importante nel mondo del Kendo e ad altri e, come avvenne per il Judo, anche il tradizionale "kenjutsu" (tecniche della spada) cambiò il nome in "kendo" (via della spada) e l'allenamento divenne un mezzo per l'educazione fisica e lo sviluppo mentale e morale.
Nel 1882, a soli 22 anni aprì il suo primo dojo, di soli 12 tatami, ( ogni tatami, così si chiamano i materassini su cui si pratica il Judo misura 1 metro per 2) in una saletta del Tempio di Eisho nel quartiere Shimoya di Tokyo.

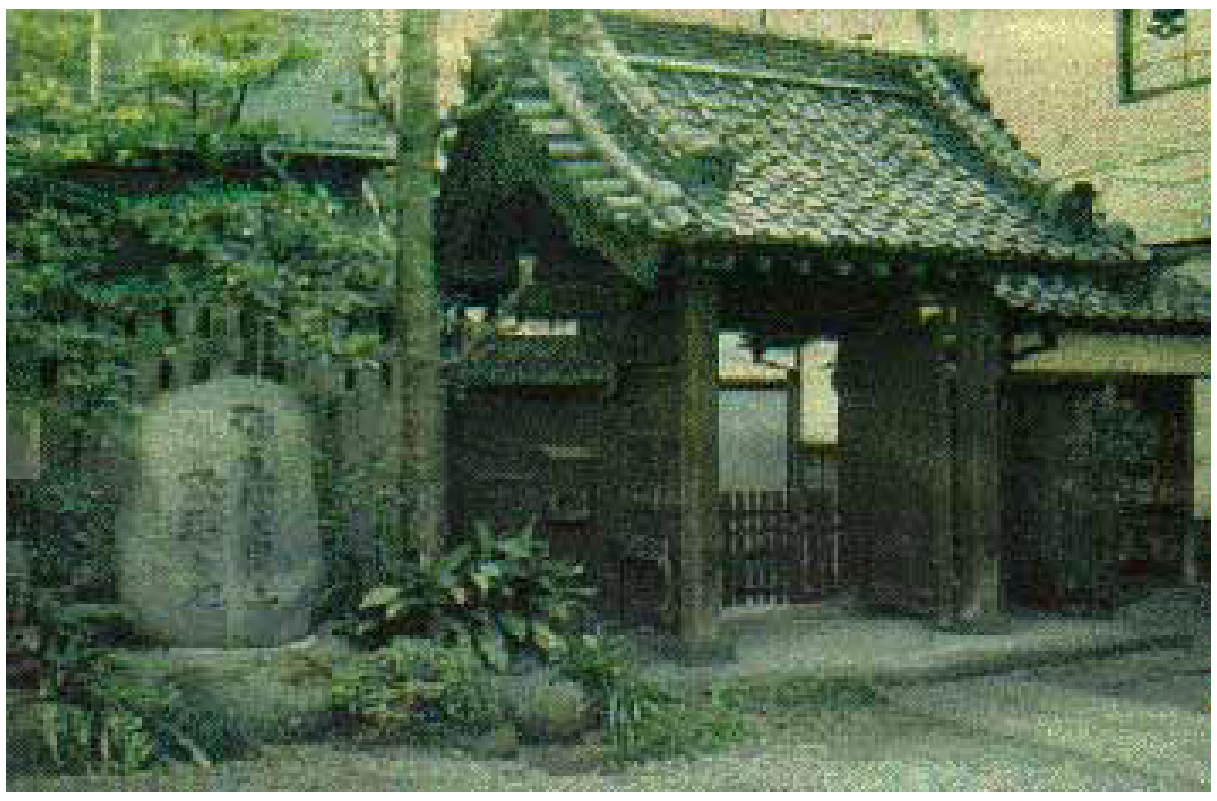

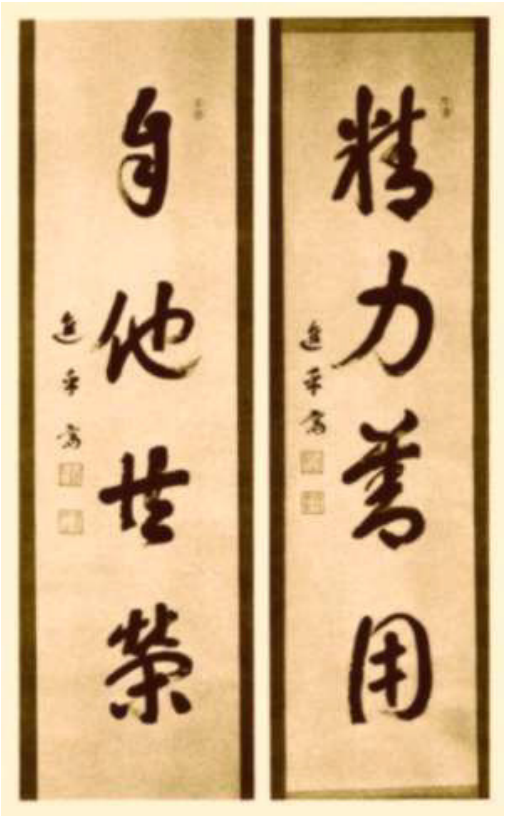

***Fig. 11 12 il tempio di Eisho nel quartiere Shimova Tokyo e due calligrafie di Kano rispettivamente da sinistra a destra Jita Kyoei ( prosperità e mutuo benessere ) e seiryoku zenyo ( massima efficacia )***

Con l'aiuto di soli nove discepoli fondò il Kodokan (il luogo dove apprendere la "Via").
Qui fece nascere il suo metodo, chiamato "Judo" (via della cedevolezza), nel quale fece convergere i metodi delle antiche scuole di arti marziali associandoli al concetto dell' ottenimento del miglior risultato col minimo sforzo, formando una disciplina efficace tanto per il fisico quanto per la mente.
Già dall'inizio come ricordato precedentemente, egli suddivise culturalmente il Judo in tre parti, *rentai-ho*, *shobu-ho*, e *shushin-ho*.
*Nel Rentai-ho si* intende l'aspetto del Judo come ginnastica o esercizio fisico, mentre *Shobu-ho* si riferisce al suo aspetto di arte Marziale. *Shushin-ho* è riferito invece alla parte filosofica e morale in cui si coltiva la virtù e l'armonia e si studia l'applicazione del judo alla vita quotidiana.

***Judo risultava quindi nell'idea del suo fondatore un mezzo di educazione globale sia fisica , sia spirituale che formava non solo gli individui singoli al sacrificio, insegnandogli a superare le difficoltà personali del "combattimento" quotidiano, ma aiutava a crescere anche la comunità dei praticanti, insegnando loro il gusto estetico ed armonico dell'applicazione "perfetta" della tecnica, con il miglior uso dell'energia, producendo l'armonizzazione di individui abituati al gusto estetico ed al rispetto reciproco che contribuivano insieme alla mutua costruzione del loro benessere fisico e morale.***

Jigoro Kano ricoprì numerosi incarichi per il governo, e riuscì a fare inserire il judo nelle materie scolastiche accanto all'educazione fisica. L'insegnamento del "Metodo Kanō" cominciò ad aver vita all'Accademia navale e nelle Università di Tokyo e Keio.
Il nuovo Judo Kodokan era al centro dell'attenzione pubblica, grazie ai lodevoli principi ed agli elevati ideali. Un'altra importante differenza fra Ju Jitsu e Judo era basata sull'applicazione dello squilibrio (Kuzushi) una teoria intuita da Jigoro Kano durante i suoi allenamenti ed usata con successo contro il maestro Tsunetoshi Iikubo della scuola Kito-Ryu. "Usando un minimo di forza è possibile proiettare l'avversario se riuscite a squilibrarlo" Con i suoi migliori allievi, Kanō nel 1895 stabilì il Gokyo, cioè un metodo di insegnamento di quaranta tecniche di proiezione, chiaro e diviso in 5 gradi ( mentre nelle antiche scuole non veniva mai fatto conoscere il completo programma di apprendimento). Sono dello stesso periodo le prime elaborazioni di Kata, con le forme delle proiezioni Nage no kata e del combattimento reale Kime no kata.
Successivamente, nel 1921, migliorò il Gokyo, con l'aiuto dei suoi allievi più esperti e con i maestri delle ultime scuole di Jujitsu assorbite dal Kodokan.
L'impostazione scientifica con cui Kano organizzò il Judo, fu sicuramente frutto anche delle sue conoscenze dei metodi di ginnastica occidentali che nello stesso periodo venivano sottoposti ad una riorganizzazione biomeccanica generale .
La prima applicazione della teoria scientifica degli esercizi ginnici e dello sport può ricondursi al lavoro svolto da P.Lesgaft, che nel 1877 egli elaborò in Russia il primo sistema organico nazionale di Educazione Fisica basato sulla teoria del movimento.
Venti anni dopo, nel 1899 Kano fu inviato in Europa dal Governo allo scopo di aggiornarsi sui metodi ginnici europei e visitò Parigi, Berlino, Brusselles, Amsterdam e Londra.
Il retaggio e l'evoluzione dei concetti acquisiti in questo viaggio possiamo leggerli, nell'aureo testo di Yokoyama ed Oshima ***Judo Kyohan*** ( Precetti di Judo) relativamente alle tre aree precedentemente introdotte *rentai-ho*, *shobu-ho*, e *shushin-ho*:
*"... Rentai-ho :* non è formato da movimenti macchinosi e senza scopo, come il mero allenamento fisico degli attuali sistemi ordinari di ginnastica. Ma judo con la sua pratica sviluppa egualmente ed in modo armonico tutte le parti del corpo … dal punto di vista dello *Shobu-ho* il judo si mostra ancora superiore. Ogni parte del corpo lavora conformemente alle leggi della fisiologia, l'impiego delle forze segue sempre i principi della Meccanica …ed infine lo *Shushin-ho* si sviluppa secondo le leggi della Psicologia … non si potrebbe trovare un metodo di combattimento più armonioso.
Dal punto di vista intellettuale il judo è un'eccellente scuola di ponderazione, di decisione nel giudizio, di buon umore e di dirittura morale, tutte qualità importanti nella vita …. Quanto alle qualità morali esse si acquisteranno gradualmente attraverso un duro lavoro e saranno, in ultima analisi, l'ideale a cui l'allievo cercherà sempre di tendere." ***Judo Kyohan pag 4-7.***
Nella vita pubblica Jigoro Kano fu una personalità di spicco in Giappone.
Nel 1881, Kano si laureeò all'Università Imperiale di Tokyo e subito ottenne il posto d'insegnante di letteratura alla Gakushuin la scuola dell'elite Imperiale del Giappone.
Kano divenne dirigente alla Gakushuin all'età di soli 25 anni. Secondo le usanze questa scuola ammetteva solo i figli della famiglia imperiale e quelli dei nobili, ma dopo che Kano iniziò a dirigere l'istituto, le iscrizioni furono allargate non solo alla borghesia ma a qualsiasi strato sociale.
Come direttore sia della Gakushuin che della scuola di allenamento degli insegnanti di Tokyo ( oggi divenuta Università di Tokyo dell'educazione) per circa un quarto di secolo, Jigoro Kano gettò le basi della moderna educazione del Giappone.
Cambiò la Gakushuin in un semi convitto permettendo agli studenti di tornare a casa solo nel fine settimana.
Rigettò la comune credenza che i figli dei nobili fossero più intelligenti per nascita ed aprì la scuola, come già accennato a tutti gli strati sociali- una mossa rivoluzionaria per i suoi tempi.
Egli decise anche che gli studenti seguissero una vita disciplinata improntata all'umiltà, in tal modo tutto l'ambiente cambiò sotto l'amministrazione di Kano, ed è comprensibile che i genitori degli alunni fossero entusiasti dei risultati ottenuti dal nostro Educatore.

Fu un uomo che credeva alla disciplina come miglioramento dell'individuo, ma fu anche un uomo molto generoso, offrendo ogni giorno ai suoi studenti di judo al tempio tè e riso misto a radici di loto e prendendosi carico dei più poveri a cui offriva vestiti pratici anche per il judo, vestiti che spesso si faceva carico di far lavare per loro.
Nel 1882, dopo aver terminato gli studi, fu , come abbiamo visto nominato professore e successivamente, nel 1884, Addetto alla Casa Imperiale, un titolo di grande prestigio.
Il 1884 fu un anno chiave per il Kodokan che fu fondato a norma di legge e Kano dichiarò", "Condividendo tutti i meriti con le varie scuole di Jujitsu da cui ho tratto le nozioni a cui ho aggiunto mie proprie scoperte ed invenzioni, Io ho fondato un nuovo sistema atto all'educazione fisica, mentale ed al combattimento, questo sistema Io ho chiamato Kodokan Judo"

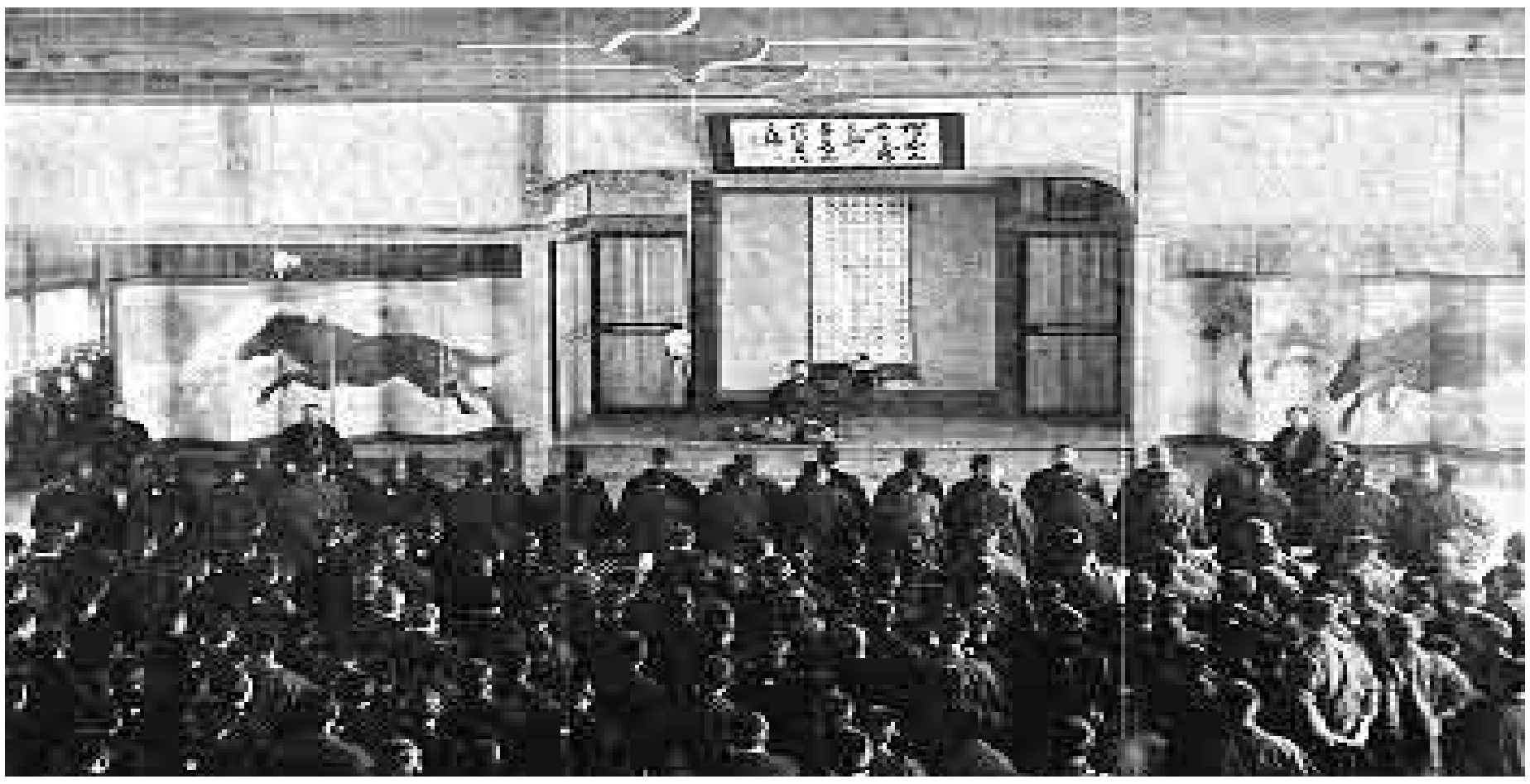

***Fig.13 cerimonia del passaggio di grado nella sala principale del Kodokan (Tokyo)***

Più tardi, nel 1891, diviene consigliere del Ministero dell'Educazione, del quale diverrà Direttore nel 1898.

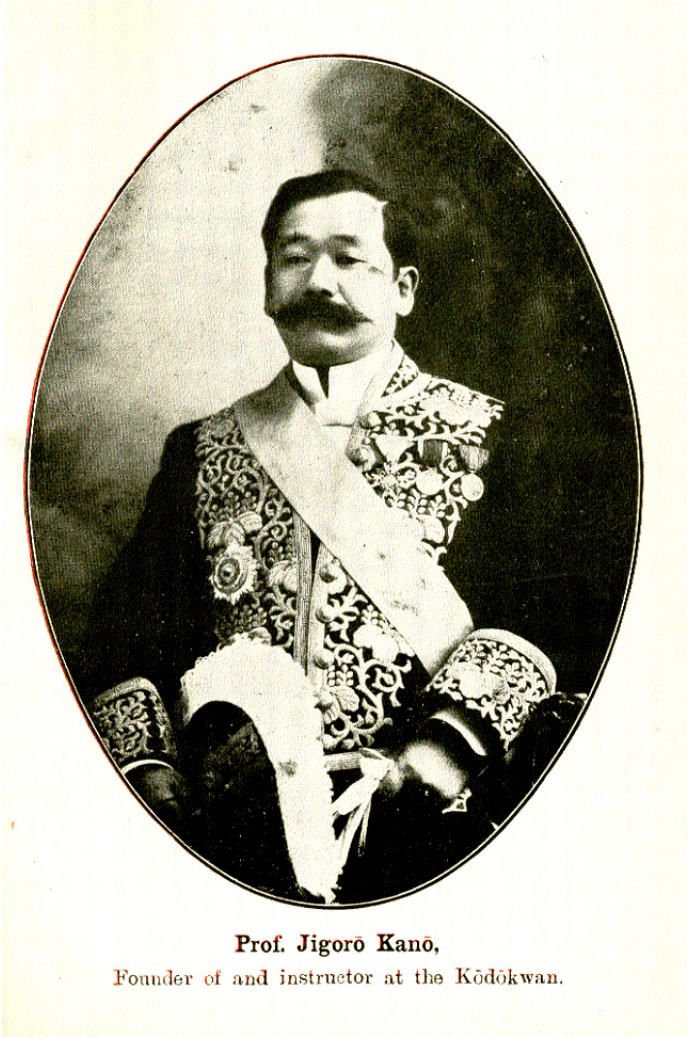

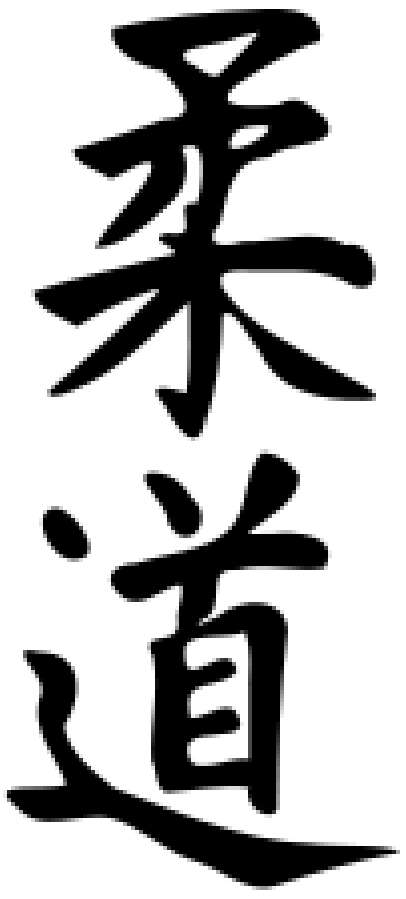

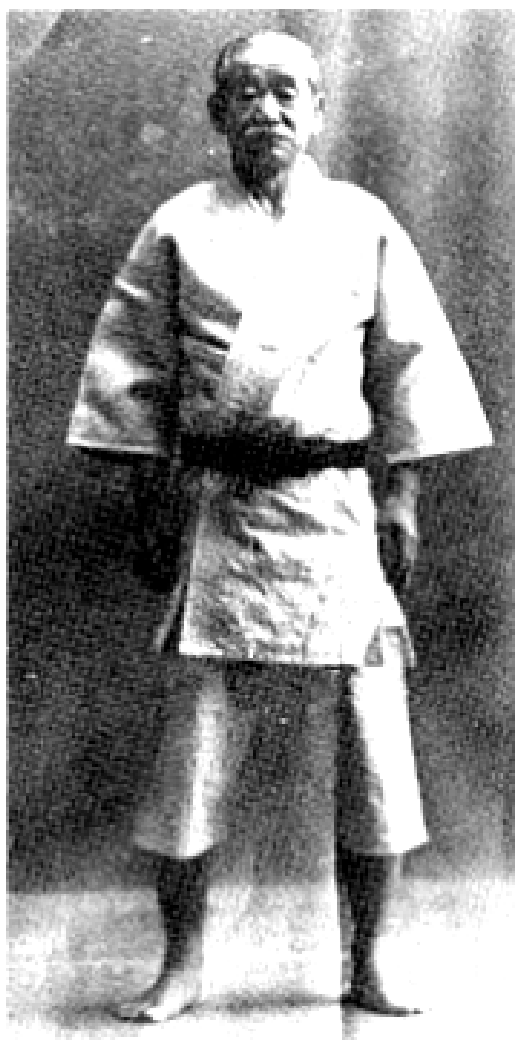

***Figg. 14-15 il Dr. Jigoro Kano in alta uniforme ed in judogi ed i kangi di Ju Do***

Sebbene Kano fosse orgoglioso del Judo e lo considerasse un mezzo di educazione importante, egli non disdegnò praticamente di interessarsi a tutti gli sport, anche in ciò egli gettò le basi di un'Educazione moderna in Giappone basata sulla Motricità ( Judo ), inoltre egli divenne anche il padre dello sport moderno in Giappone, fondando il primo club di Baseball che oggi è sport nazionale in Giappone .

Nel 1911 egli fondò la Associazione Atletica Giapponese e ne divenne il primo presidente e come membro del CIO partecipò alla Quinta Olimpiade in Stoccolma nel 1912 –Prima Olimpiade a cui partecipò il Giappone.
Nello stesso anno fu anche eletto e divenne il primo membro asiatico ad essere ammesso al CIO e vi rimase per ben 29 anni fino cioè alla sua morte nel 1938.

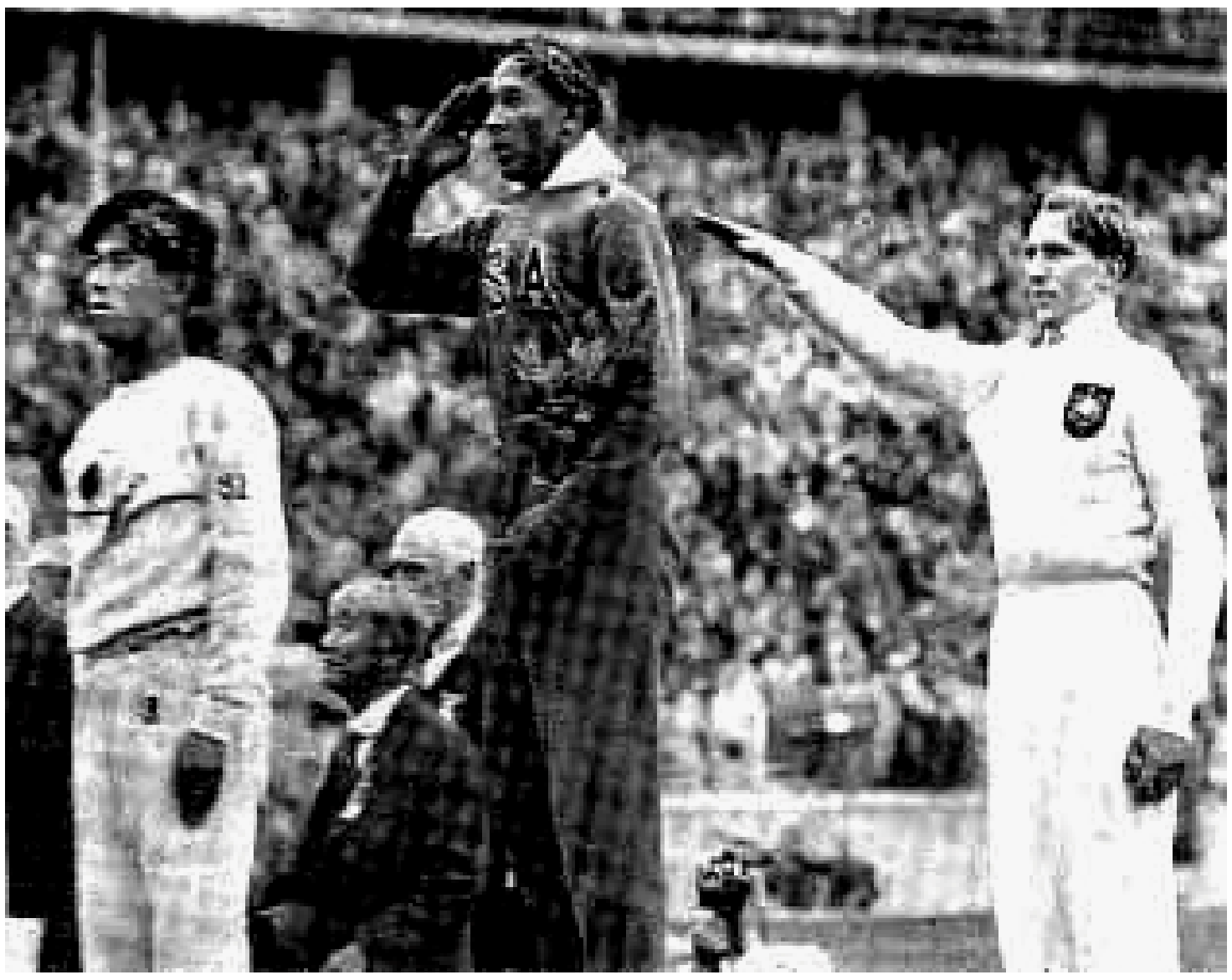

***Fig.16 Jigoro Kano in un momento storico 1936 Olimpiade di Berlino , l'americano Jesse Owens vince la medaglia d'oro nel lungo dopo aver sconfitto la Germania.***

Nel 1928 e nel 1934 fu in Italia, e visitò i centri judoistici creati da Carlo Oletti. Qui di seguito lo vediamo in una rarissima foto dell'epoca.

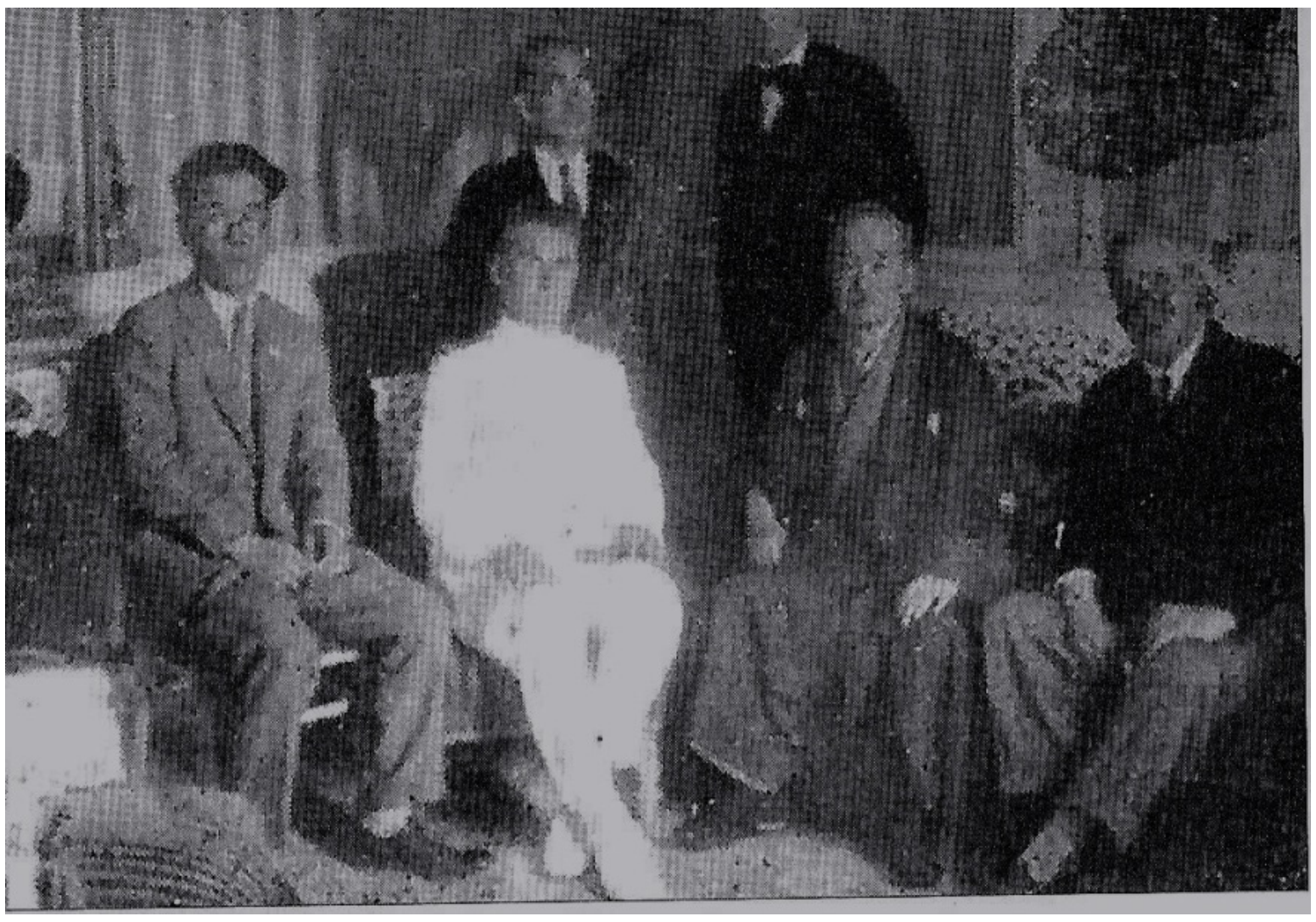

***Fig 17 Jigoro Kano e Carlo Oletti a Roma nel 1934 con il Mestro Mori ( Primo Inviato del Kodokan in Italia) alla destra di Oletti***

Nel 1935 Kano ricevette il premio Asahi per i suoi eccezionali contributi nei campi delle arti, scienze e sport. Nel corso della sua vita  Jigoro Kano compì più di trenta viaggi oltremare per il CIO e per diffondere il judo nel mondo;

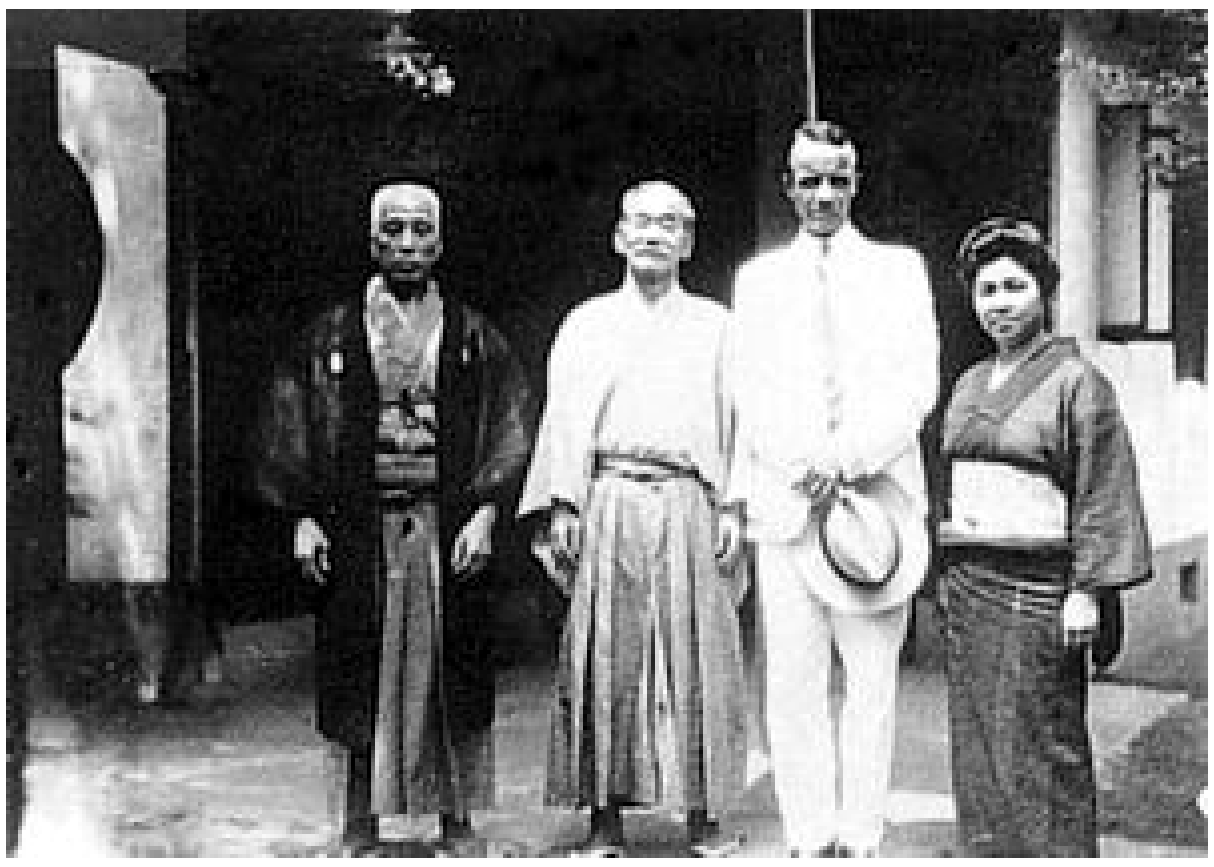

***Fig 18 Yamashita e signora, jigoro Kano e K. Roosevelt***

Nel 1938 venne inviato in rappresentanza del Giappone al 12° Convegno del CIO (Comitato Olimpico Internazionale) al Cairo, che approvò la proposta di far svolgere i Giochi Olimpici a Tokyo. Nonostante non considerasse il judo semplicemente uno sport ma piuttosto un mezzo di Educazione, si adoperò per portarlo gradualmente verso i giochi olimpici, poiché in questo modo sarebbe stato possibile far conoscere la sua disciplina nel mondo. Ed infatti judo fu proposto come sport dimostrativo per l'Olimpiade del 1940.

Oggi, più di 23 milioni di persone praticano Judo  nelle 199 nazioni associate alla International Judo Federation, prova vivente della frase profetica di Kano.

" Quando io morirò, Kodokan Judo non morirà con me, poiché tutto può essere studiato se i suoi principi (massima efficacia con il miglior uso dell'energia, e prosperità e mutuo benessere ) saranno studiati"

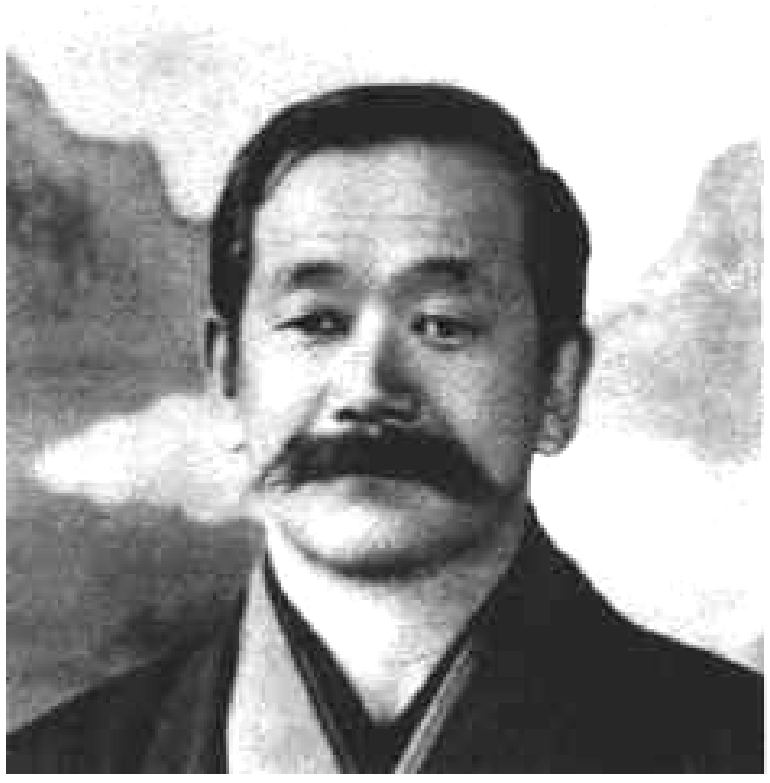
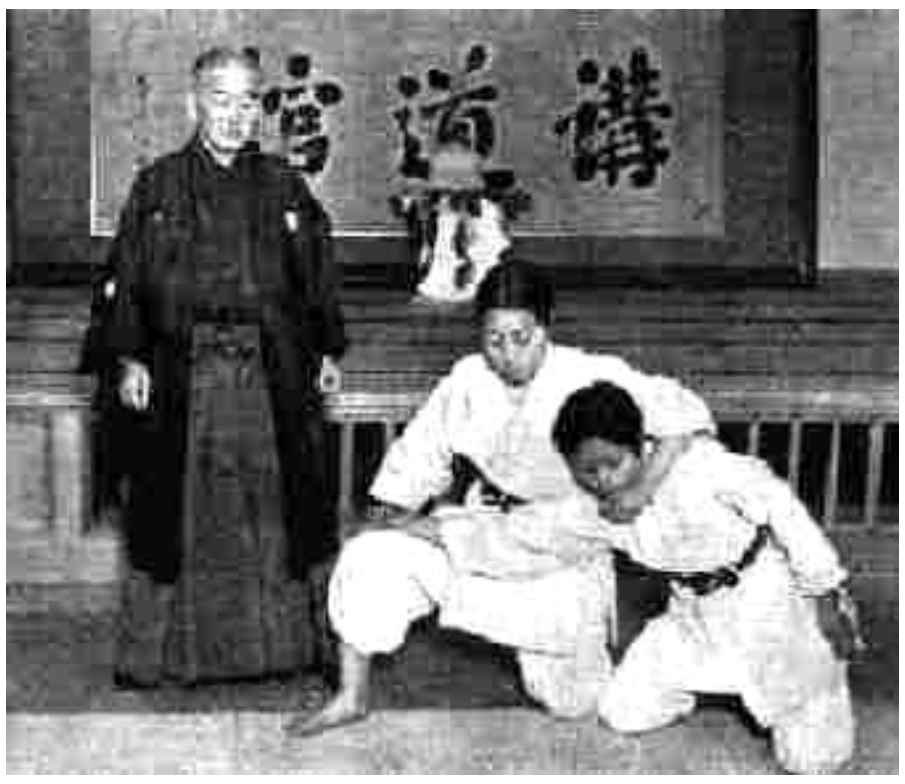
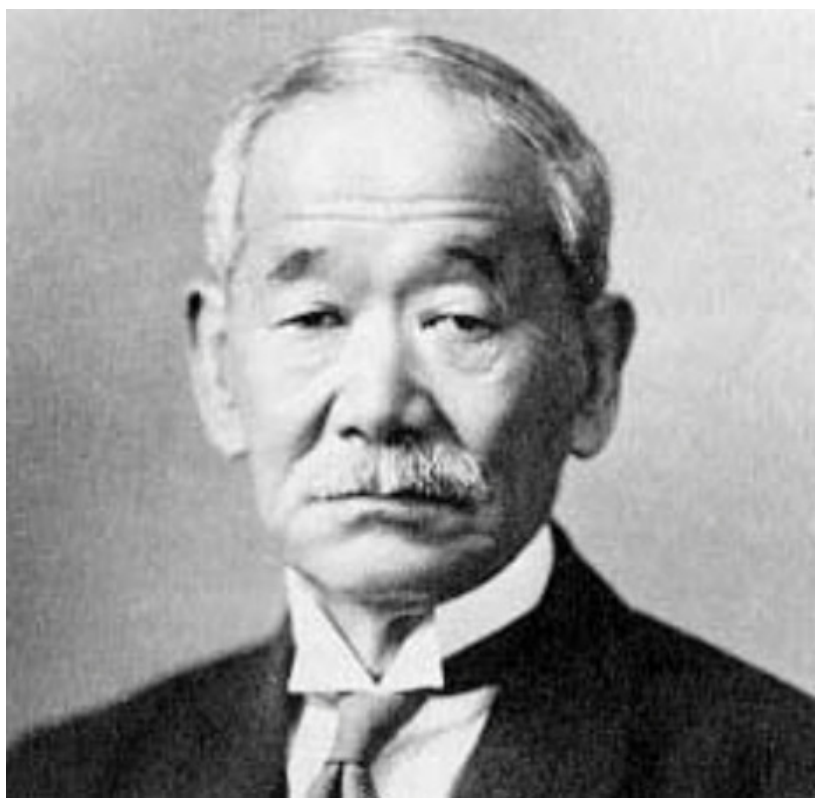

***Fig 19-21 Jigoro Kano Educatore e fondatore del Judo***

## *NAPOLI - La Biomeccanica del Judo: un esempio applicativo della capacità interpretativa dei moderni concetti scientifici*

---

### *L'incontro fortuito con Jigoro Kano*

Per l'autore di queste note, in ritardo come sempre, è nato di 29 e purtroppo non di **28** come Borelli e Kano, l'incontro con il Judo fu precedente alla scoperta della biomeccanica e di Borelli.
Ma andando con ordine: judo fu una soluzione ( in realtà di ripiego) trovata all'Università di Napoli.
Infatti nel 1967 l'unico sport del CUS Napoli, ch' era compatibile con le mie personali necessità di studio, si praticava dalle 19 alle 21 presso la palestra di scherma sita nell'istituto di Matematica a Via Mezzocannone e si chiamava judo.
Senza aver alcuna idea di cosa fosse, l'iscrizione fu fatta e così entrai nel mondo dei *lottatori in pigiama.*
L'inizio fu entusiasmante e l'apprendere questi nuovi gesti sportivi, l'abitudine a combattere ed a difendersi, superare gli avversari ed insegnare fu un'esperienza interessantissima, così dopo quattro anni raggiunsi il grado di cintura nera e fui la prima cintura nera universitaria del CUS Napoli, poi dopo uno sfortunato campionato universitario che vinsi lussandomi una spalla, finì la mia breve carriera agonistica ed iniziai quella arbitrale, giungendo al grado di arbitro internazionale Europeo, ma l'interesse per il judo come Sport andava sempre più affievolendosi, non trovando in esso più stimoli d'interesse, ne culturali, ne sportivi.
Ma come avviene spesso nella vita, le sorprese sono dietro l'angolo e così un giorno mi fu prestato un libro definito dal suo proprietario il libro "della matematica del Judo (?!)".
Io risposi che ciò non aveva senso compiuto perché non poteva esistere la matematica del judo, ma quando ricevetti il libro rimasi stupito nell'osservare che il testo era un tentativo di applicare le leggi della fisica al Judo. Il risultato mi sembrò ( in vero) non molto soddisfacente e con nuovo ardore decisi di rivedere Judo Sportivo come fenomeno fisico, e così iniziò in Napoli …. quel nodo tra Jigoro Kano ed Alfonso Borelli o meglio tra Tokyo e Napoli che oggi nel mondo è conosciuto come Biomeccanica del Judo.

---

### *L'incontro voluto con Alfonso Borelli*

L'incontro con Borelli è stato un incontro voluto infatti in occasione del I° Simposio Internazionale di Scienza Sportiva applicata al Judo tenutosi in Pamplona il **28** maggio 1988 ( sempre il numero ventotto) poiché ero stato invitato a tenere la relazione d'apertura del congresso dal titolo " La Biomeccanica Sportiva applicata al Judo" decisi di approfondire la mia conoscenza di Alfonso Borelli e mi recai alla Biblioteca Nazionale per cercare di consultare il suo "*De Motu Animalium* " il numero di catalogo piuttosto anonimo era 28 ( ancora lui)- C-32-33, ma il mio stupore sorse quando mi vidi presentare ben due tomi di notevoli dimensioni e si accrebbe ancor di più quando realizzai che uno era il " *De Motu Animalium*" e l'altro era un testo non richiesto dal titolo "*De vi percussionis"* che trovai corredato da questo sottotitolo "*De vi percussionis et motibus naturali bus a gravitate pendenti" Introductiones et Illustrationes Physico-Mathematicae apprime necessarie ad opus intelligendum: De Motu Animalium.*

Posto davanti allo sforzo intellettuale fatto dal Borelli compresi, con stupore, che egli aveva prima dovuto trovare e definire le leggi fisico-matematiche necessarie e successivamente le aveva applicate al corpo umano ed al suo moto.
Ma solo addentrandomi nella lettura del *De Motu Animalium* potei comprendere a pieno la portata realmente storica del pensiero del Borelli.Infatti si legge nel suo Proemio:

**"*Aggredior arduam physiologiam de Motibus Animalium quae licet a plurimis antiquor tentata fuit…hanc igitur mihi operam suscepi ut haec Physices pars demonstrationibus mathematicis ornata fuit. In primo copiore disceptabimus de motionibus Animalium, …, artuum flexionibus, extensionibus, tandem de gressu, saltu, volatu, natatu et annexis…In secunda de causis motus Muscolorum et motibus interni nempe de humorum, qui per vasa viscera Animalium fiunt… inquirendo musculorum fabrica .. postea exponemus musculi modus operamdi…*"**

Così ho ritrovato le due famose leggi sperimentali ( dette appunto di Borelli) sulla struttura morfologica dei muscoli. Ma non solo, infatti Borelli presentava il tutto con un sano ed encomiabile equilibrio critico. Si legge infatti nei titoli. *Dei moti e delle operazioni della contrazione meccanica dei muscoli. Delle false cause della contrazione dei muscoli adottate da altri. Delle probabili cause della contrazione vitali dei muscoli.* Si legge inoltre nel capitolo *De usu respirationis* :
*Attraverso la respirazione le particelle d'aria si mescolano al sangue. Si da ragione per cui durante il moto concitato ed il lavoro dei muscoli, il respiro si affanna e si affretta. Ragione perché il lavoro muscolare effettuato nell'aria rarefatta produce l'affanno. Progetto di una macchina che permette ad uomini immersi nell'acqua, di poter respirare e vivere per diverse ore.*

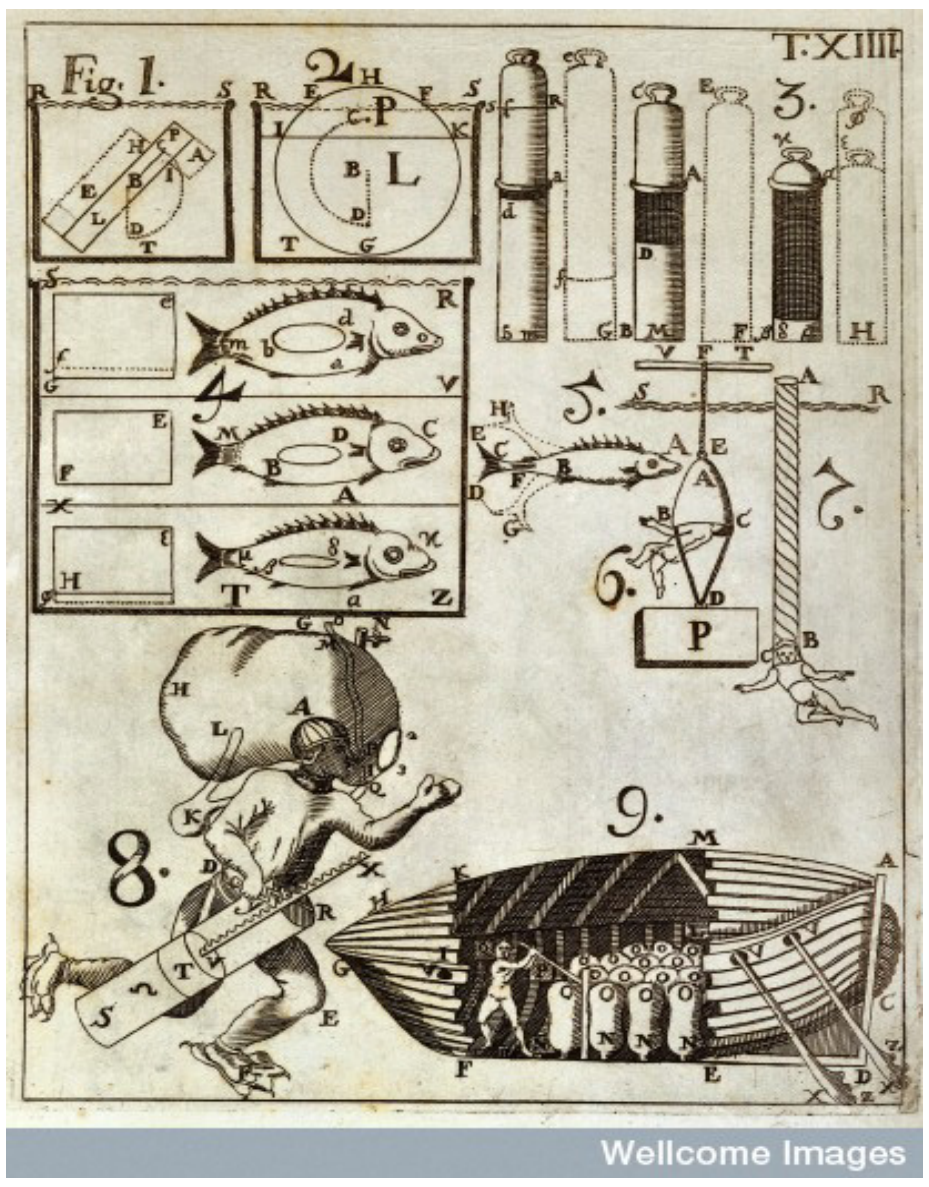

***Fig 22 progetto di uno scafandro personale e di un sommergibile.***

Perché vi mostro il progetto di scafandro? Perché nella mia avventura si era inserito un altro personaggio legato a Kano e Borelli durante le mie ricerche delle fonti del Judo mi imbattei nel testo

***Judo – Manuel de Ju-jitsu de l'école Kano à Tokyo*** *Par le professeurs Yokoyama et Oshima Traduit du japonais par l'E.d.v. Le Prieur, 1911 in-8°, XI-209 pagine, figure, tavole e ritratti, copertina illustrata a colori. Berger-Levrault Editeurs (Paris – Nancy)*

I protagonisti di questa avventura letteraria sono in regola:

Yves Le Prieur (Lorient Morbihan 1885 – Nizza 1963). Comandante di Marina, ricercatore e inventore nell'ambito della Marina, dell'Aviazione e dell'Aeronautica Navale. Di fatto comandante nientemeno che di Jaque Cousteau ed inventore dello scafandro da sommozzatore, brevetto che donò alla sua morte a Cousteau, suo attendente, fu di fatto ***il primo traduttore dell'unico libro con prefazione di Kano sul Judo.***

**Il Comandante di vascello Yves Le Prieur ci fa sapere** nell'introduzione**:**
*"Mandato in Giappone dal Ministero della Marina in qualità di apprendista-interprete, ho approfittato di due anni di soggiorno a Tokyo (1908-1910) per seguire i migliori corsi di jiu-jutsu.*
*Ho cominciato col maestro Matsui, di Akasaka, ma dopo un anno sono passato al Kodokan (che è un po' l'Università del jiu.jutsu) sotto l'alta direzione del suo eminente fondatore, professor Kano.*
*Fin dall'inizio ho notato la grande differenza che esisteva con i principi del jiu-jitsu che mi erano stati insegnati in Francia o che avevo potuto leggere sui libri.*
*Quindi ho pensato che fosse interessante per i cultori dello sport di difesa sapere in cosa consiste esattamente il jiu-jitsu insegnato a Tokyo; e che non si potesse avere informazioni più sicure di quelle direttamente fornite dai manuali giapponesi. M'informai presso i miei insegnanti di quale fosse il miglior saggio esistente ed essi me ne segnalarono uno che era appena uscito (8 Maggio 1908). Che costituiva, secondo loro, quanto di più serio fosse mai apparso in Giappone. Così cominciai a tradurlo.*
*Si tratta dell'opera dei maestri Yokoyama e Oshima, due aiutanti del professor Kano. Questo manuale ha avuto un grande successo, perché sul jiu-jitsu esistevano solo opuscoli poco illustrati.*
*E' con la speranza di aver fatto un lavoro utile mostrando il jiu-jitsu sotto la sua vera luce che presento la mia fatica di traduttore. E approfitto per ringraziare delle loro lezioni i professori Matsui e Kano, che mi hanno sempre trattato con simpatia.* Lorient, 1° Marzo 1911

**Prefazione del professor Kano, capo-istruttore del Kodokan:** *"Recentemente mi hanno fatto visita i signori Yokoyama Sakujiro (7° dan) e Oshima Eisuke per mostrare il manoscritto di Judo Kyohan (Precetti di Judo) chiedendomi di correggerlo e di aggiungervi una prefazione. Scorse le pagine ho constatato che si trattava delle lezioni che ho impartito al Kodokan fino ad oggi. Ho segnalato agli autori qualche inesattezza, ma questo manuale può essere raccomandato, per qualità e correttezza delle nozioni, a tutti i praticanti.*
*In molti mi hanno chiesto di scrivere un manuale di judo, ma la complessità dell'argomento e gli impegni mi hanno trattenuto dal sedermi a tavolino. Così questo libro è arrivato a proposito per soddisfare l'ardente desiderio dei miei allievi*
*Questo scritto è essenziale; e io non posso escludere che, nonostante la revisione a cui l'ho sottoposto, contenga ancora qualche difetto. Ma vista la competenza degli autori sono sicuro che le future edizioni saranno perfette. E mi auguro dunque che, per l'ortodossia del judo, queste pagine siano conosciute da tutti i praticanti".* Tokyo 1908

Ed è proprio in questo testo, prima edizione dell'aureo Judo Kyohan, che trovai le prime conferme di quanto ho affermato prima
Cioè più analizzavo l'impostazione del judo nella sua strutturazione e più la trovavo scientificamente impostata. Si può proprio affermare, senza tema di smentite che l'impostazione originale fornita dal Dr Kano era realmente di tipo Proto-Biomeccanico.
Certamente il connubio Scienza e Judo è di lontana origine infatti nel già citato primo libro del Judo **Judo Kyohan** si possono leggere queste parole illuminanti relative allo squilibrio dell'avversario :

"*… riguardo al concetto di squilibrio si potrebbero dare sapienti dimostrazioni matematiche ma non essendo esse alla portata di tutti preferiamo fare l'esempio più comprensibile di un bastone che viene rovesciato ....*" .

Esse mostrano quanto la scienza fosse ben presente nella mente del fondatore e dei suoi seguaci. Ma scomparso Kano la scienza scompare ( almeno Ufficialmente dal Judo) ed in occidente si parlerà solo di tecnica e della sua effettuazione fino al 1960 anno in cui appare *"My Study of Judo"* di Koizumi a cui seguono nel 1961 "*Mechanics of Judo*" di Blanchard e *"The secret of judo"* di Watanabe ed Awakian, e il meno noto *"Planning de Judo-Base"* di Hamot & Tissier del 1967 o l'ancor meno "*Judo Analise Mecanica das tecnicas de Projeccao do Gokio*" del 1980 di Almada .
In realtà esisteva presso il Kodokan *"The Association for the scientific study on Judo"* associazione che in preparazione dell'Olimpiade di Tokyo nel 1964 in cui Judo apparve per la prima volta, come sport Olimpico, sviluppò con l'aiuto delle migliori università Giapponesi una enorme mole di lavori scientifici e praticamente tutta l'analisi biomeccanica del fenomeno, partendo dal 1958.
La maggior parte di questi lavori editi nei bollettini pluriennali del Kodokan, furono raccolti poi in un testo, mai tradotto e praticamente sconosciuto in occidente, dal nome inglese di "*Judo coaching"* di Y. Matsumoto del 1984.
Comunque tutti questi lodevoli sforzi, pur nel loro valore intrinseco, hanno peccato di frammentarietà. Mancava infatti una revisione critica del lavoro del Dr. Kano e l'analisi scientifica globale del fenomeno Judo nella sua interezza.
Di fatto questa visione unitaria ed unificante è stata da me portata avanti fino a costituire un corpus unico che ha trattato in modo sistematico di tutti gli aspetti del fenomeno, sia come attività motoria, sia come sport olimpico e non solo del Judo.
Ma ricordando la massima di F. S. Marvin che afferma **"L'essenza della scienza è scoprire l'identità nella differenza"**, la trattazione è stata estesa fruttuosamente a tutti gli sport di situazione sia di coppia, sia di squadra.
L'intero lavoro di revisione e sistematizzazione può esser racchiuso tra le mie due pubblicazioni relative all'argomento tra il 1988 "***Biomeccanica del Judo***" ed. Mediterranee ed il 2010 "***Advances in Judo biomechanics research***" Muller Verlag.
In Figg sono mostrate le pagine di copertina dei due testi

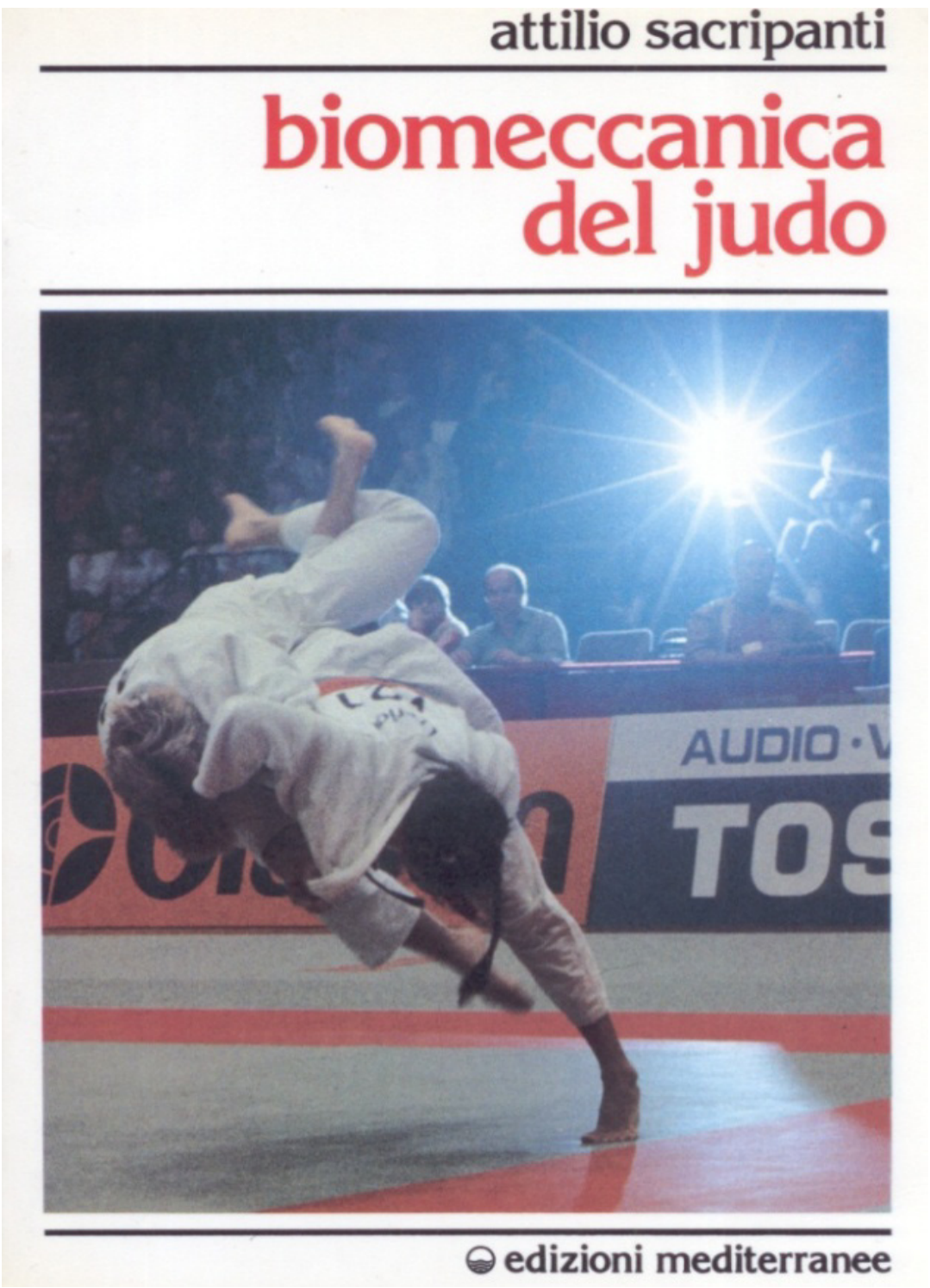

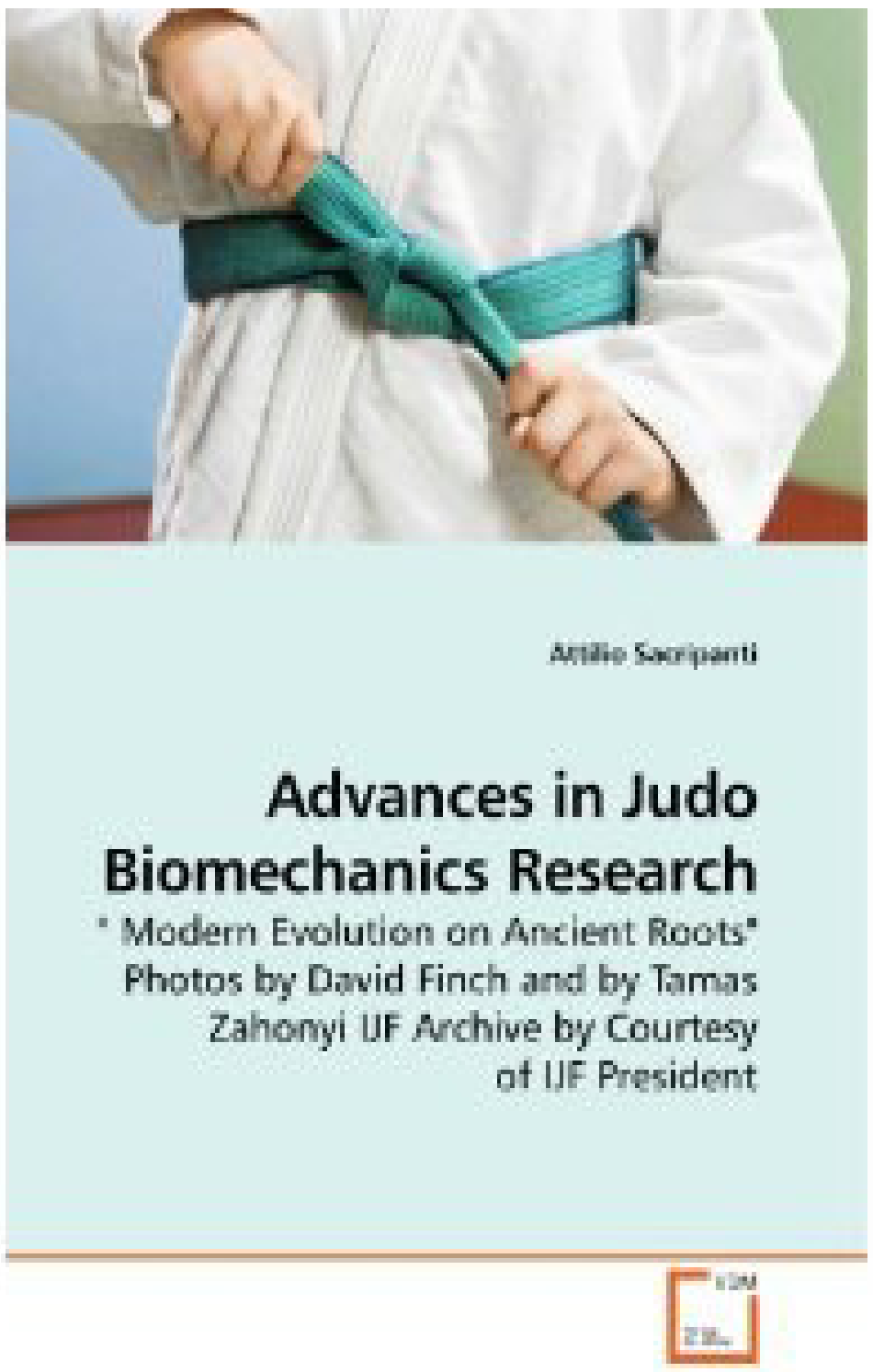

***Figg. 23-24 Copertine dei due testi di Biomeccanica del Judo che racchiudono la completa sistematizzazione del judo alla luce delle conoscenze Biomeccaniche più avanzate.***

## *I Risultati*

Per prima cosa è questo il luogo ove bisogna sottolineare la grandezza e la modernità del Dr. Kano nell'impostazione didattica della materia, egli ha dapprima Classificato le tecniche secondo principi "biomeccanici" per agevolare la loro comprensione razionale e permetterne uno studio sistematico.
Ad esempio: le tecniche di controllo delle cadute furono classificate in base alla direzione di caduta; le tecniche di lancio classificate in base alla parte del corpo che svolge il ruolo dominante nel trasferimento della forza; le tecniche di immobilizzazione classificate secondo il modo di effettuare il controllo; le tecniche di strangolamento secondo il sistema usato per soffocare ed infine le tecniche di lussazione al gomito secondo i meccanismi di iperestensione e pronazione / supinazione della catena cinetica superiore.
Poi ha organizzato la materia secondo una progressione di 40 tecniche di lancio secondo cinque gruppi di lezioni ( detta Go Kyo) strutturata sulla base "dal più semplice al più complesso" sia come movimento dell'attaccante (Tori), sia come caduta di chi subiva (Uke).
Ha inoltre introdotto come già accennato il metodo differenziale di analisi da applicarsi allo studio di ogni tecnica di lancio, suddividendo il movimento in tre fasi distinte: Squilibrio, Posizionamento, Proiezione, ( Kuzushi, Tsukuri, Kake). Egli infine sempre dal punto di vista pratico ha sempre considerato Judo anche come un esercizio ginnico più completo degli esercizi calistenici dell'epoca, dimostrando anche in ciò una capacità intuitiva superiore come vedremo nel seguito.
Facendo riferimento invece alla mia revisione biomeccanica del Judo questa ha permesso di mettere a punto, per le prima volta ed in modo definitivo, la metodica generale di analisi per tutti gli sport di situazione. Essi, nonostante la loro intrinseca complessità, si possono studiare in due fasi separate che per il Judo saranno:

1. studio dell'interazione nel sistema coppia di atleti
2. studio del moto del sistema coppia di atleti in competizione.

---

## *Studio dell'Interazione*

Lo studio dell'interazione può essere affrontato con l'ausilio della Meccanica Newtoniana in condizioni statiche perché l'applicazione della relatività Galileana permette, con buona approssimazione, l'estensione alla situazione di competizione in cui il sistema si muove praticamente di moto uniforme a velocità irrisorie ( tra 0.3 m/s e 0.5 m/s).
Questa fase affrontata con sistemi classici di analisi, pur nella complessità della materia, è stata proficua di risultati unificanti che si riassumono brevemente:
Nelle fasi di squilibrio –posizionamento ( Kuzushi – Tsukuri) :

1. Individuazione delle classi di spostamenti relativi al posizionamento - squilibrio, dette ***Invarianti generali d'azione.***
2. Individuazione delle classi di movimenti delle catene cinetiche detti ***Invarianti specifici d'azione***
3. individuazione delle tecniche di lancio che non necessitano dello squilibrio.
4. individuazione delle tecniche che necessitano di un tempismo particolare usando solo ***invarianti generali d'azione*** ( movimenti non complessi implicano un minor consumo energetico).

5. Individuazione delle tecniche che necessitano di una capacità coordinativa maggiore usando sia gli ***invarianti generali*** che gli ***invarianti specifici*** (movimenti complessi implicano un maggior consumo energetico).

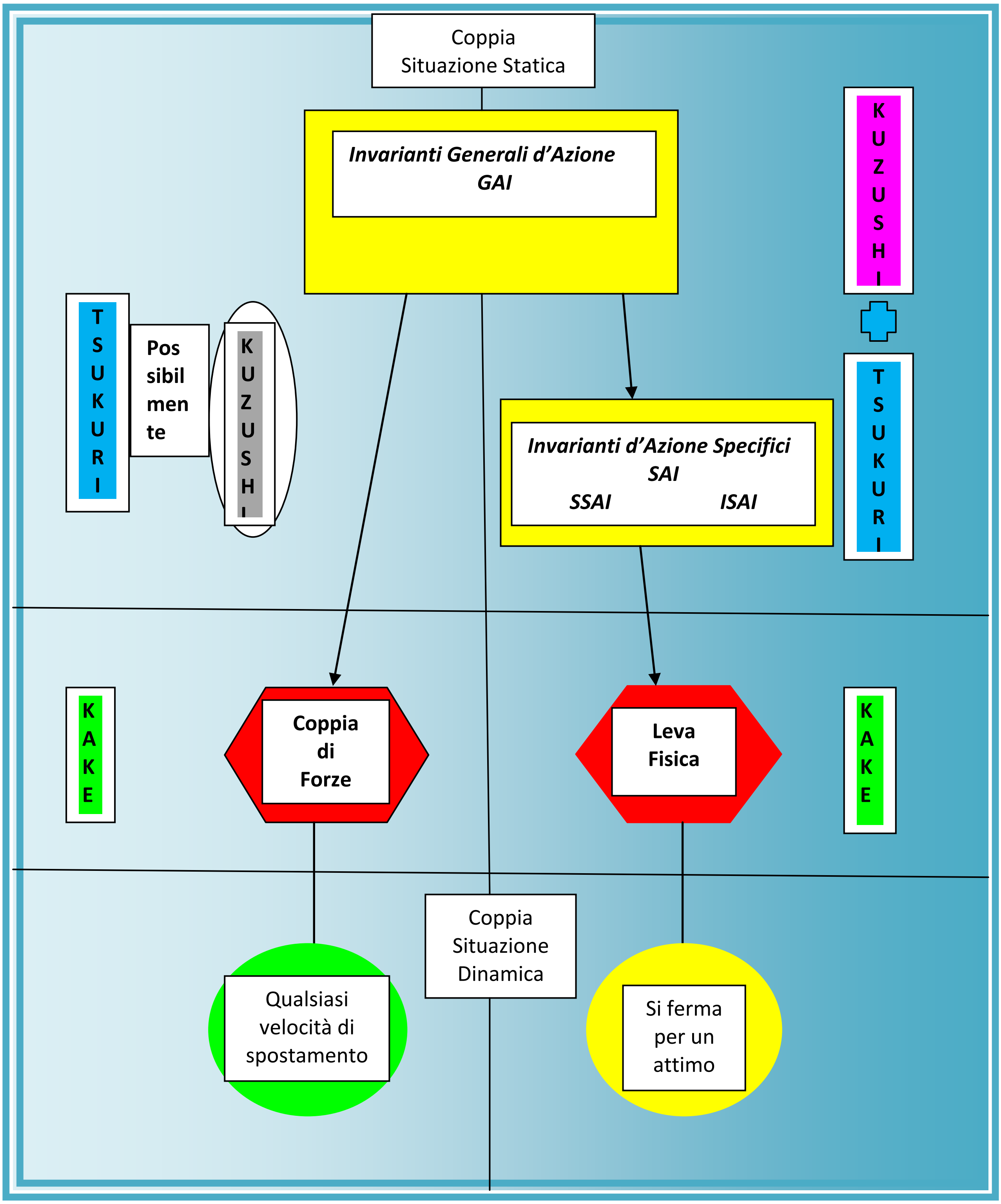

***Diag 1 Prospetto degli invarianti d'Azione del Kuzushi_Tsukuri connesse alla fase di Kake delle tecniche di proiezioni.***

Per la fase di proiezione (Kake ):

1. Individuazione dei due principi fisici che sono alla base di tutte le tecniche di proiezione.

2. Identificazione dei movimenti di base ( fondamentali) da applicarsi per ogni gruppo di tecniche di lancio.

3. Effettuazione di una nuova classificazione in soli due gruppi ( nei confronti dei cinque di Kano) evidenziandone i principi fisici di base ed il ruolo giocato dalle catene cinetiche.

4. Definizione dei corollari statici e dinamici sull'utilizzo delle forze per proiettare.

5. Individuazione dei meccanismi anatomici che giocano un ruolo essenziale nell'effettuazione dei lanci.

6. Identificazione delle traiettorie di base del corpo lanciato, che sono le geodetiche di particolari simmetrie, e rappresentano quindi le traiettorie di minor consumo energetico per chi lancia

   .

7. Identificare i meccanismi fisico- fisiologici attivati nelle tecniche di controllo delle cadute.

Pertanto tutte le tecniche di proiezione possibili possono essere classificate sotto solo due principi fisici di base:

***A) Applicazione di una coppia di forze***
***B) Applicazione di una leva fisica***

Nelle figure che seguono vengono mostrati esempi di alcune proiezioni di Judo appartenenti ai due gruppi menzionati : della coppia di forze e della leva fisica

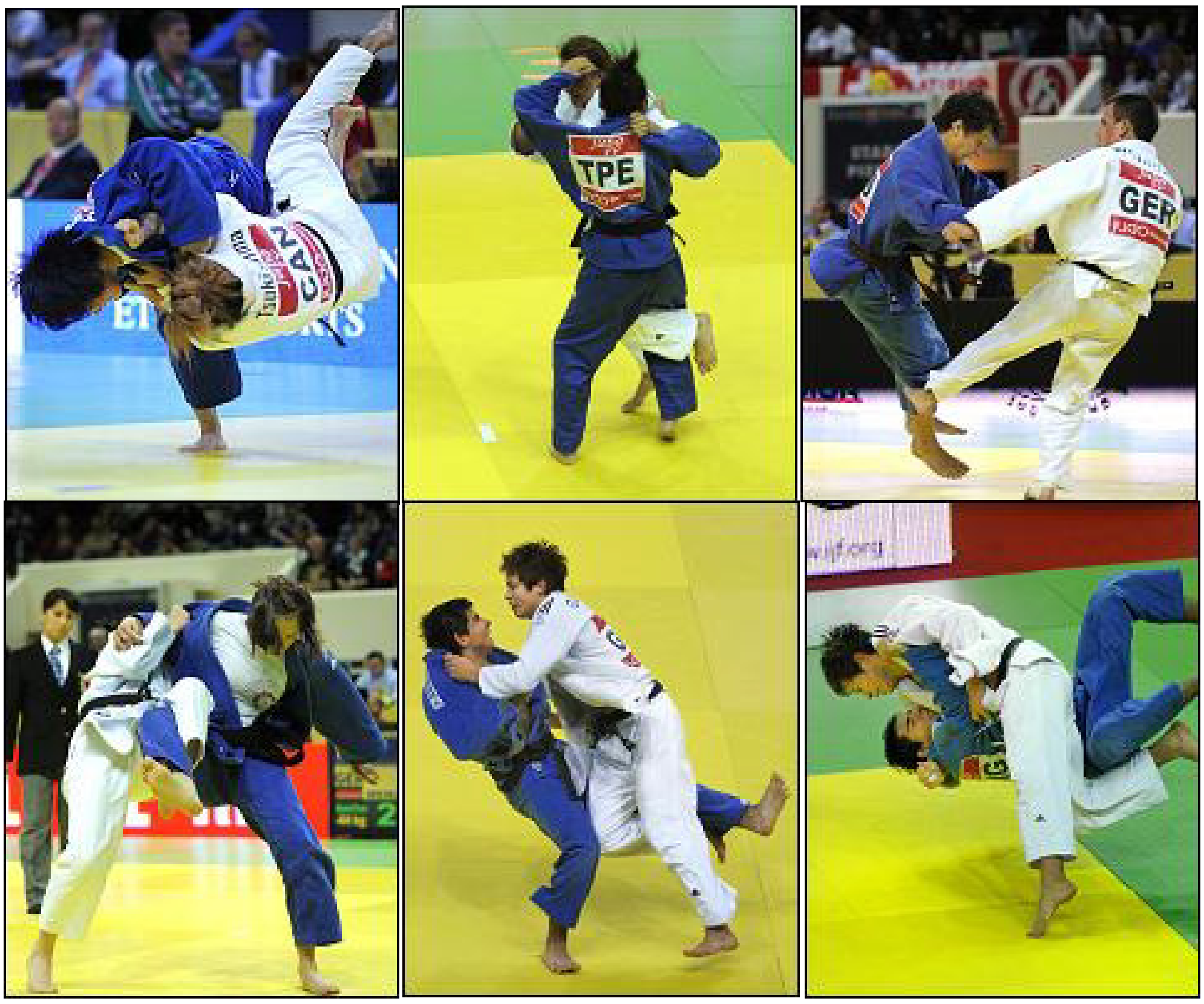

*Fig 25-30 Varie Applicazioni della coppia di forze nei tre piani di simmetria del corpo*

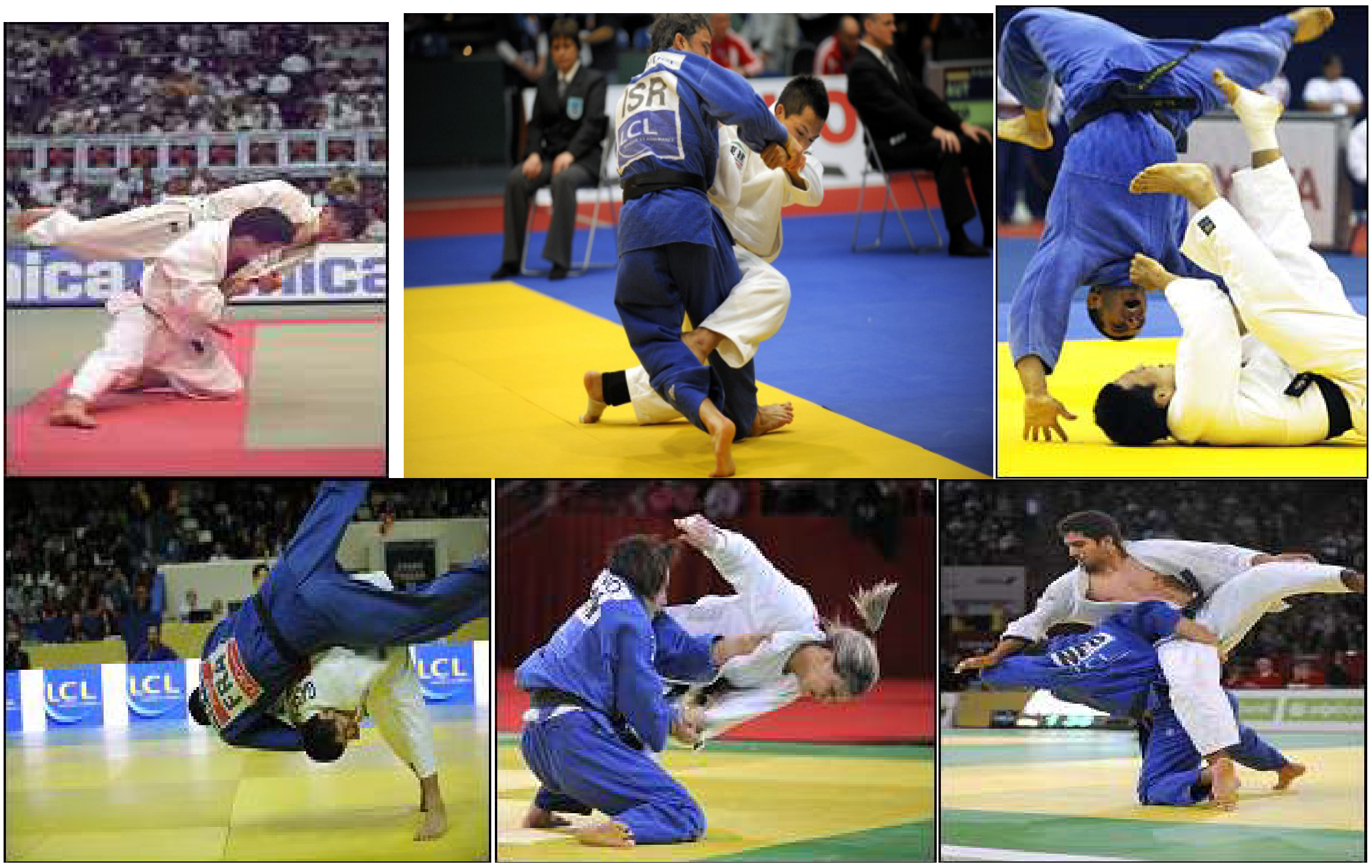

*Fig 31-36  Varie Applicazioni  della leva fisica.*

Per la fase di lotta a terra:

1. Classificare le tecniche di controllo ed immobilizzazione in funzione della pressione esercitata.

2. Determinare il ruolo attivo/passivo delle catene cinetiche nelle immobilizzazioni.

3. Identificare i meccanismi di base di liberazione dalle immobilizzazioni.

4. Analizzare le basi fisico biomeccaniche delle tecniche di leva articolare.

5. Determinare i meccanismi meccanico compressivi delle tecniche di strangolamento.

Nelle figure che seguono vengono mostrate alcune fasi di lotta a terra due immobilizzazioni uno strangolamento ed una leva articolare

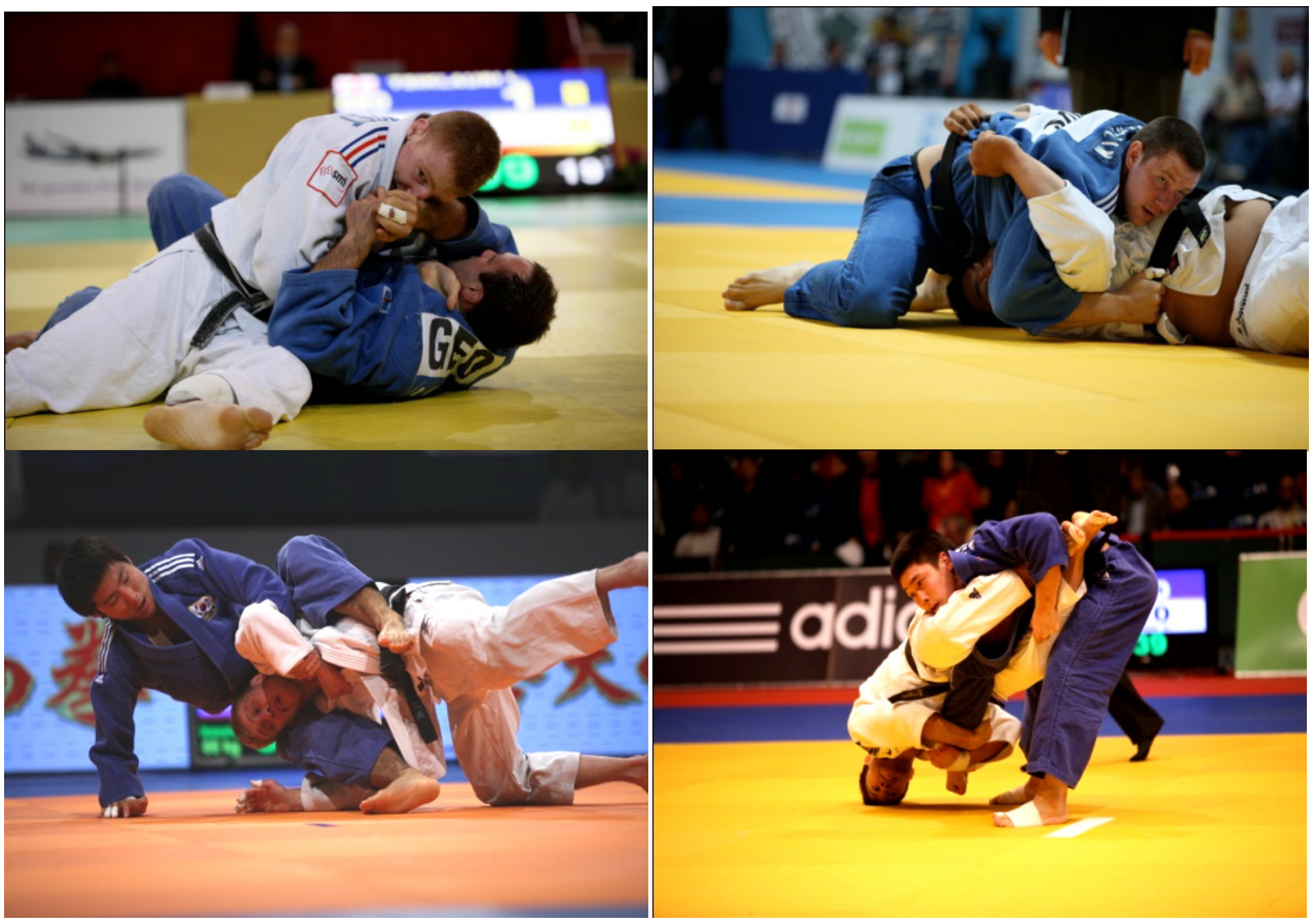

***Figg37-40  Tecniche varie di lotta a terra.***

## *Studio degli spostamenti in competizione.*

Se lo studio dell'interazione ha presentato difficoltà notevoli, infatti si sono dovute analizzare più di 197 tecniche diverse o loro composizioni lineari, lo studio degli spostamenti in competizione si trova a ordini di grandezza superiore nella scala delle difficoltà.
Il problema basico è che negli sport di situazione una competizione non è mai uguale ad un 'altra. Questo era stato l'ostacolo maggiore che aveva impedito l'analisi scientifica di questo e di fenomeni analoghi, negli anni precedenti.
In effetti l'applicazione allo sport delle moderne metodiche fisico-matematiche era ancora limitata , per cui l'aver dimostrato nel 1990, mediante la fisica statistica e la sinergetica, sia teoricamente, sia sperimentalmente ( utilizzando dati sperimentali ottenuti dai Giapponesi nel 1978) che il moto della coppia di atleti apparteneva alla classe dei Moti Browniani è stato davvero un lavoro profondamente pionieristico.

### Il sistema Coppia di Atleti: definizioni e caratterizzazione fisica.

La caratterizzazione fisica dell'ambiente di competizione ci porta facilmente ad individuare le forze agenti sugli Atleti:

1) la forza di gravità;
2) la forza d'impatto e/o di spinta prodotta dall'avversario;
3) le reazioni vincolari prodotte dal tappeto e trasmesse mediante l'attrito.

Se si definisce il sub-sistema Atleta in termini operativi di "atleta biomeccanico" cioè un solido a geometria variabile ed a simmetria cilindrica, che può assumere differenti assetti e che mediante gli snodi articolari è capace di compiere solo determinate rotazioni, possiamo poi, passare alla definizione del sistema globale d'interesse della nostra analisi della competizione, cioè il sistema Coppia di Atleti, che definiremo come: un sistema snodato a simmetria cilindrica, formato dall'unione semirigida di due atleti biomeccanici.
Un tale sistema potrà avere solo due di quelli che definiremo come " stati energetici" a cui sono legati un numero diverso di gradi di libertà . L'analisi della meccanica della competizione, come fenomeno formato da "situazioni" praticamente non ripetibili, che si verificano in maniera "casuale" con una certa probabilità di frequenza su un gran numero di combattimenti; non può essere affrontato con gli strumenti deterministici della meccanica newtoniana.
Appare pertanto opportuno analizzarlo con le metodiche proprie della meccanica statistica, che sono in condizioni di fornire una serie d'indicazioni utili su grandezze che possono essere valutate sperimentalmente.
Lo sviluppo successivo si fonda sull' estensione formale della teoria statistica all'analisi del sistema "unico" ed isolato la coppia, in cui non si vuole analizzare l'interno (interazione).
Pertanto la massa, la velocità, l'energia e le altre grandezze devono sempre intendersi come proprietà dell'intera coppia e non del singolo atleta.
Considerando la "Coppia di Atleti " che si sposta sul tappeto, è lecito affermare che essa compie spostamenti "casuali" prodotti dall'aumento o dalla diminuzione di velocità di scivolamento della coppia , o dal cambio di direzione prodotto dalla risultante delle forze generate dai due atleti, al fine di creare una opportuna "situazione" già interiorizzata da uno dei due, perché studiata in allenamento, per applicare opportunamente ( interazione) la tecnica "risolutiva" contro l'altro .
Ove il termine "casuale" sottintende la condizione che su di un gran numero di combattimenti ( al limite infiniti ) non deve esistere una direzione privilegiata di spostamento.

**Coppia di Atleti chiusa.**

Il sistema Coppia di Atleti chiusa ( con prese stabilizzate), compie dunque spostamenti “casuali” al fine di creare una “situazione” opportuna, che permetta di applicare la tecnica risolutiva.
Questo moto è possibile, grazie all’attrito presente al “contatto” tra piedi e tappeto, in base al III Principio della Dinamica.
L’equazione generale che descrive questa situazione dinamica è la ben nota seconda legge di Newton $m\dot{v} = F$.
Nella forza generalizzata F compariranno sia i contributi “attrito” che quelli “spinta/ trazione”.
La componente attrito è proporzionale alla velocità $F_a = -\mu v$. I cambi di velocità e direzione determinati da opportune spinte/trazioni sono prodotti dalla risultante delle forze sviluppate dai due atleti. Essi rappresentano, rispetto all’intera durata dell’incontro, impulsi agenti su brevissimi intervalli di tempo. Pertanto la singola variazione può esser espressa dalla **δ** di Dirac dell’impulso **u** della forza elementare. Dove **u** rappresenta in effetti la variazione del la quantità di moto **m Δv.**

$$\varphi(t) = u\sum_j \delta(t - t_j)$$

La risultante è data dalla somma algebrica delle spinte/trazioni in cui dovranno considerarsi anche i cambi casuali di direzione ( descritti dalla variazione di segno $(\pm 1)_j$ della forza elementare).

La forza totale è : $\varphi(t) = u\sum_j \delta(t - t_j)(\pm 1)_j$ = **F’**.

Pertanto la forza generalizzata è $F = F_a + F'$ e l’equazione generale del moto acquista la forma di un’equazione di tipo Langevin ( Sacripanti) e potrà scriversi:

$$\dot{v} = -\frac{\mu}{m}v + \frac{u}{m}\sum_j (\pm 1)_j \delta(t - t_j) = F_a + F'$$

Essendo la risultante delle spinte/trazioni di tipo “casuale” non è possibile predire la traiettoria in un singolo incontro, ma l’analisi statistica estesa ad un numero significativo di competizioni, permetterà di trarre informazioni importanti sul comportamento del sistema.

1) Poiché i cambi di direzione sono equiprobabili, cioè su di un gran numero di combattimenti, non esiste una direzione privilegiata allora il valor medio nel tempo di **F’** su una sequenza casuale di direzioni è nullo $\langle F' \rangle = 0$

2) Se si considera il prodotto di due spinte/trazioni e si media sul tempo e sulle direzioni, la funzione prodotto fornirà informazioni sulla variazione della forza nel tempo

$$\langle F(t)F'(t) \rangle = \frac{u^2}{m^2}\langle (\pm 1)_j (\pm 1)_i \delta(t - t_j)\delta(t - t_i) \rangle = \frac{u^2}{m^2 t_0}\delta(t - t')$$

L a verifica di queste condizioni ci permette di affermare che il moto degli atleti, può esser descritto in termini di meccanica statistica, come un moto appartenente alla classe dei moti Browniani su una superficie infinita.

Se questo è vero, allora detta **f(x,t) dx** la probabilità di ritrovare la Coppia di Atleti, o più correttamente il suo baricentro generale, nella posizione **x** dell'intervallo **dx** al tempo **t** ; è possibile dimostrare che la probabilità soddisfa l'equazione di Fokker-Planck, che, com'è noto, descrive la variazione della distribuzione continua di probabilità di presenza durante il tempo dell'incontro.
Ovvero in formule:

$$\frac{df}{dt} = -\frac{d}{dx}(Kf) + \frac{1}{2}D\frac{d^2}{dx^2}f$$

dove **K= -µx** è il coefficiente di spinta/trazione e **D** è il coefficiente di diffusione.
Ora ricordando la famosa relazione di Einstein, esso può esser messo in relazione con l'evoluzione statistica della posizione del baricentro della Coppia nel tempo, nel limite di intervalli di tempo molto piccoli o molto grandi rispetto al tempo di ritardo, ovvero con la media temporale del quadrato dello spostamento e quindi con l'energia che per tempi molto grandi rispetto al tempo di ritardo, risulta uguale :

$$< x^2 >= 2Dt = \frac{4}{\mu}\eta O_2 t$$

Mentre in analogia per tempi molto piccoli si potrà scrivere

$$\langle x^2 \rangle = 2Dt = 2\eta O_2 t^2$$

Così il coefficiente di diffusione per tempi grandi è direttamente proporzionale al doppio dell'energia spesa ed inversamente proporzionale al coefficiente di attrito.
Mentre per tempi piccoli è direttamente proporzionale al doppio dell'energia per il tempo.
La funzione f fornisce, nel tempo, la traiettoria di spostamento più probabile, che viene individuata dall'insieme dei punti di massima probabilità della funzione stessa nel tempo. Fig. .

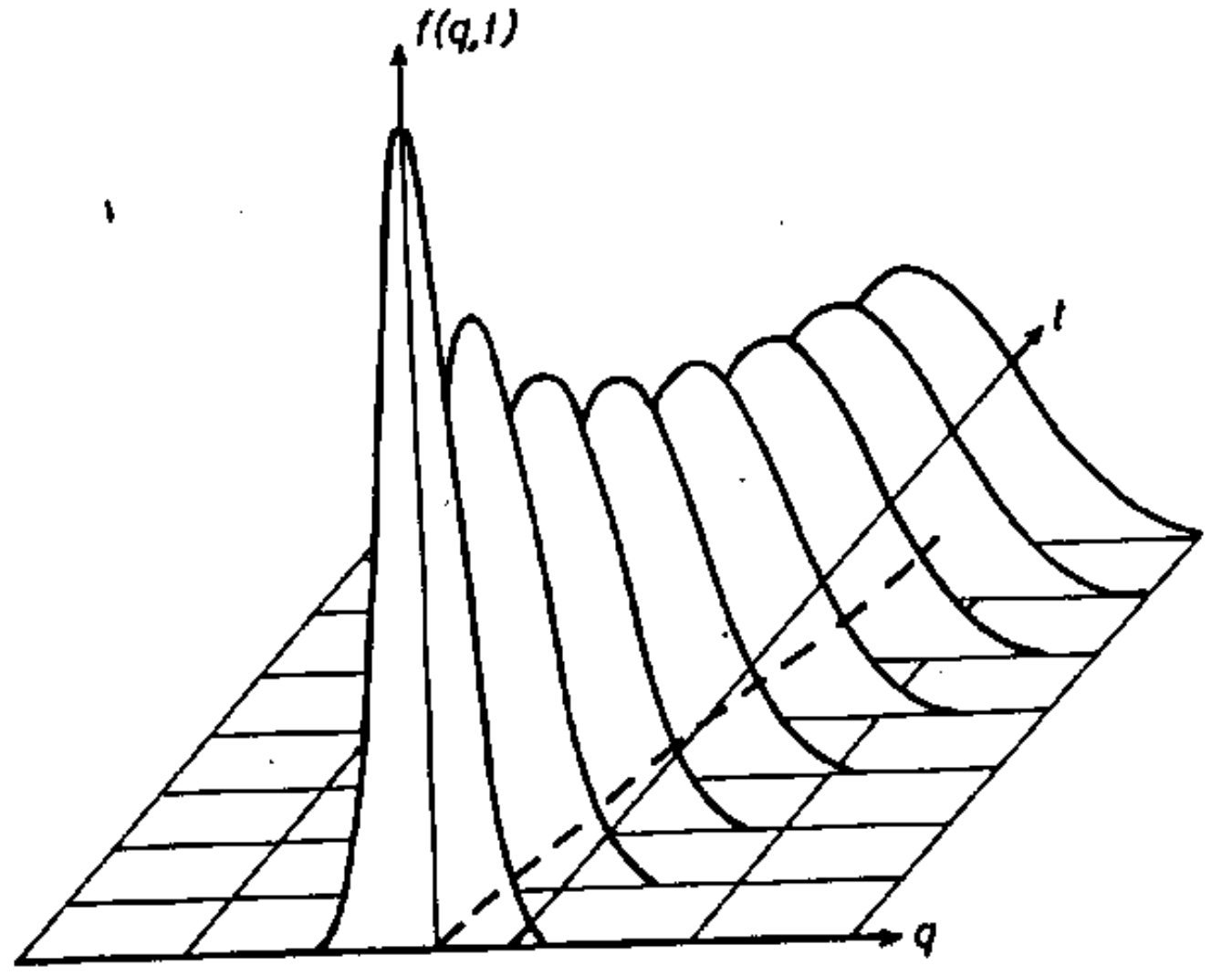

***Figura 41***
***Traiettorie di spostamento più probabili della coppia di atleti***

**Validazione sperimentale.**

Le risultanze dello studio delle traiettorie di spostamento del sistema Coppia di Atleti durante il combattimento, fin qui effettuato con le tecniche della meccanica statistica, in quanto si analizza una classe particolare di eventi, formati da situazioni non ripetibili, che si verificano in maniera "casuale" con una certa probabilità di frequenza su un gran numero di eventi, potrebbero apparire un mero esercizio teorico, se non fossero corroborate da una verifica sperimentale ottenuta per di più da dati ricavati sperimentalmente per altri fini.

Se ricordiamo che il nocciolo della dimostrazione, che il moto è Browniano si basa sul fatto che valgano le relazioni 1 e 2, allora si comprenderà che dimostrare che su base statistica nel tempo non vi sia una direzione privilegiata di moto del sistema "Coppia di Atleti" ma che tutte le direzioni siano equiprobabili significherà dimostrare l'assunto.

Nelle figure si mostrano i dromogrammi ( tracce delle traiettorie di spostamento del sistema "Coppia di Atleti " ) detti anche variogrammi rilevati da studi della competizione di judo ( a fini strategici ) effettuati in Giappone ( 1971 ) ; già il dromogramma somma di soli 12! incontri (quantità statisticamente irrilevante ) mostra chiaramente che non vi è una direzione privilegiata nel moto, (ovvero tutta la superficie risultando coperta dalle tracce appare nera ) pertanto essendo il valor medio nel tempo delle forze di spinta trazione nullo se ne deduce che il moto appartiene alla classe dei moti Browniani.

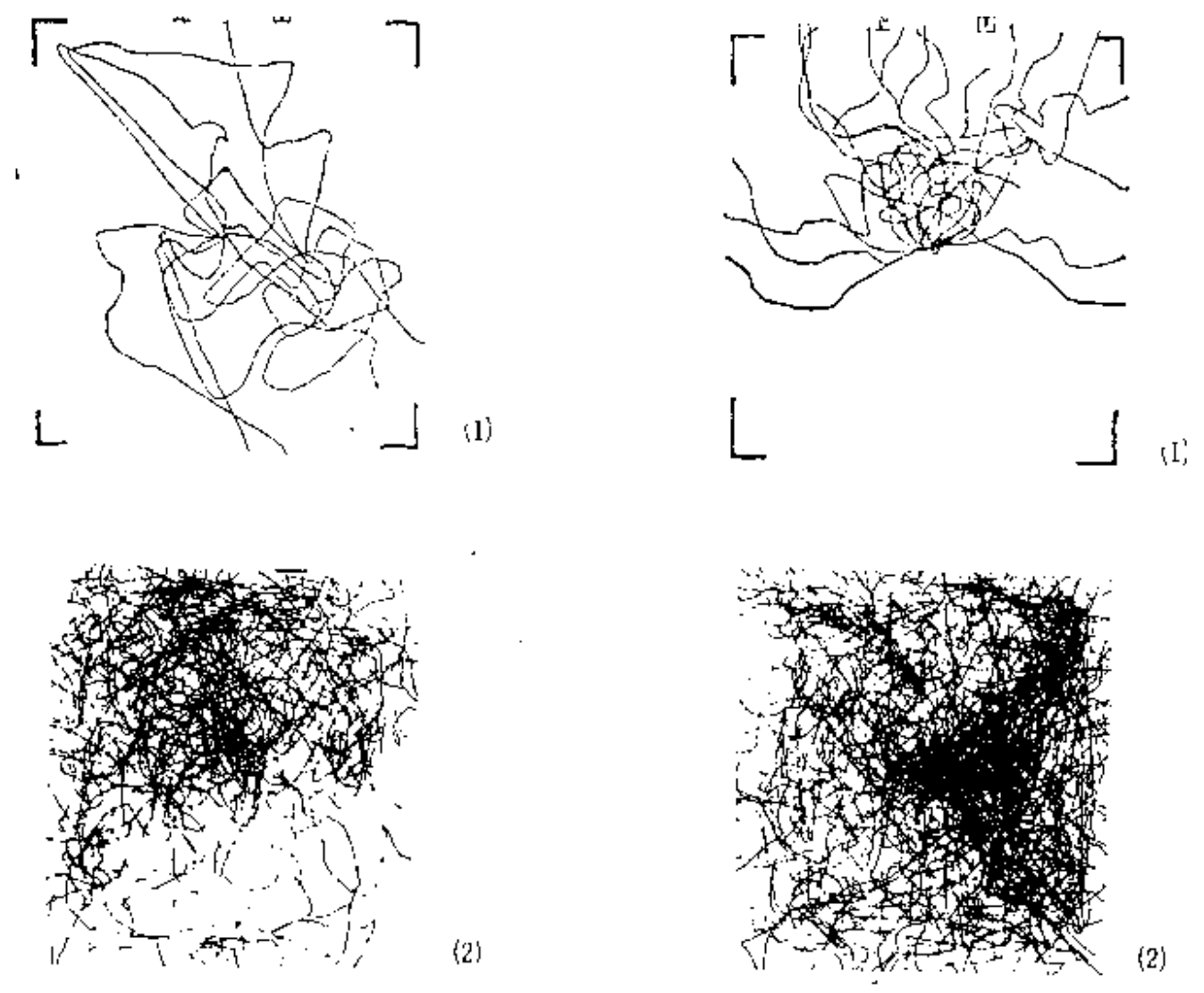

***Fig 42 Dromogrammi o Variogrammi somma di 7 e 12 combattimenti di judo***

***Moderni risultati a supporto delle idee innovative introdotte***

Certo le idee innovative introdotte possono apparire strane, se non fantasiose, riflettendo sul concetto che appare improbabile che nel mondo macroscopico due atleti possano produrre un moto ben conosciuto solo a livello microscopico.
Questo dubbio (lecito) è stato fugato dall'evolversi delle conoscenze fisiologiche, a livello microscopico e dall'introduzione di nuovi concetti matematici a livello sia mesoscopico, che macroscopico.
Nel seguito sarà mostrata in una rapida carrellata come l'avanzarsi delle conoscenze in varie branche quali la fisiologia, la fisica, e la matematica hanno confermato e giustificato le innovazioni proposte e testate negli anni 90 dall'autore.
Mostriamo nel seguito, in modo estremamente sintetico, una panoramica dei risultati delle ricerche degli ultimi trent'anni nel campo della biologia,fisiologia,fisica e matematica applicate al corpo umano a varie scale dimensionali di analisi, che non solo suffragano ma permettono persino di giustificare le risultanze precedenti.
E' prassi consolidata sia per la Biofisica, sia per la Scienza Medica moderna studiare il corpo umano come "sistema complesso", pertanto l'introduzione del concetto di frattalità negli studi ha permesso di approfondire la comprensione della struttura "*Chaotica*" interna del corpo umano.
Ricordando che i frattali sono strutture geometriche complesse, che godono della proprietà di essere auto simili ed invarianti di scala, con dimensione non intera, generati da semplici regole iterative. E' comune distinguere all'interno del corpo umano due tipi di strutture frattali: tutte le strutture geometriche complesse ( definite " frattali statici" ) e tutti i segnali di risposta temporale dei vari organi del corpo, che sono serie temporali non lineari ( definite " frattali dinamici ") .
I sistemi non lineari vengono studiati con la teoria del "*Chaos*" che fornisce informazioni sulla loro evoluzione nel tempo attraverso lo studio di particolari enti geometrico - matematici chiamati "*Attrattori*" tra cui quelli classificati come"*Strani*" hanno dimensione frattale.
La connessione tra questi tipi di strutture statiche e dinamiche è il Moto Browniano Frazionario che di fatto origina tali strutture o permette di analizzare tali risposte
Nel figure seguenti vengono mostrati esempi dei due tipi di frattali presenti nel corpo.

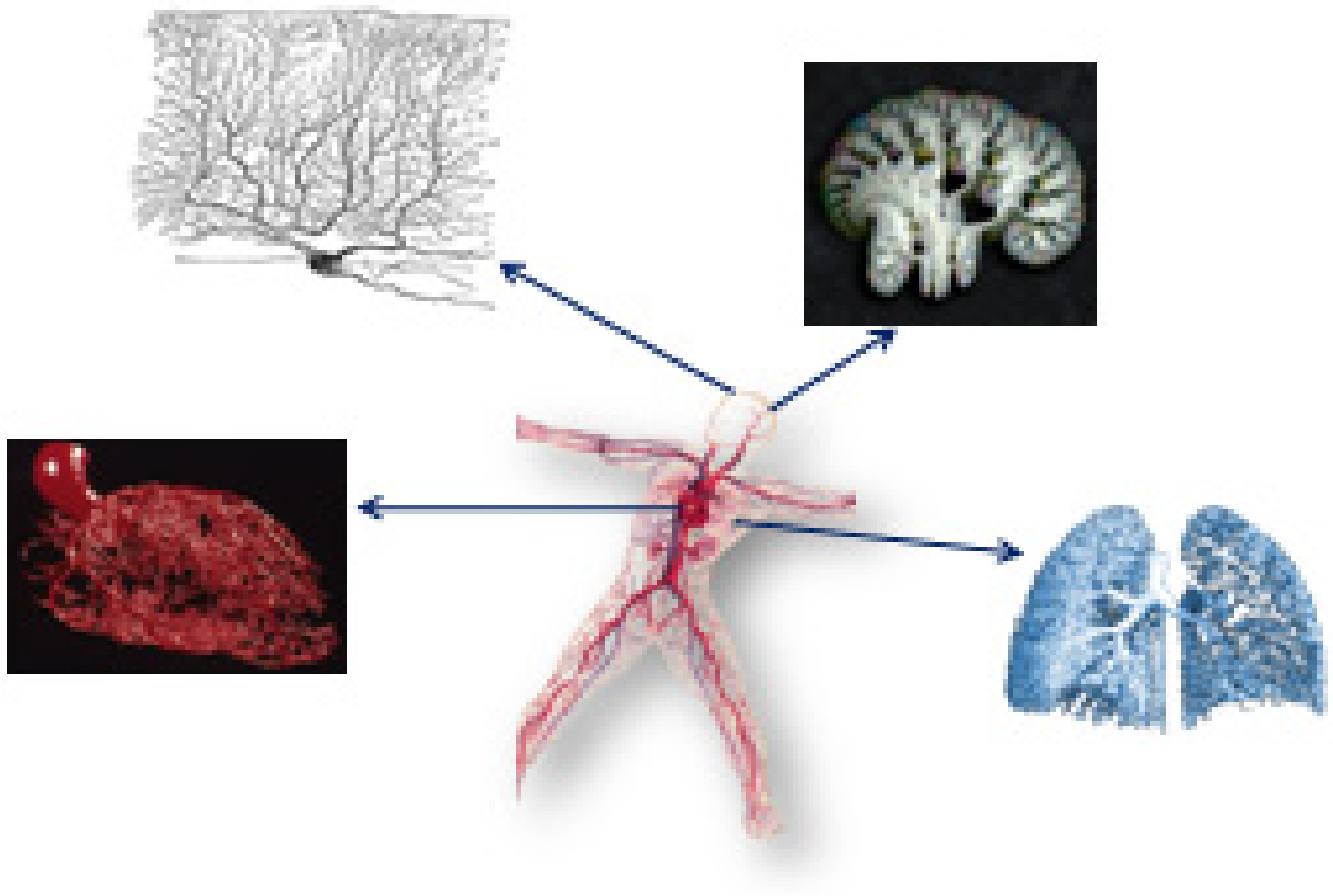

***Fig. 43 Frattali geometrici ( statici) nel corpo umano***

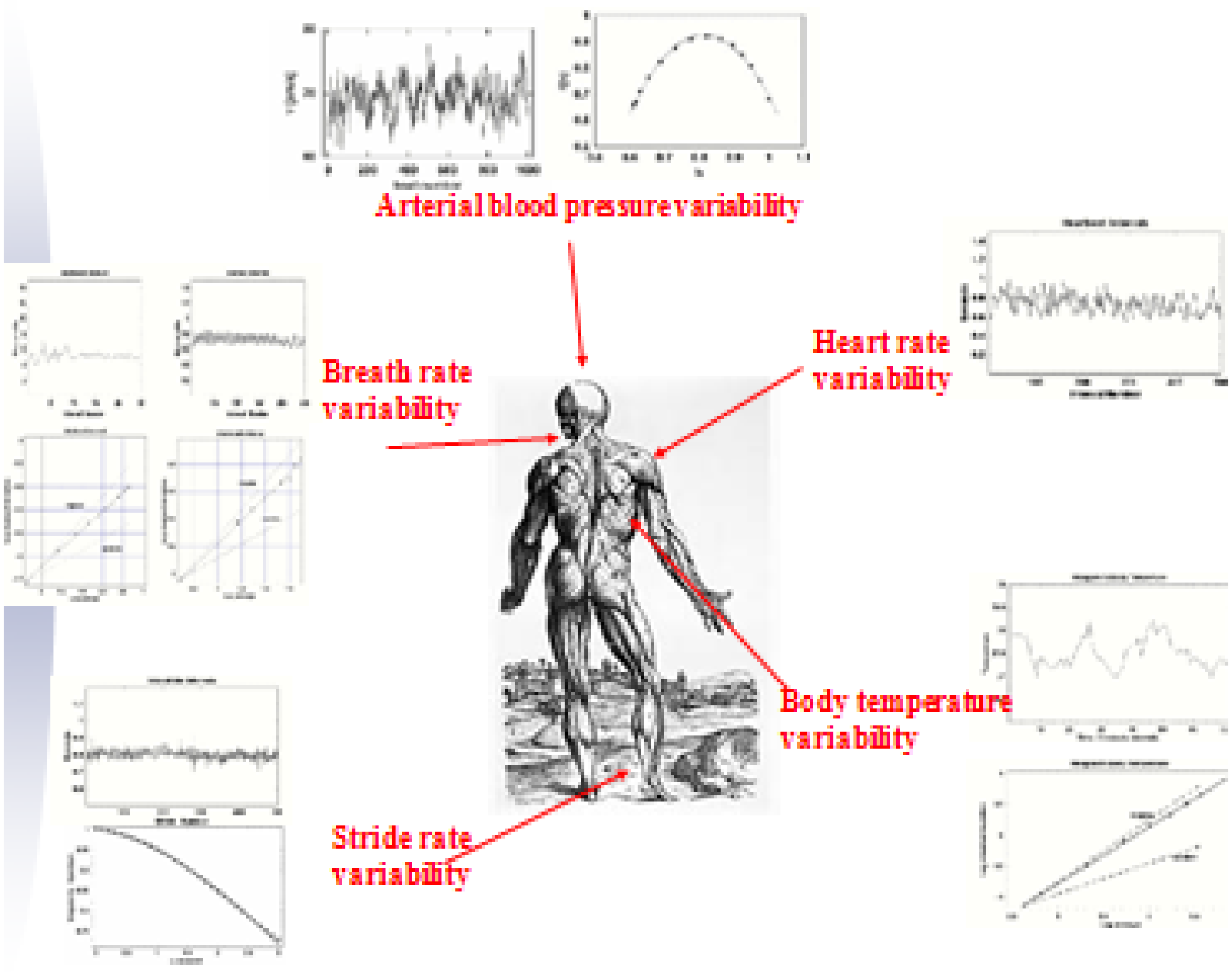

*Fig.44  esempi di risposte temporali del corpo umano ( frattali dinamici)*

## Partendo dal Microscopico

**Il motore non processivo della Miosina II e la contrazione muscolare.**

E' oggi sperimentalmente accettato che il Motore Browniano da conto, in modo accettabile  del meccanismo non  processivo della Miosina II che usando sia le fluttuazioni termiche, sia l'energia chimica conservata nella struttura dell'ATP , "tira" il filamento di Actina. Producendo la contrazione muscolare.
Il modello a motore Browniano non processivo ha dato conto ( Yanagita et al 2000) che manipolando una singola testa di miosina le distanze percorse lungo il filamento di Actina sono dell'ordine dei 5.5nm  per un singolo step meccanico (misure valide anche per il modello classico), ma possono raggiungersi gli 11-30 nm se gli step da due a cinque avvengono in rapida successione (non prevista dal modello classico). Simili sub step osservati sono costanti 5.5 nm sul monomero di Actina, indipendenti dalla forza prodotta e poiché alcuni sono in direzione opposta a quella predetta dal modello classico appare più probabile che la testa della Miosina II si muova lungo il monomero di Actina sulla base di un Moto Browniano.
Nella figura successiva vengono mostrati vari step della Miosina II e V rispettivamente motore non processivo e motore processivo.
Quindi alla base della contrazione muscolare, fenomeno macroscopico studiato dal Borelli vi è un moto microscopico di tipo Browniano.

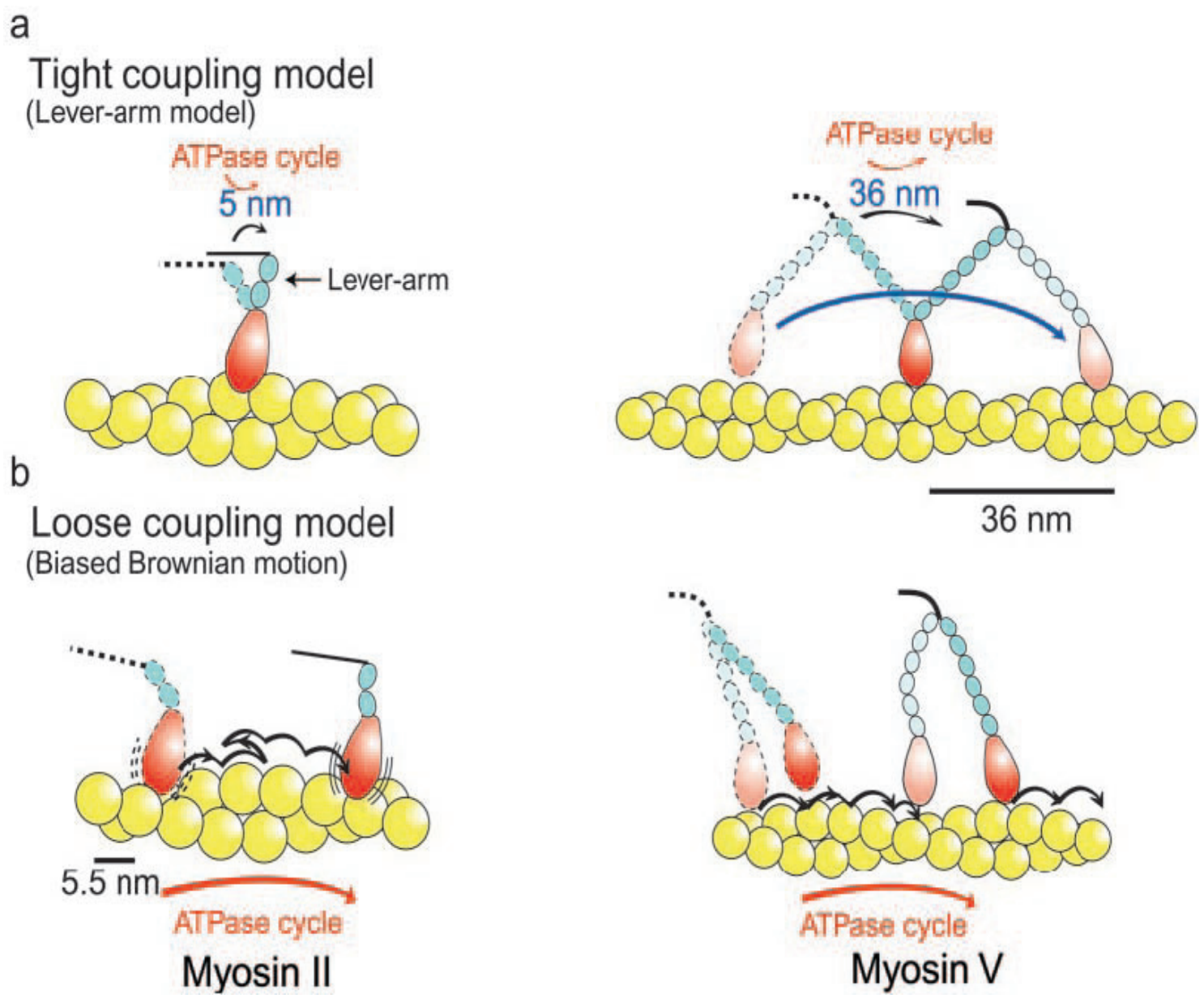

***Fig. 45 Motori Browniani : Miosina II non Processivo, Miosina V Processivo***

# Passando per il Mesoscopico

## Fluttuazioni superficiali della temperatura

Se la superficie del corpo, in condizioni di equilibrio termico, è analizzata nella sua emissione all'infrarosso, mediante una termo-camera, se ζ (t) è la fluttuazione random della radiazione incidente assorbita e ΔT(t) la differenza della temperatura superficiale nel tempo, allora l'equazione che descrive topologicamente la variazione locale di temperatura è del tipo Langevin

$$c_v \frac{d\Delta T(t)}{dt} + h\Delta T(t) = \zeta(t)$$

equazione di tipo Langevin ( Sacripanti)

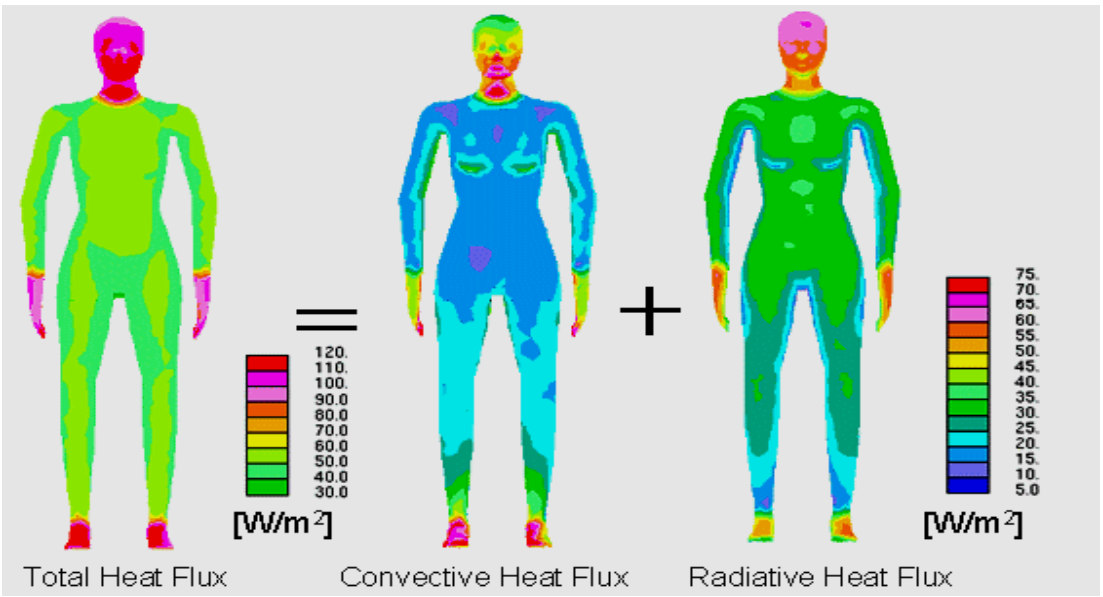

***Fig 46 temperatura superficiale del corpo***

La equazione precedente mostra che la fluttuazione superficiale della temperature del corpo umano è di tipo Browniano.
Ma la topologia della temperatura superficiale presa dalla termo camera mostra facilmente che la temperatura superficiale è una funzione non solo del tempo, ma anche della posizione come si vede facilmente dalla figura successiva.

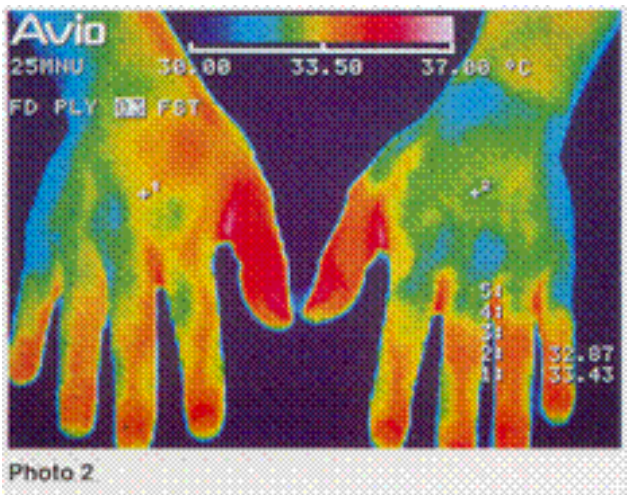

***Fig 47  Distribuzione topologica della temperatura superficiale***

Se però il corpo effettua un lavoro , esempio una tecnica di judo, allora l'equazione variando nel tempo, prende il seguente aspetto:

$$c_v \frac{d^2 \Delta T(t)}{dt^2} + \left( h(t) + \frac{dc_v}{dt} \right) \frac{d\Delta T(t)}{dt} - \frac{d\zeta(t)}{dt} = 0$$

In realtà  il coefficiente termico h che spesso viene trattato come costante,  non è costante, ma la sua dipendenza temporale è molto complessa , essa infatti per lavori come tecniche di lancio di Judo deve soddisfare la seguente  relazione sperimentale di Sacripanti:

$$S\sigma\varepsilon\left(\frac{T_s^4 - T_a^4}{t-t_0}\right) + 0.6n\frac{kS\,\mathrm{Re}^{0.8}\,\mathrm{Pr}^{0.33}}{l}\frac{T_i - T_a}{t-t_0} + \left\{ e^{-\frac{4S}{lh}\frac{T_s-T_b}{T_b-T_a}} \right\}$$

$$\left\{ \left[ 0.132\varepsilon_h \frac{4S^2 k\,\mathrm{Re}^{0.8}\,\mathrm{Pr}^{0.33}}{hl^2}\frac{(T_s - T_a)^{1.2}}{T_a^{0.2}(t-t_0)} \right] + \left[ 0.16(1-\varepsilon_h)\frac{4S^2 D\lambda\,\mathrm{Re}^{0.8}\,Sc^{0.33}}{Rl^2 h}\left(\frac{M_s e_s}{T_s} - \frac{M_a e_a}{T_a}\right)\frac{(T_{vs} - T_{va})^{1.2}}{T_{va}^{0.2}(t-t_0)} \right] \right\}$$

$$\left\{ e^{-(0.2\varepsilon^2 + 0.5\varepsilon - 0.7)\frac{\lambda P - \sum}{\lambda P}} - 1 \right\} = \frac{dO_2}{dt}$$

I termogrammi seguenti fanno parte di una ricerca pionieristica sviluppata alla fine degli anni ottante inizio novanta  dall'autore e collaboratori  tra CONI;ENEA e FILPJK,, in essa venivano connessi l'emissione termica ed il consumo d'ossigeno degli atleti ( che è proporzionale al lavoro effettuato), mediante la precedente relazione messa a punto dall'estensore di queste note.
Nelle figure successive è mostrata una tecnica di Judo analizzata con la termo camera.
Da essa si rilevano non solo i muscoli maggiormente impegnati, per visione diretta , ma anche il consumo d'ossigeno come valutazione indiretta ottenuta mediante l'equazione.

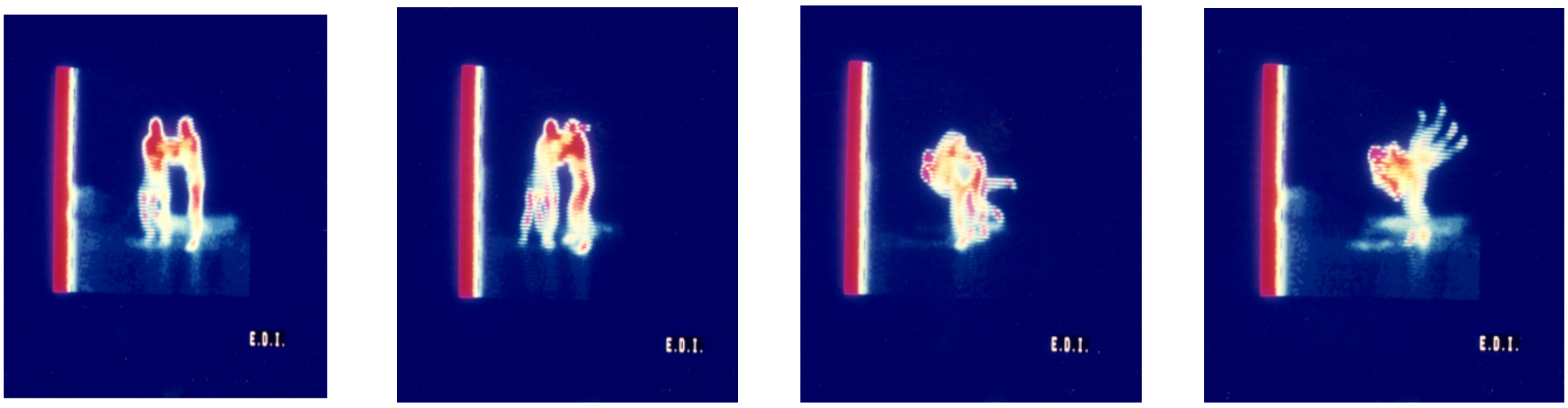

***Fig 48  Termogrammi di una tecnica di judo***

## Giungiamo al Macroscopico ( Statico)

### Equilibrio del corpo umano  Centro di Pressione ( COP )  vs. Centro di Massa ( COM)

Nella condizione di stazione eretta il COM ed il COP si trovano sulla stessa verticale ed il COP può coincidere con la perpendicolare baricentrale al suolo.
E' cosa nota in neurofisiologia che il diagramma prodotto nel tempo dalle coordinate del COP è un moto Browniano, anzi meglio secondo Collins e de Luca, se si considera anche il tempo di stasi del punto, esso è un Moto Browniano frazionario (che è geometricamente legato alle strutture frattali come Mandelbrot  e van Ness dimostrarono nel 1968).
La meccanica classica non spiega il Moto browniano delle coordinate del COP , esse possono essere spiegate invece dall'equazione di  Hastings & Sugihara 1993  che combina il termine stocastico con l'attrito come nell'equazione di tipo Langevin

$$dx(t) = -rx(t)dt + dB(t)$$ equazione di tipo Langevin ( Hastings & Sugihara)

Dove dB è un rumore gaussiano con media zero.

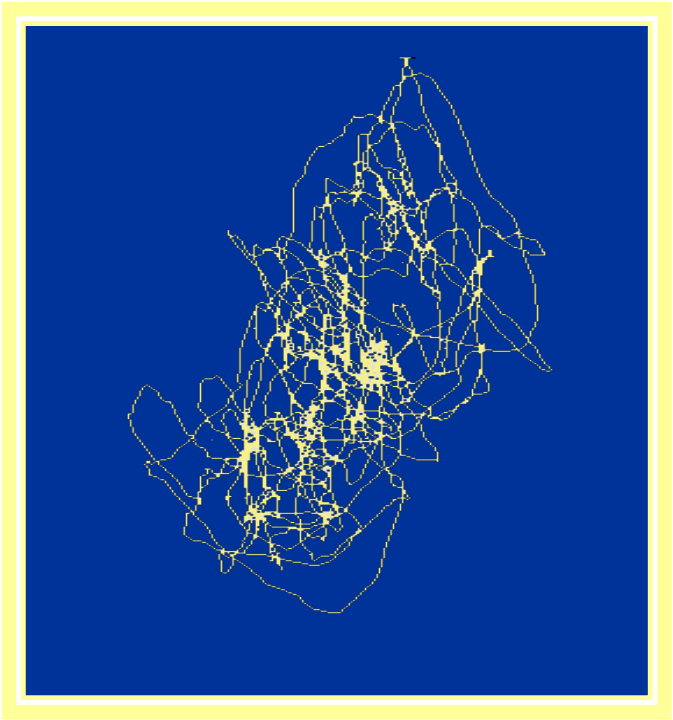

*Fig 49  Posturogramma- Moto Browniano  delle coordinate del COP.*

## Ed al Macroscopico (Dinamico)

### Multifrattali Nella locomozione umana

La locomozione è un'attività volontaria molto complessa, le strutture tipiche mostrano che le serie temporali degli intervalli tra i passi suggeriscono particolari meccanismi  neuromuscolari che possono essere modellati matematicamente.
La natura frattale delle serie temporali della frequenza del passo fu incorporata in un modello dinamico da Hausdorff , usando un modello stocastico esteso successivamente da Askhenazi e collaboratori,  in modo da descrivere l' evoluzione dinamica della locomozione con l'età.
Il modello era essenzialmente un moto Browniano su di una catena neurale Markoviana, in ogni nodo della quale agisce un potenziale d'azione con una intensità particolare quando è interessato da un  moto browniano.
Questo meccanismo genera un processo frattale con aspetti multi frattali, in cui il tempo dell'esponente di Holder, dipende parametricamente dall'ampiezza del passo variabile con variazione Brownianae dal tempo variabile di sosta del piede, anch'esso Browniano.

L'analisi multi frattale viene spesso utilizzata per studiare la dinamica della locomozione per pazienti con affezioni tipo Parkinson o emiparesi.

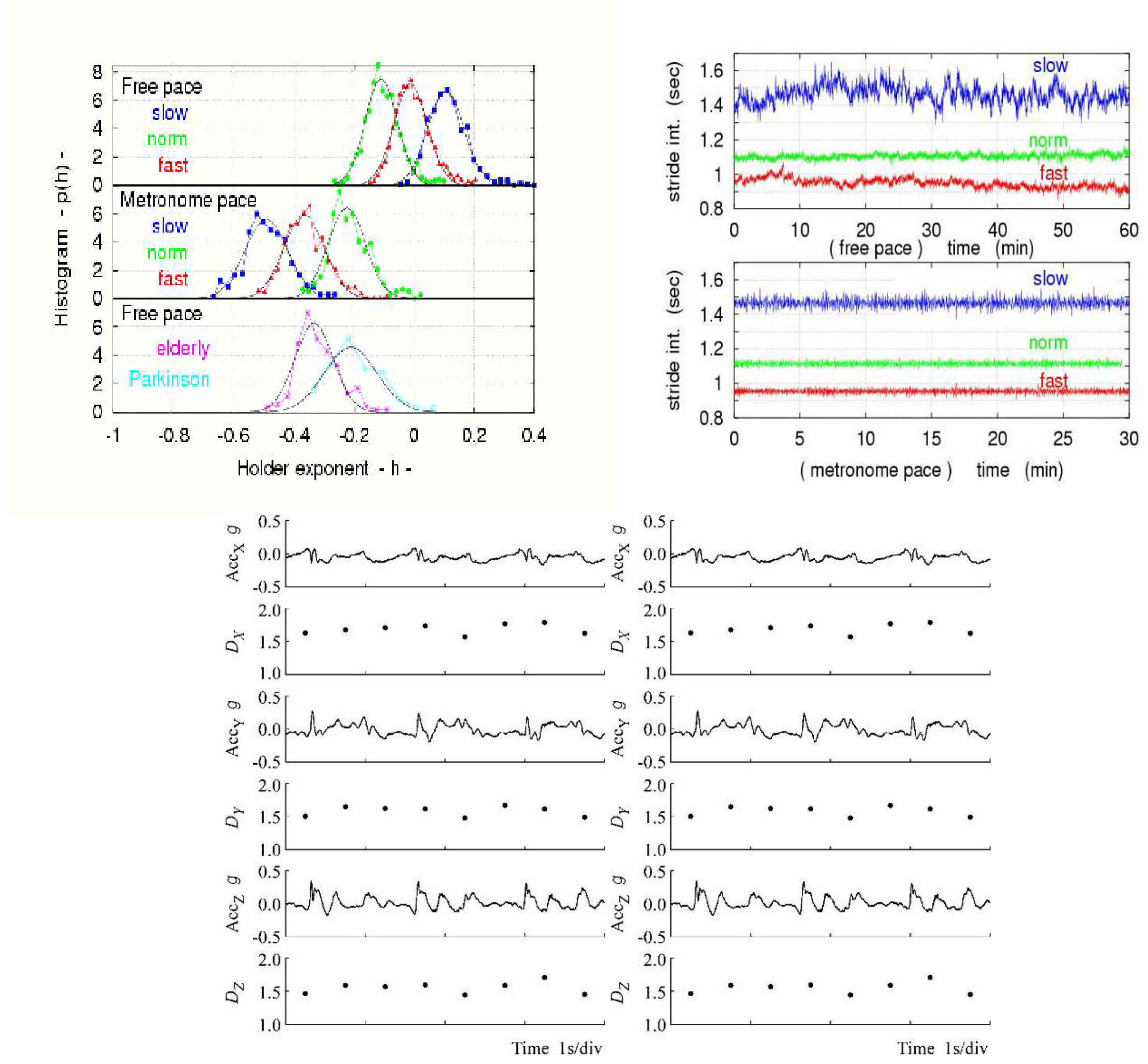

***Fig.50 locomozione di individui normali, con Parkinson o affetti da emiplasia***

***Moto della proiezione del Baricentro della Coppia di Atleti***

Dopo questi brevi accenni di alcune delle ultime risultanze della fisiologia e neurofisiologia ottenute negli ultimo anni, sulla base delle più avanzate conoscenze della meccanica non lineare e della geometria frattale, ritorniamo sui risultati ottenuti più di un decennio fa ed alla luce dei più moderni studi in matematica cerchiamo di validare i risultati ottenuti anche per via teorica indiretta. Si è visto così che la base della contrazione muscolare è retta da un moto browniano, mentre nella condizione ferma di stazione eretta il centro di pressione del peso sotto la pianta del piede non è un punto come potrebbe attendersi, ma le sua coordinate descrivono un moto complesso che è stato individuato come Moto Browniano Frazionario.
Anche la locomozione non risulta monotonamente uguale , ma le variazioni di lunghezza del passo e di tempo di stasi del piede durante il movimento sono legati da una relazione multifrattale o in altri termini sono rette da un Moto Browniano Frazionario.

Nel caso del sistema coppia di Atleti, dopo aver stabilizzato le prese, il moto viene descritto dalla traccia della proiezione del baricentro totale del sistema sul tappeto.
Ricordando che il baricentro del sistema è legato ai baricentri dei singoli atleti, possiamo affermare che esso appartiene alla classe dei moti Browniani.
E' ovvio che il Moto Browniano frazionario bidimensionale del COM del sistema è connesso con i Moti Browniani frazionari delle due proiezioni dei singoli baricentri degli atleti.
Il Moto Browniano Frazionario è un moto complesso, legato ai frattali, che mostra proprietà di auto similarità.
Ricordando che il posturogramma mostrato precedentemente per condizioni statiche , viene chiamato variogramma quando esso mostra l'evoluzione del tempo del posturogramma.
Il variogramma in questo caso è dato dalla formula:

$$\left\langle [X(t)-X(0)]^2 \right\rangle \approx t^{2H}$$

che richiede una descrizione bilogaritmica.
Nella figura seguente viene mostrata la decomposizione di una traccia di combattimento che è un Moto Browniano Frazionario bidimensionale ( variogramma) in due monodimensionali

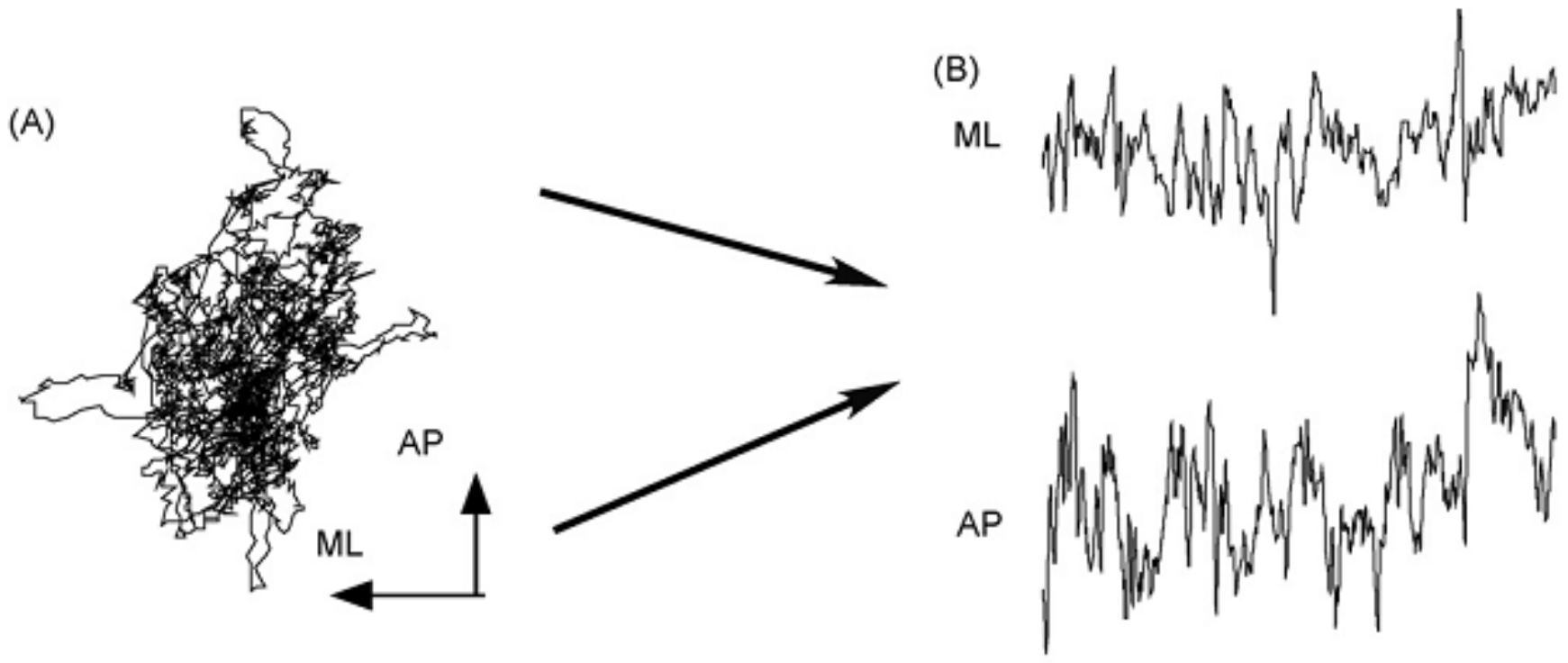

***Fig.51 Decomposizione di un moto Browniano frazionario Bidimensionale in due monodimensionali***

In questo caso la dimensione frattale frazionaria D lungo ciascun asse è legata all'esponente di scala H ( esponente di Hurst) , nel caso mostrato D=1-H.
Come mostrato il variogramma di un combattimento è un Moto browniano frazionario bidimensionale che corrisponde alla proiezione del baricentro dell'atleta che si muove nello spazio di un moto tridimensionale.
Anche se questa affermazione appare logica e banale, essa è molto lontano dall'evidenza scientifica perché fino ad oggi non esistevano valutazioni teoriche che legassero le proprietà 3D di un Moto Browniano frazionario a quelle 2D.

## *Il Teorema delle proiezioni*

Solo recentemente un avanzatissimo studio teorico ha mostrato che il parametro H di un frattale isotropo n-dimensionale è connesso con un frattale (n-1) dimensionale attraverso la seguente funzione di auto similarità:

$$H_{(n-1)D} = H_{nD} + 0.5 \quad \textit{che nel nostro caso è} \rightarrow H_{2D} = H_{3D} + 0.5$$

La figura successive che mostra le risultanze del teorema, permette anche di definire corretta l'intuizione teorico sperimentale proposta qualche decennio fa, ovvero:

***Teorema delle Proiezioni - (Sacripanti 2009) -***
***In Competizione, il Moto Browniano Frazionario, tracciato sul tappeto dalle coordinate della perpendicolare del baricentro dell'atleta, è proiezione bidimensionale del Moto Browniano Frazionario tridimensionale che il Centro di Massa dell'atleta effettua nello spazio .***

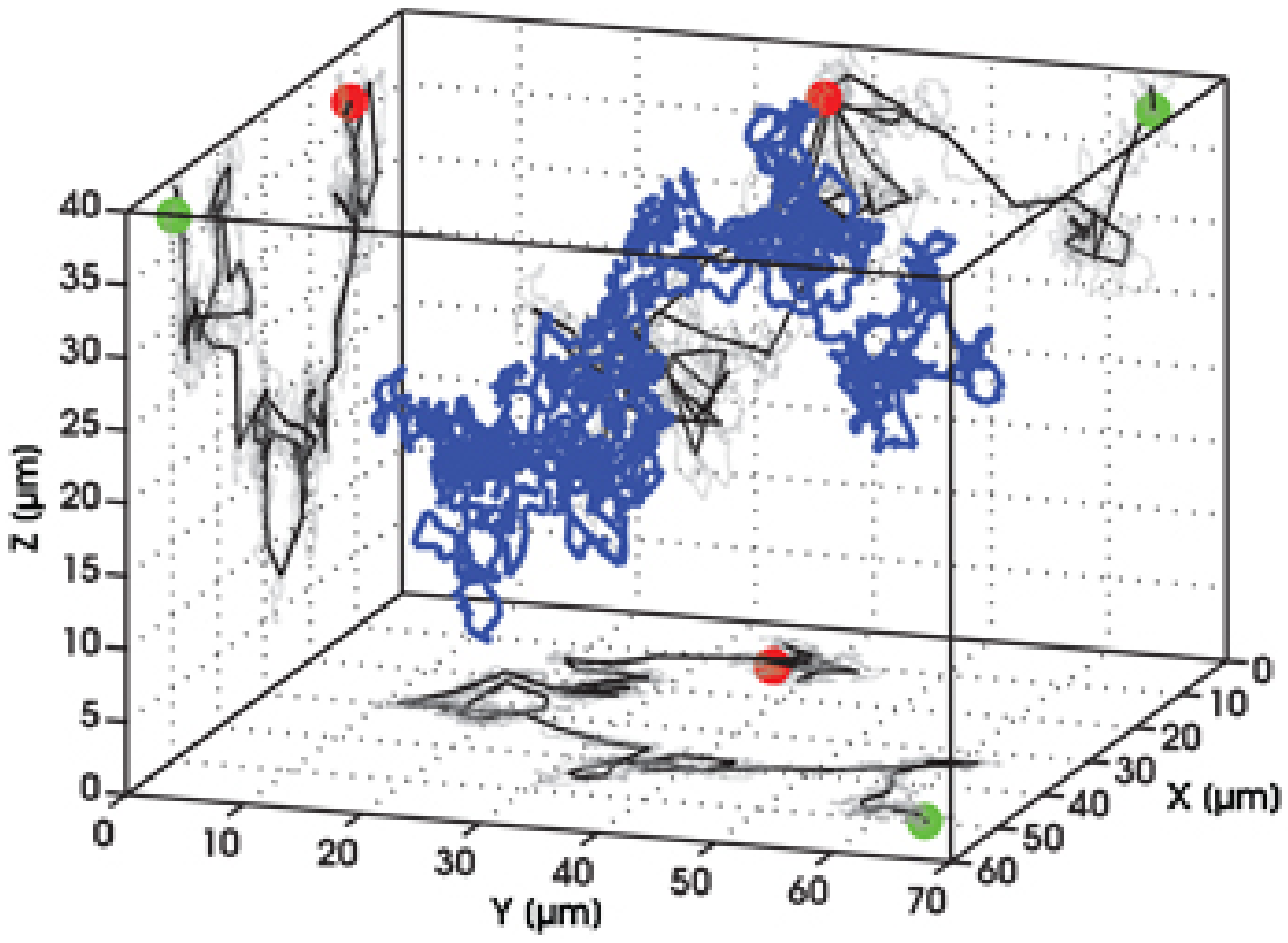

***Fig.52 Illustrazione geometrica del Teorema delle Proiezioni ( Sacripanti)***

***Dall'approccio Browniano a quello Newtoniano.***

Normalmente nella vita quotidiana e nell'uso comune valgono l'approccio lineare e la dinamica Newtoniana, come è pensabile connettere la visione Browniana con quella newtoniana?
Senza indulgere in tediose dimostrazioni matematiche lontane dal livello di questo viaggio, diciamo che è possibile dimostrare che l'approccio newtoniano è connesso con la media di tempi e spazi lunghi, mentre il Moto Browniano Frazionario caratterizza tempi molto piccoli e spazi microscopici di osservazione.

Ma traiettorie microscopiche e macroscopiche sono connesse da ovvi fattori di scala che si basano sull'auto somiglianza e matematicamente sull'auto-similarità .
Una ulteriore e definitiva dimostrazione degli assunti si fonda sulla valutazione computazionale dell'equazione del primo modello proposto dall'autore, per ritrovare tracce auto simili a quelle sperimentali. Le seguenti figure mostrano i risultati di questa indagine definitiva.

Fig. 53a
**N-W** asymmetry
one game
a **1°** realization

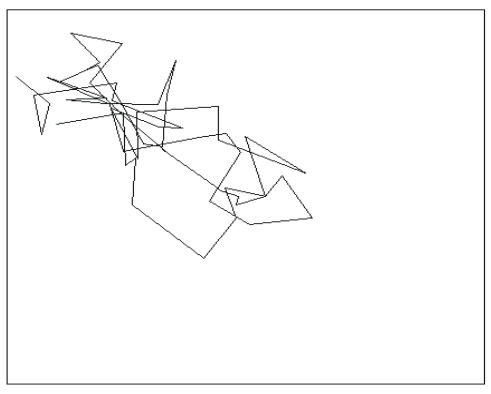

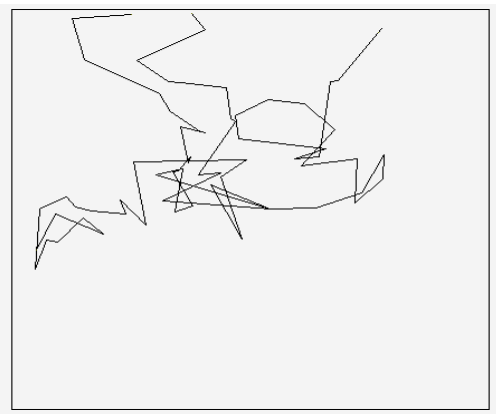

Fig. 54a
**S-N** asymmetry
one game
a **1°** realization

Fig. 53b
**N-W** asymmetry
one game
a **2°** realization

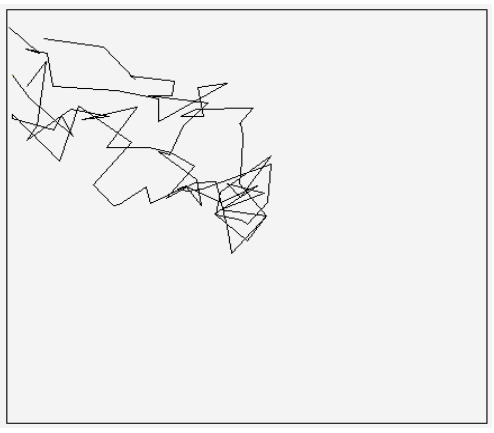

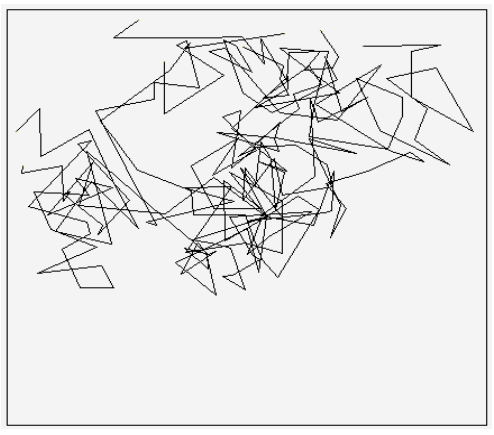

Fig. 54b
**S-N** asymmetry
one game
a **2°** realization

Fig. 53c
one game
**experimental** tracks

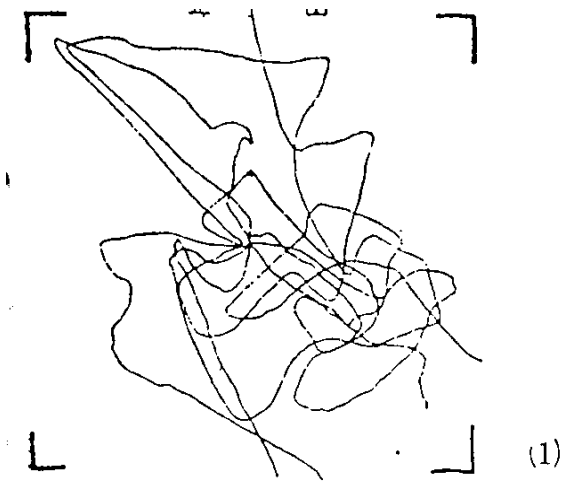

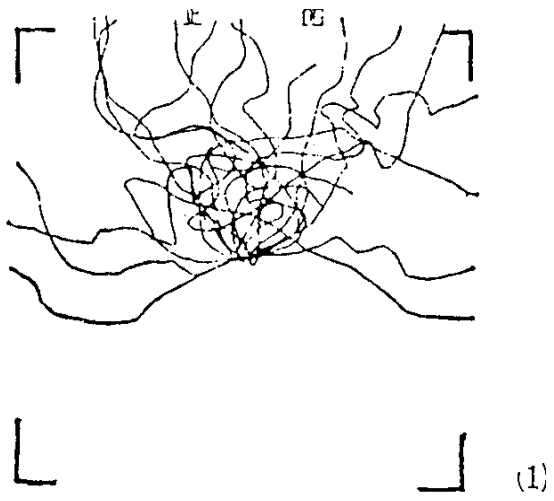

Fig. 54c
one game
**experimental** tracks

Fig. 55a
7 games
**experimental** tracks

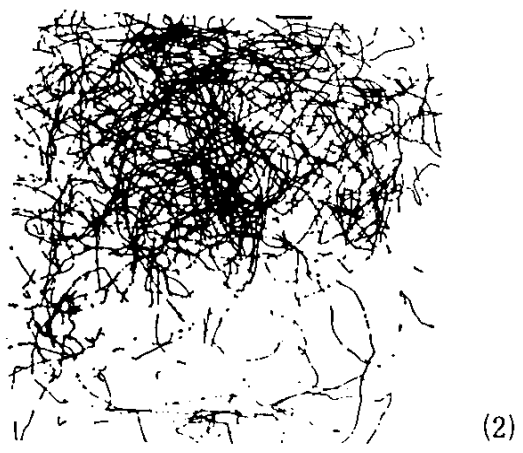

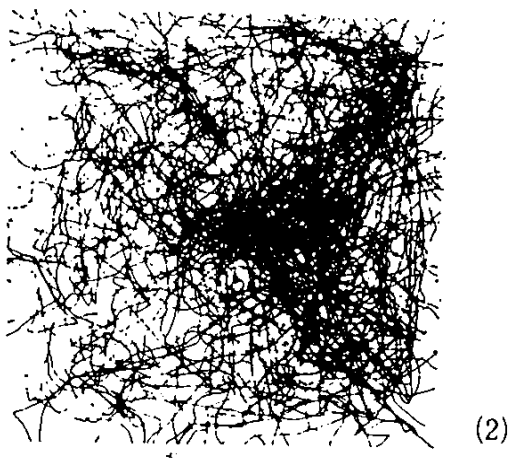

Fig. 56a
12 games
**experimental** tracks

Fig. 55b
**N-W** asymmetry
7 games
a **1°** realization

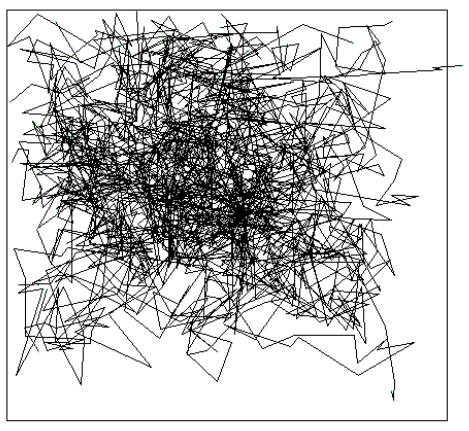

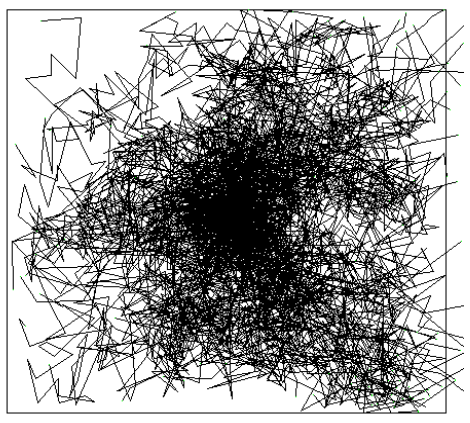

Fig. 56b
**N-E** asymmetry
12 games
a **1°** realization

Fig. 55c
**N-W** asymmetry
7 games
a **2°** realization

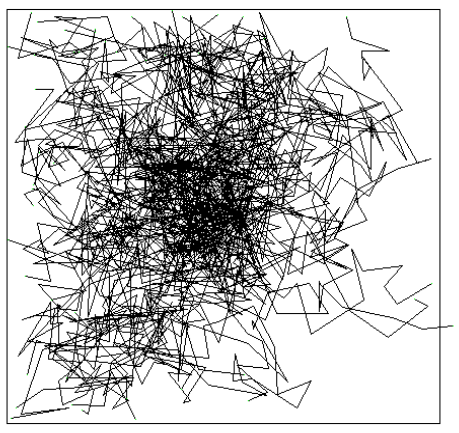

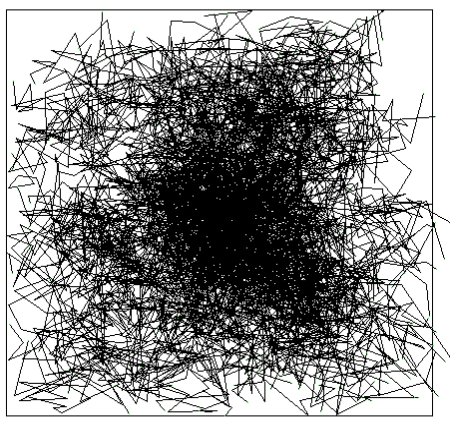

Fig. 56c
**N-E** asymmetry
12 games
a **2°** realization

***Fig 53-56 Ricostruzione computazionale mediante l'equazione di Sacripanti delle tracce sperimentali di spostamento.***

***La più recente sistematizzazione : Match Analysis per il Judo.***

Da un punto di vista biomeccanico, la competizione di Judo è un sistema complesso, non lineare, con intriganti aspetti " chaotici" e frattali.
La competizione è in realtà il banco di prova dove sono valutate sia le capacità dell'allenatore, sia quelle dell' atleta.

La competizione è il momento della verità relativamente al tempo speso nel condizionamento fisico o nella preparazione tecnica, dall'atleta, ed è anche il climax dal punto di vista dell' insegnamento. Inoltre, essa è la più importante sorgente di informazioni tecniche.

Per cui il suo studio è essenziale agli allenatori per ottenere informazioni utili per le loro metodiche di allenamento.

La Match Analysis ( studio della competizione) deve essere considerata come la chiave di volta della comprensione degli sport di situazione (duali o di squadra) come il Judo.
Essa è uno strumento essenziale di aiuto nel compito difficile dell' allenatore per la preparazione Nazionale o di livello Olimpico.

Vengono ora condensate le più importanti attuazioni metodologiche e la sistematizzazione che l'autore ha prodotto in questi ultimissimi tempi.
Sulla base del più moderno sviluppo tecnologico, la Match Analysis è una fonte preziosa di quattro livelli di informazioni:

1. Dati fisiologici dell'atleta
2. Dati tecnici
3. Strategie
4. Conoscenza dell'avversario

Questi mezzi di analisi della competizione risultano essenziali nella gestione tecnica di una squadra nazionale.

Primo l'analisi competitiva, va applicata al sistema "Coppia di Atleti", questa riflessione ha permesso di individuare gli osservabili che sono costanti in competizione.

Questi osservabili furono classificati in termini dei principi biomeccanici connessi con le prese sull'avversario.

Tale analisi ha fatto si che potessero essere individuate sei classi di posizioni della Coppia definite Invarianti di Competizione, vedi figura successiva.
(Fig.57)

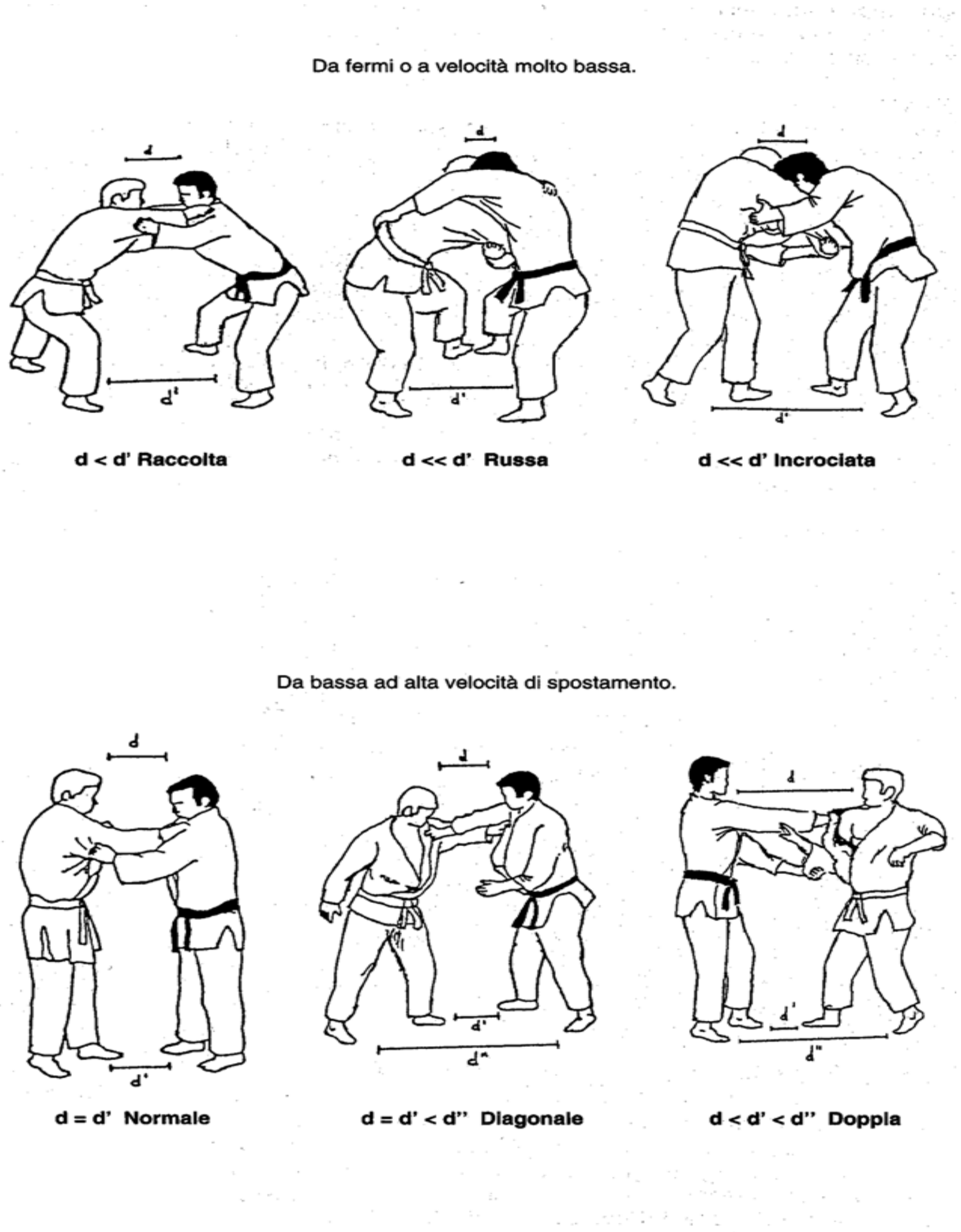

***Fig 57 le sei classi di invarianti di competizione***

Ora come esempio finale delle potenzialità insite nell'individuazione del moto della Coppia di Atleti in competizione verrà mostrato il loro utilizzo in match analysis
Le tracce del sistema Coppia di Atleti sono uno strumento utile pieno di informazioni nascoste.
Esse infatti sono sorgente di informazioni nascoste , ma il prezzo da pagare è una non banale analisi di queste serie temporali con strumenti non lineari.
Nel Moto Browniano Frazionario inizialmente presentato da Mandelbrot e van Ness nel 1968 le serie di temporali possono essere considerate una combinazione di meccanismi deterministici e stocastici.
Il concetto sviluppato attraverso Moto Browniano frazionario è una generalizzazione del lavoro di Einstein che mostrò che un processo stocastico è caratterizzato da una relazione lineare tra lo spostamento quadratico medio $<x^2>$ ed il tempo crescente t in formule::

$$\left\langle x^2 \right\rangle = 2D\Delta t$$

Il principio generale della struttura del Moto Browniano Frazionario è che l'aspetto della traiettoria, espressa come una funzione di tempo può essere quantificata da un numero non intero o da una dimensione spaziale frazionaria, offrendo in tal modo una misura quantitativa della uniformità della traiettoria.

È possibile scrivere in forma matematica:

$$D_t^{\alpha}\left[X(t)\right]-\frac{X(0)}{\Gamma(1-\alpha)}t^{-\alpha}=\xi(t)$$

Il primo termine è una derivata frazionaria, il secondo è connesso alla condizione iniziale del processo, e il terzo è sempre l'azione random della forza sul COM.
In questo caso è importante conoscere lo spostamento quadratico medio

$$\left\langle\left[X(t)-X(0)\right]^2\right\rangle=\frac{\left\langle\xi^2\right\rangle}{(2\alpha-1)\Gamma(\alpha)^2}t^{2\alpha-1}\propto t^{2H}$$

Da questa espressione è possibile capire che siamo in presenza di processi di diffusione diversi, identificati dal parametro di Hurst.
In particolare questo parametro è indipendente dal tempo e descrive i moti Browniani :anti-correlati per $0 < H < 1/2$ ; e correlati per ½ < il H <1. Se il H = ½ parleremo di moto Browniano puro.

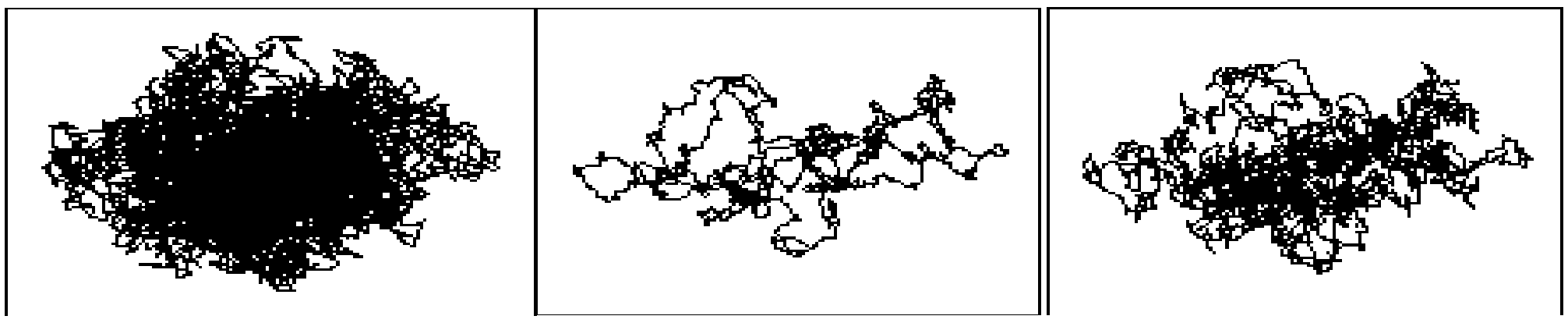

A B C

***Fig 58 a,b,c Moti Browniani***

***A) Persistente; B)Antipersistente; C) Puro***

**Riflessione sulle informazioni strategiche ottenibili .**

I tracciati dell'atleta (Dromogrammi o Variogrammi ) sono l'evoluzione nel tempo della proiezione del COM sull'area del tappeto. Normalmente nella vecchia Match Analysis ogni azione tecnica o lancio si riteneva che appartenesse alla classe dei Sistemi Markoviani questo significa che l'azione era dipendente solo dall'istante precedente, senza alcuna correlazione con i movimenti passati.
L' approccio matematico più avanzato che è stato utilizzato, è invece capace di superare questa limitazione concettuale.
Infatti una caratteristica importante della modellazione con il Moto Browniano Frazionario è la presenza, per ogni combattente, delle correlazioni a lungo termine tra passato ed incrementi futuri. Questo vuol dire che il sistema è non Markoviano, ma essendo correlato al passato risulta più simile alla situazione reale.

***Quindi una traiettoria di combattimento può mostrare, se è analizzata correttamente, se il combattente ha una specifica strategia di combattimento o no (moto casuale) durante la sua competizione.***

Questa informazione può ottenersi attraverso l'analisi del coefficiente H ( coefficiente di Hurst).

Infatti un valore medio di 0.5 per il coefficiente di Hurst indica che non c'è nessuna correlazione, così la traiettoria mostra una distribuzione casuale (Moto Browniano puro).
D'altra parte se H differisce da 0.5 ; H positivo (H> 0.5) o negativo ( H < 0.5) può essere dedotta la correlazione col modo di combattere, indicando che una dato aspetto dell' iniziativa è sotto controllo. Il tutto dipende da come H è posizionato, riguardo al valore medio 0.5, per cui si può dire che il soggetto controlla o meno la traiettoria (e dunque l'evoluzione della lotta nel tempo):. Più vicino H è a 0.5, più importante è il contributo dei processi stocastici (attacchi casuali senza la strategia). Se H è maggiore o meno di 0.5, possono essere dedotti due comportamenti generali:

1. persistente (attacco) o
2. antipersistente ( difesa) .

In altre parole, se la proiezione del centro di massa , ad un certo tempo si sposta con probabilità maggiore verso una direzione determinata, la probabilità maggiore è che in questa direzione persista un comportamento di attacco o al contrario se arretra nella direzione opposta, potremo dire che persiste un comportamento difensivo.
L'uguaglianza tra queste due probabilità indica che non c'è una strategia definita nella lotta, come il semplice moto casuale o processo stocastico.
Queste informazioni ottenute da una "lettura matematica" pura delle traiettorie; possono essere accresciute aggiungendo alla "lettura matematica avanzata" informazioni biomeccaniche, come le prese, gli invarianti di competizione usati, gli invarianti d'azione preferiti, ecc.
E 'possibile, con questo arricchimento di dati, ottenere molte informazioni utili per la strategia sotto forma di albero di probabilità e trattare successivamente questo albero di informazioni con opportuni algoritmi di " Data Mining" per ottenere una classificazione delle strategie potenzialmente efficaci connesse con le traiettorie di spostamento e le altre informazioni biomeccaniche. Tali informazioni, ordinate per importanza ed efficacia, sono utili sia per gli atleti , sia per i coach. Questo è un esempio delle informazioni avanzate che possono ottenersi per mezzo di questo strumento a tutt'oggi sottostimato: le traiettorie di combattimento degli atleti.

## Conclusione

In questo breve viaggio siamo partiti un giorno 28 del 1608 da Napoli, in cui è nato Alfonso Borelli grazie alla cui eclетticità è stata evidenziata non solo la forza del metodo sperimentale e la capacità della matematica di essere l'alfabeto dell'universo, ma anche il corretto modo di affrontare i sistemi biologici, come sistemi complessi e la necessità quindi di utilizzare per essi tecniche e metodi multidisciplinari.
Siamo poi giunti un giorno 28 del 1860 nel lontano Giappone dove nasceva un altro uomo eccezionale Jigoro Kano che ha introdotto un concetto innovativo nel campo dell'"Educazione" : " la motricità vista non solo come mezzo di miglioramento del singolo, ma addirittura come sistema di crescita ed armonizzazione globale di una società composta da singoli, affratellati dal gusto e dall'armonia e mitigati dalla disciplina e dal rispetto reciproco".
Di questo retaggio Ju Do, oggi come la punta di un iceberg , è all'attenzione del mondo e dei mass media solo la parte sportiva ( circa il 10% del volume totale, l'altro 90% è nascosto e sconosciuto ad occhi profani) .
Siamo poi ritornati a Napoli il 28 del 2010 per annodare definitivamente l'approccio interdisciplinare ai sistemi complessi, introdotto dal Borelli, alla "parte emersa" del judo lo Sport Olimpico, sistema fisico complesso per la cui completa trattazione non ci si può arrestare alla Fisica Newtoniana, ma bisogna mutuare aiuti dalla matematica del Chaos e dei Frattali e dalla fisica moderna dei sistemi stocastici non lineari.
Questo piccolo nodo, noto come " Biomeccanica del Judo", scienza sperimentale atta a studiare moti complessi, non può certo affrontare il lato educazionale del Judo.
Pur tuttavia il "Fenomeno" Judo risulta tanto complesso che già dall'inizio Kano e discepoli gli hanno dato, come abbiamo visto, una impostazione fondata su salde basi scientifiche perché solo la scienza permette di strutturare un sistema cognitivo in modo chiaro ed univoco.
Ma se il metodo educativo resta saldo sulla base delle due massime di Kano ( massima efficienza, e benessere e mutua prosperità), la comprensione del fenomeno fisico e della sua evoluzione nel tempo, necessita di un sempre più accurato approfondimento, mediante l'utilizzo dei mezzi scientifici che man mano vengono messi a disposizione dei ricercatori.
La comprensione del gesto sportivo permette di elaborare vari modelli fisico- fisiologico - matematici del sistema motorio. Sulla base di essi si ricerca la variante ottimale da sperimentare e modificare poi alla luce dei problemi insorti nell'esecuzione pratica.
Judo come Sport appartiene alla classe complessa degli sport di situazione duali.
Gli sport di situazione sono quegli sport in cui, l'espletazione della prestazione sportiva non può individuarsi in una periodizzazione semplificativa dei movimenti che vengono effettuati, a causa della presenza di uno o più avversari.
Non vi è quindi uno schema fisso nel tempo ed al moto non può applicarsi il teorema della indipendenza delle azioni simultanee.
Pertanto questi sport dovranno essere studiati con le metodiche proprie della fisica statistica ed utilizzando le tecniche delle teorie del "Chaos" e dei "Frattali".
La Biomeccanica degli Sport di Situazione Duali, ( Sport di Combattimento) di fatto è rimasta per lunghissimo tempo un' area della biomeccanica sportiva non sistematizzata, fino alla completa trattazione del Judo, sviluppatasi tra il 1985 ed il 2010.
Nei tempi attuali è ben noto, che il successo sportivo non è frutto di magia, ma è il risultato di una stretta collaborazione tra scienza e sport, che si produce a monte di ogni prestazione sportiva di alto livello.
Questo fatto oggi noto, era già nella mente di Kano che con la fondazione nel 1932, presso il Kodokan, della Association for the Medical Study of Judo evolutasi poi nel 1948 in the Association for the Scientific Studies on Judo, ha mostrato l'intima connessione tra scienza e Judo Sport .

A  maggior  ragione oggi, vista l'estrema evoluzione delle capacità tecnico-prestative degli atleti, si è evidenziato quanto sia necessario, per gli allenatori degli atleti d'elite, di esser sempre aggiornati per captare i più intimi aspetti evolutivi di questo sport complesso.
Basti pensare alla moderna teoria dell'allenamento che inizia a tener conto della non linearità della risposte degli atleti.
Questa verità appare tanto evidente che nei circoli degli addetti ai lavori si parla oggi di  CKB (Coach Knowledge Based ).  Per essi è in cantiere addirittura un Master Universitaro  Internazionale organizzato dalla Federazione e dall' Università di Roma 2  sotto l'egida dell EJU ( European Judo Union).

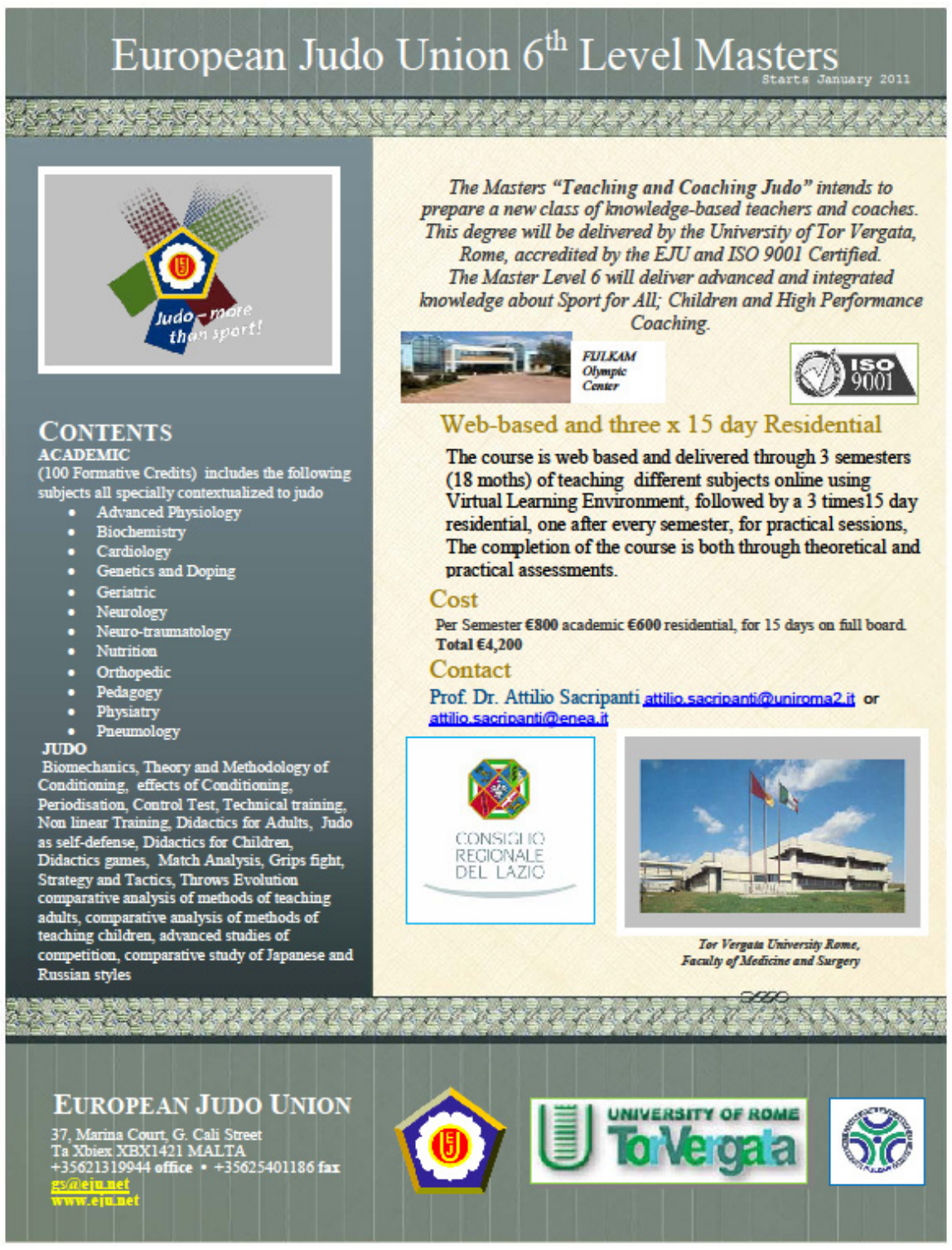

***Fig. 59  Brochure del Master Internazionale  web based "Teaching and Coaching judo"***

Sorge a questo punto spontanea la domanda, ma judo ha perduto il suo smalto Educativo ?
Se ne prendiamo isolatamente il mero aspetto Sportivo, Judo come molti altri Sport Olimpici è divenuto una vera e propria specializzazione di vita per gli atleti, visto l'incremento vertiginoso delle capacità tecnico-agonistiche che si richiedono, ad esempio, per un'Olimpiade, ma ciò non ha nulla a che fare con l'Educazione dell'individuo, come Kano la intendeva.
Proprio su queste riflessioni il Master prevede l'insegnamento avanzato non solo del Coaching per high performance, ma anche metodiche per bambini ed adulti ed in futuro per diversamente abili, al fine di fornire una panoramica completa delle potenzialità di questo sistema Educativo.
Judo, pur se la società ha subito drastici mutamenti,  ha conservato tutto  il suo contenuto educativo in quanto le basi filosofiche individuate da Kano sono ancora valide a tutt'oggi.  Ma tutto ciò è connesso  ovviamente alla necessità di insegnare in ogni palestra il sistema completo, che non è solo la specializzazione allo shiai (combattimento) che viene naturalmente richiesta  per gli atleti di alta qualificazione.
Quindi solo se si riuscirà a conservare le due anime del Judo ( Sport ed Educazione) ogni palestra ritornerà ad essere una fucina sia di Atleti, sia di Uomini adatti alla Società armonicamente strutturata, che Jigoro Kano sognava.

## *Bibliografia*